\documentclass[12pt]{article}

\usepackage{latexsym,amsmath,amscd,amssymb,graphics, amsthm}
\usepackage{enumerate,framed,color}

\usepackage{graphicx}
\usepackage{url}
\usepackage[all]{xy}

\usepackage{relsize}
\newcommand{\bfi}{\bfseries\itshape}

\newcommand{\rem}[1]{}

\newcommand{\pa}{\partial}

\DeclareMathOperator{\ad}{ad}

\DeclareMathOperator{\Emb}{Emb}

\makeatletter

\@addtoreset{figure}{section}
\def\thefigure{\thesection.\@arabic\c@figure}
\def\fps@figure{h, t}
\@addtoreset{table}{bsection}
\def\thetable{\thesection.\@arabic\c@table}
\def\fps@table{h, t}
\@addtoreset{equation}{section}

\makeatother

\newcommand{\quotebox}[1]{\vspace{2 mm}\par \noindent
\begin{center}
\framebox{\begin{minipage}[c]{0.8 \textwidth}
 #1 \end{minipage}}
\end{center}
\vspace{2 mm}\par}

\begin{document}

\newtheorem{theorem}{Theorem}[section]
\newtheorem{definition}[theorem]{Definition}
\newtheorem{lemma}[theorem]{Lemma}
\newtheorem{remark}[theorem]{Remark}
\newtheorem{proposition}[theorem]{Proposition}
\newtheorem{corollary}[theorem]{Corollary}
\newtheorem{example}[theorem]{Example}

\newcommand{\pair}[2]{{\big\langle {#1}\, , \, {#2}\big\rangle}}
\newcommand{\Pair}[2]{{\Big\langle {#1}\, , \, {#2}\Big\rangle}}
\newcommand{\ppair}[2]{{\bigg\langle {#1}\, , \, {#2}\bigg\rangle}}
\newcommand{\Lpair}[2]{{\big( {#1}\, \big| \, {#2}\big)}}
\newcommand{\LPair}[2]{{\Big( {#1}\, \Big| \, {#2}\Big)}}
\newcommand{\Lppair}[2]{{\bigg( {#1}\, \bigg| \, {#2}\bigg)}}

\def\below#1#2{\mathrel{\mathop{#1}\limits_{#2}}}

\newcommand{\lform}[2]{{\big( {#1} \big|\, {#2}\big)}}
\newcommand{\Lform}[2]{{\Big( {#1} \Big|\, {#2}\Big)}}
\newcommand{\scp}[2]{{\Big\langle {#1}\, , \, {#2}\Big\rangle}}
\newcommand{\prt}{\partial}
\newcommand{\sig}{\sigma}
\newcommand{\mR}{{\mathbb{R}}}
\newcommand{\id}{{\mathrm{id}}}
\newcommand{\ti}{\times}
\newcommand{\Ad}{\text{Ad}}
\renewcommand{\ad}{\text{ad}}
\newcommand{\de}{\delta}
\newcommand{\om}{\omega}
\newcommand{\al}{\alpha}
\newcommand{\ga}{\gamma}
\providecommand{\th}{}
\renewcommand{\th}{\theta}
\newcommand{\la}{\lambda}
\newcommand{\be}{\beta}
\newcommand{\Om}{\Omega}
\newcommand{\mg}{{\mathfrak g}}
\newcommand{\CX}{{\mathcal X}}
\newcommand{\cst}{\text{cst}}
\newcommand{\ze}{\zeta}
\newcommand{\gd}{\dot{g}}
\newcommand{\deta}{\dot{\eta}}
\newcommand{\dxi}{\dot{\xi}}
\newcommand{\gp}{g^\prime}
\newcommand{\ginv}{{g}^{-1}}
\newcommand{\bfu}{\mathbf{u}}
\newcommand{\bfw}{\mathbf{w}}
\newcommand{\zhat}{\hat{z}}
\newcommand{\beq}{\begin{equation}}
\newcommand{\beqs}{\begin{equation*}}
\newcommand{\beqa}{\begin{eqnarray}}
\newcommand{\beqas}{\begin{eqnarray*}}
\newcommand{\eeq}{\end{equation}}
\newcommand{\eeqs}{\end{equation*}}
\newcommand{\eeqa}{\end{eqnarray}}
\newcommand{\eeqas}{\end{eqnarray*}}



\title{Geometric dynamics of optimization}
\author{Fran\c{c}ois Gay-Balmaz$^{1}$, Darryl D. Holm$^{2}$,  and Tudor S. Ratiu$^{3}$}
\addtocounter{footnote}{1}
\footnotetext{Control and Dynamical Systems, California Institute of Technology 107-81, Pasadena, CA 91125, USA and Laboratoire de M\'et\'eorologie Dynamique, \'Ecole Normale Sup\'erieure/CNRS, Paris, France.
\texttt{fgbalmaz@cds.caltech.edu}
\addtocounter{footnote}{1}}
\footnotetext{Department of Mathematics, Imperial College London. London SW7 2AZ, UK. \texttt{d.holm@imperial.ac.uk}
\addtocounter{footnote}{1}}
\footnotetext{Section de
Math\'ematiques and Bernoulli Center, \'Ecole Polytechnique F\'ed\'erale de
Lausanne.
CH--1015 Lausanne. Switzerland.
\texttt{Tudor.Ratiu@epfl.ch}
\addtocounter{footnote}{1} }

\pagestyle{myheadings}

\markright{Gay-Balmaz, Holm and Ratiu \hfill  Geometric optimization dynamics \hfill
 \qquad}

\date{}
\maketitle

\makeatother

\maketitle



\begin{abstract} 
This paper investigates a family of dynamical systems arising from an evolutionary re-interpretation of certain optimal control and optimization problems. We focus particularly on the application in image registration of the theory of \emph{metamorphosis}. Metamorphosis is a means of tracking the optimal changes of shape that are necessary for registration of images with various types of data structures, without requiring that the transformations of shape be diffeomorphisms, but penalizing them if they are not.  This is a rich field whose possibilities are just beginning to be developed. In particular, metamorphosis and its related  variants in the geometric approach to control and optimization can be expected to produce many exciting opportunities for new applications and analysis in geometric dynamics.
\end{abstract}

\tableofcontents

\section{Introduction}
 
With the advent of new devices capable of seeing objects and structures not previously imagined, the realm of science and medicine has been extended in a multitude of different ways. The impact of this technology has been to generate new challenges associated with the problems of formation, acquisition, compression, transmission and analysis of images. These challenges cut across the disciplines of mathematics, physics, computational science, engineering, biology, medicine, and statistics. 
{}
For example, in computational anatomy (CA) biomedical images are compared quantitatively by calculating the ``distance'' between them, along a path that is \emph{optimal} in transforming one such image to another. The optimal path is traversed along a curve of deformations in the group of smooth invertible maps with smooth inverses (i.e., the diffeomorphisms) and it is governed by a partial differential equation (PDE) called the EPDiff equation. In particular, EPDiff governs the geodesic flow on the group of diffeomorphisms, with respect to any prescribed metric.   This flow from one shape to another also has an evolutionary interpretation that invites ideas from the analysis of evolutionary equations. In particular, the \emph{momentum map} for EPDiff identified first in \cite{CaHo1993} and explained more completely in \cite{HoMa2004} yields the canonical Hamiltonian formulation of the dynamics of the singular evolutionary solutions of EPDiff. Moreover, in an optimization sense, this momentum map also provides a complete representation of the landmarks and contours (outlines) of images to be matched, in terms of the canonical positions and momenta associated with the evolutionary interpretation \cite{HoRaTrYo2004}. In addition, it provides a natural strategy for finding the optimal path between two configurations of either landmarks or contours \cite{YoArMi2009}. Thus, the momentum map (a concept from Hamiltonian systems) is crucial in the construction of an isomorphism between the data structures used in the optimal  matching of images and the evolutionary singular solutions of the EPDiff equation. This isomorphism has already suggested new dynamical paradigms for CA, as well as new strategies for assimilation of data in other image representations, for example, as gray-scale densities \cite{HoTrYo2009,YoArMi2009}. The converse benefit may also develop, in which methods of optimal control and optimization of data assimilation used in image matching for CA may suggest new strategies for investigating  dynamical systems of evolutionary PDE. In short, the variational formulations, Lie symmetries and associated momentum maps encountered in applications of EPDiff have led to a convergence in the analysis of both its evolutionary properties and its optimization equations.
{}

This paper focuses on the evolutionary aspects of the PDE that are summoned by adopting a dynamical interpretation of the optimal control and optimization methods used in the registration of various types of images. The paper does not perform any applications of optimization methods to image registration, nor does it develop any numerical algorithms for making such applications. Instead, the paper re-interprets the endeavor of image registration from a dynamical systems viewpoint.
In particular, as we shall explain, a recent development in the large deformation diffeomorphic matching methods (LDM), in an approach for image registration called \emph{metamorphosis}%
\footnote{Although the term ``metamorphosis'' has a precise mathematical definition that will be given below, it also satisfies its proper dictionary definition, as ``a change of physical form, structure or substance''. This paper interprets the change as a type of evolution.}
\cite{MiYo2001,TrYo2005,HoTrYo2009} introduces a new type of  evolutionary equation that may be called \emph{optimization dynamics}. In following this line of reasoning, the geometric mechanics approach for evolutionary PDE provides a framework that we hope will inform both optimization and dynamics. 
{}
The primary example in the line of reasoning leading to optimization dynamics is the EPDiff equation \cite{HoMaRa1998a,HoMaRa1998b,YoArMi2009}.   

\subsection*{A brief history of  the EPDiff equation} EPDiff stems from the recognition by Arnold in \cite{Arnold1966} that incompressible fluid dynamics could be characterized as geodesic flow in the group of volume preserving diffeomorphisms, with respect to the kinetic energy metric ($L^2$ norm of the fluid velocity). A few years later, the one-dimensional compressible version of EPDiff reappeared as the dispersionless limit of the Camassa-Holm (CH) equation \cite{CaHo1993}. The CH equation is a completely integrable evolution equation for shallow water waves, whose soliton solutions develop sharp peaks in the dispersionless limit. Its peaked soliton solutions (peakons) correspond to concentrations of momentum into delta-function singularities and are solutions of EPDiff in one dimension with the $H^1$ kinetic energy metric. Slightly later, the incompressible version of EPDiff with the $H^1$ kinetic energy metric was generalized to higher dimensions in \cite{HoMaRa1998a,HoMaRa1998b} by using its symmetry-reduced variational principle, and was interpreted as Euler's fluid equations, averaged following Lagrangian particle trajectories. This interpretation soon led to the introduction of viscosity and some interesting applications of the resulting viscous equations as a turbulence model by Chen et al. \cite{Chen-etal1998,Chen-etal1999}. 

Around the same time, EPDiff arose independently in a completely different context. Namely, it arose as the governing equation in the optimization problem for large deformation diffeomorphic matching (LDM) in image registration \cite{Trouve1995,Trouve1998,Yo1998}. The recognition that EPDiff was arising in these two different contexts provided a fruitful  opportunity for dual interpretations of the solutions of the same equation. In particular, the ``peakons'' of the CH equation in the water wave context were soon recognized to be the ``landmarks'' in images in the LDM context. Since then, the two types of problems have continued their optimization-dynamics interplay and have been found to inform each other, while also showing  intriguing differences and similarities that arise in their dual formulations as initial value problems on one hand and boundary value problems on the other. In particular, the concept of symmetry reduction and momentum maps from geometric mechanics that had previously been applied so effectively in fluid dynamics \cite{Arnold1966} and shallow water soliton theory \cite{CaHo1993},  has recently been recognized as a unifying  approach for developing multi-mode LDM methods for images whose  data structure may comprise arbitrary tensors, or tensor densities \cite{BrGBHoRa2011}. This is a rich and rapidly developing area of science, for which a complete literature review would be beyond our scope here. 

A convergence of these two independent endeavors has led to dual interpretations of the same equation and the same key ideas in such different but complementary contexts. This convergence is fascinating, and we continue our investigation of it here. In the present paper, we emphasize the dynamical interpretations of the equations and approaches that are applied in optimal image matching. This is not to say that we solve optimal matching problems for images at all in this paper. Rather, being cognizant of the ideas and variational formulations underlying the optimal matching approach, we shall apply these formulations to study certain classes of equations that arise in the problem of image registration, not from the viewpoint of optimization, but rather from the evolutionary viewpoint of \emph{geometric mechanics} \cite{HoScSt2009,MaRa1999}.

The geometric mechanics approach emphasizes Lie group actions on manifolds, momentum maps, and reduction by symmetry. This approach leads to an understanding of certain  classes of control and optimization problems as systems of evolutionary equations. In particular, the Lie symmetry ideas underlying the process of optimal image assimilation known as \emph{metamorphosis} 
\cite{MiYo2001,TrYo2005,HoTrYo2009} in combination with the evolutionary geometric mechanics viewpoint leads the family of EPDiff equations into the realm of \emph{optimization dynamics}. Optimization dynamics extends the previous association of image matching ideas with soliton theory \cite{HoRaTrYo2004} to produce new results, such as the derivation and re-interpretation of the two-component CH system (CH2) as a equation for the dynamics of metamorphosis of gray-scale images \cite{HoTrYo2009}. The CH2 system is a completely integrable evolutionary system of equations that was recently discovered using isospectral methods for solitons \cite{ChLiZh2006}. Its inverse scattering transform is discussed in \cite{HoIv2011}. Recognizing that some systems of equations arising in optimization dynamics for image analysis may be associated with soliton theory raises many questions about the mathematical properties of these systems and their solutions, particularly when the equations are nonlocal. For example, the initial value problems for some of the nonlocal equations obtained in optimization dynamics investigated here allow \emph{emergent} singular solutions, in which the evolution of a smooth, spatially confined, initial condition becomes singular by concentrating itself  into delta function distributions.  In particular, EPDiff has that property and so does the corresponding system of equations for the optimization dynamics of metamorphosis. See \cite{HoMaRa1998a,HoMaRa1998b,HoScSt2009} and \cite{Yo2010,YoArMi2009}, respectively, for further discussions of EPDiff from the different but complementary viewpoints of geometric mechanics and image matching.
{}

\subsection{LDM approach, EPDiff, and momentum maps}
The LDM approach is based on minimizing the sum of a time-integrated \emph{kinetic energy metric} whose value defines the length of an optimal deformation path, plus a {\it penalty norm} that ensures an acceptable tolerance in image mismatch. (The matching cannot be exact because of the unavoidable errors that arise in real applications.) LDM approaches were introduced and systematically developed in Trouv\'e \cite{Trouve1995,Trouve1998}, Dupuis et al. \cite{DuGrMi1998}, Joshi and Miller \cite{JoMi2000}, Miller et al. \cite{MiYo2001,MiTrYo2002}, Beg \cite{Beg2003}, and Beg et al. \cite{Beg-etal-2005}. The LDM  approaches of those papers are based on Grenander's \emph{deformable template} paradigm for image registration \cite{Gr1993}. Grenander's paradigm, in turn, is a development of a biometric strategy introduced  by D'Arcy Thompson \cite{Thompson1917} of comparing a template image $I_0$ to a target image $I_1$ by finding a smooth invertible transformation of coordinates th
 at maps one image to the other. This transformation is assumed to belong to a Lie group $G$ of diffeomorphisms that acts on the set of templates containing $I_0$ and $I_1$. The effect of the transformation on the data structure that is encoded in the set of templates is called the \emph{action} of the Lie group $G$ on the set of images. The optimal path in the transformation group is the one that costs the least in time-integrated kinetic energy for a given tolerance. This concept of optimization summons a control theory approach into the analysis and registration of images.
\medskip

In applications of the LDM approach, the optimal transformation path is often sought by using a variational optimization method such as the one developed in \cite{DuGrMi1998,Trouve1995,Trouve1998}.
Using this method, the optimal path for the matching transformation in this problem is obtained from a gradient-descent algorithm based on the Euler-Lagrange equation arising from stationary balance between kinetic energy and tolerance. This gradient-descent approach does indeed determine an optimal matching path. However, from the viewpoint of dynamical systems theory, it misses the following potentially interesting question: 
\quotebox{\it
What information and perspective might be obtained by interpreting the Euler-Lagrange equations associated to the LDM approach from a dynamical systems viewpoint?}

The answer to this question may be sought by interpreting the variational optimization method in the LDM approach as a form of  Hamilton's principle. Hamilton's principle for the variational construction of optimal paths with minimal kinetic energy for a given tolerance in image mismatch yields an associated set of Euler-Lagrange equations that may then be given an evolutionary interpretation. The optimal solutions of these equations have been investigated as evolutionary motion on the Lie group of diffeomorphisms in the \emph{absence} of additional penalty terms by Arnold \cite{Arnold1966,Arnold1989}, Holm et al. \cite{HoMaRa1998a,HoMaRa1998b}, Marsden and Ratiu \cite{MaRa1999}, and for the particular application to template matching in Miller et al. \cite{MiTrYo2002}.  As mentioned earlier, the optimal paths in these cases are geodesics with respect to the metric provided by the  kinetic energy. The kinetic energy for LDM is invariant under right translations on the diffeomorphism group. Reducing Hamilton's principle with respect to this symmetry and then invoking the Euler-Poincar\'e theory applied to diffeomorphisms produces an evolution equation known as the \emph{EPDiff equation} \cite{HoMaRa1998a,HoMaRa1998b}, whose derivation in the present context is explained in Section \ref{sec: EPDiff}. 

\medskip

The solution of the EPDiff equation yields the spatial representation of the
geodesic velocity, i.e., the tangent vector to the optimal path of deformations along which the minimal distance from one image to another is measured. The geodesics themselves may be obtained from the solutions of EPDiff for the velocity by a reconstruction process that inverts the previous reduction by symmetry after the solution to the EPDiff equation for velocity has been obtained.  This is analogous to the reconstruction process in classical mechanics that recovers the symmetry coordinate conjugate to a conserved momentum as the final step in the solution, after the other degrees of freedom have been determined in the reduced space. 
\medskip

Composing the evolutionary solutions of EPDiff with the reconstruction process provides an important representation of diffeomorphisms that relates the endpoint of a geodesic to the initial value for \emph{momentum} in the EPDiff equation. This relation is the momentum representation of the deformation. The long-time existence of this representation is based on conservation by EPDiff of the kinetic energy norm, which may be chosen so that its boundedness affords enough smoothness on the velocities to ensure the long-time existence of solutions of EPDiff. In this case, EPDiff admits emergent weak momentum solutions; for example, delta-function distributions of momentum that emerge from smooth, spatially confined initial conditions \cite{CaHo1993,HoMa2004}.  This singular behavior is well understood analytically only in certain one-dimensional cases. In particular, it is understood for the completely integrable case of the Camassa-Holm equation, see, e.g., \cite{deLeKaTo2007,Mi2002} and references therein. 

\medskip

The EPDiff equation is of central importance in computational anatomy \cite{YoArMi2009}. This is because the optimal paths sought by LDM on the image template space defined on a manifold $M$ are inherited from the geodesics on ${\rm Diff}(M)$, the Lie group of diffeomorphisms acting on the manifold $M$. These, in turn, are governed by EPDiff. Consequently, any solution of the LDM problem for optimal geodesics must involve EPDiff \cite{YoArMi2009}. Conversely, solving the LDM problem directly produces the momentum representation of the optimal diffeomorphism. The momentum representation arising from this evolutionary interpretation is then available for analyzing anatomical data sets. In any case, despite the disparate forms that the geodesic equations may take for the various data structures in the various types of images, all of them are instances of EPDiff with the corresponding representation for momentum. The specific representation for momentum in terms of the image data structure in a given case is called the \emph{momentum map}. The momentum map for images is another dynamical systems concept that emerges as a central feature in this paper.  The EPDiff equation and its associated momentum map for various image data structures are discussed in Section \ref{sec: EPDiff}.
\medskip

An interesting example of the momentum map relating solutions of LDM to solutions of EPDiff arises for the case of \emph{landmark} data structure, in which the momentum is singularly concentrated at points. The relation between these singular geodesic solutions and evolutionary soliton solutions, called \emph{peakons} for a shallow water wave equation introduced in Camassa and Holm \cite{CaHo1993}, has been examined in the context of computational anatomy in Holm et al. \cite{HoRaTrYo2004}. A numerical analysis of the stability of these equations is also given in McLachlan and Marsland \cite{McLaMa2007}. See also Micheli \cite{Micheli2008} for other recent developments involving the curvature of the space of landmark shapes. Holm and Marsden \cite{HoMa2004} explain that two independent momentum maps for EPDiff are available in the case that the image  data structure comprises the manifold $\Emb(S^1, \mathbb{R}^2)$ of embedded closed curves (embedded images of $S^1$) in the plane $\mathbb{R}^2$. The left action of the group of diffeomorphisms $\operatorname{Diff}(\mathbb{R}^2)$ of the plane  deforms the curve by a smooth invertible  transformation of the coordinate system in which it is embedded, while leaving the parameterization of the curve invariant. The right action of
the group of diffeomorphisms $\operatorname{Diff}(S^1)$ of the circle
corresponds to smooth invertible reparameterizations of the 
domain $S^1$ of the coordinates of the curve.
In this case, one momentum map corresponds to action from the left by the diffeomorphisms on $\mathbb{R}^2$, the other to their action from the right on the embedded curves. Optimal control and  reparameterization methods for matching closed curves in the plane using these two momentum maps for the space of closed curves in the plane have recently been developed in Cotter and Holm \cite{CoHo2010}. 


In summary, LDM  image analysis  is based on optimization methods that are formulated as  boundary value problems. However, the re-interpretation of their governing equations as evolutionary systems by using symmetry reduction of the corresponding Hamilton's principle allows various concepts from dynamical systems theory to be profitably applied in the solution and interpretation of image analysis problems. Thus, the transfer of concepts and ideas between these two fields  in the context of image registration has the potential to enrich them both.

\subsection{Distributed optimization dynamics, or evolutionary metamorphosis}
As we have been discussing, the paper focuses on the geometric dynamics interpretation of the optimization problems designed for image registration.
However, rather than concentrating on the development of \emph{solutions} of optimization problems, the treatment here focuses on the \emph{dynamics} that are produced in applying the method of reduction by Lie group symmetry to families of optimization problems posed in a geometric setting. This is a new arena for geometric dynamics and several new departures are being taken. Among these new departures is the investigation of the evolutionary dynamics that arises when distributed or nonlocal \emph{penalties} are imposed in Hamilton's principle, rather than local constraints.  For lack of a better name, we call this sort of problem \emph{distributed optimization dynamics}.
It is the evolutionary counterpart of the \emph{metamorphosis} approach in imaging science \cite{MiYo2001,TrYo2005,HoTrYo2009}, which, in turn, is a modification and development of LDM 
{}
that allows the evolution $n(t)$ of the image template to deviate from pure deformation. That is, metamorphosis only \emph{penalizes} the spatial average of the deviation away from the infinitesimal action of the vector fields on an image manifold, rather than \emph{enforcing} it as a local pointwise constraint. This approach, in turn, modifies the EPDiff equation and thereby introduces a wealth of new structure and new examples that we shall investigate in this paper. 

An explicit comparison for the case that the image templates are gray-scale density distributions  may help to understand the difference between the LDM approach and the metamorphosis approach.\medskip

\noindent
{\bf LDM approach:} Given the source and target templates for the images characterized as scalar densities $n_0$ and $n_T$ at the initial time $t=0$ and the final time $t=T$, respectively, minimize the quantity 
\begin{equation}\label{LDM_functional} 
\int_0^T\ell(u(t))dt+ \frac{1}{2 \sigma ^2}\|  n_0\circ \eta _T^{-1} - n_T \| ^2_{L^2},
\end{equation} 
over the time dependent vector field $u(t)$, where $ \eta _T $ is the flow of $u(t)$ evaluated at time $t=T$, and the formula
\[
\dot n (t) + {\rm div} \big( n(t) u(t) \big) = 0
\]
is its infinitesimal action on a smooth density $n(t)=n_0\circ \eta _t^{-1}$ defined over time $0\le t\le T$ on the domain of flow.\medskip

\begin{center}
\begin{figure}
\includegraphics[width=0.15\textwidth]{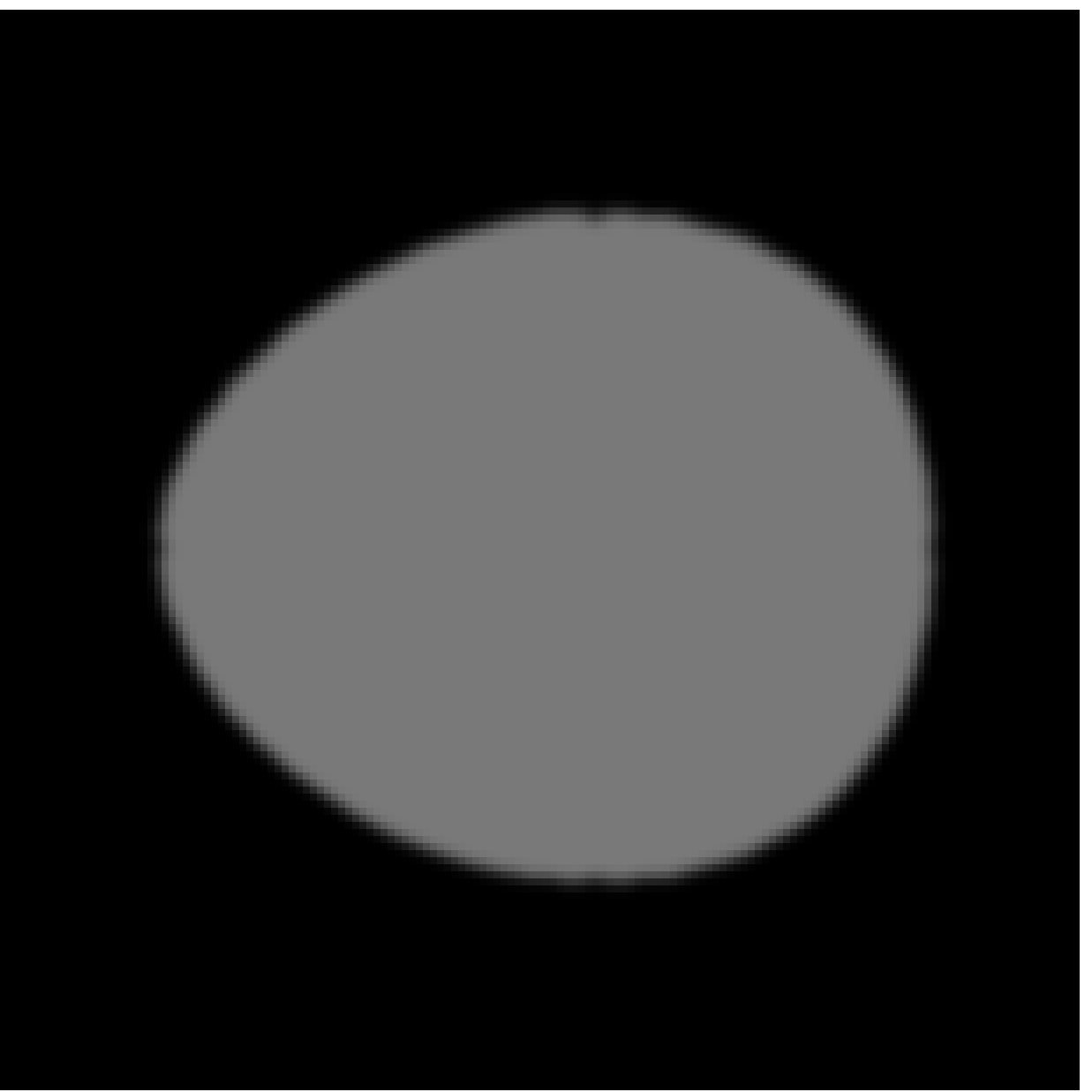} \includegraphics[width=0.15\textwidth]{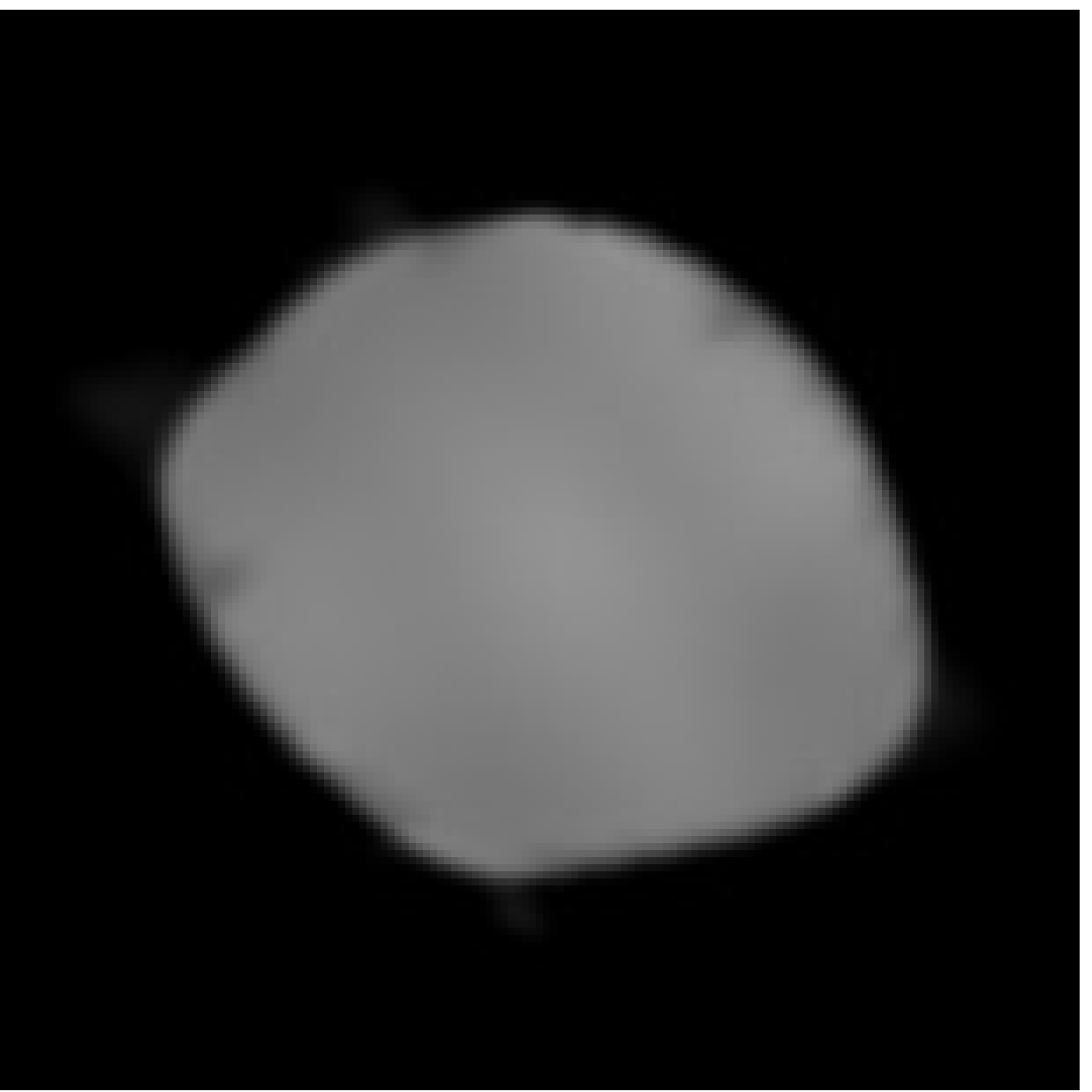} 
\includegraphics[width=0.15\textwidth]{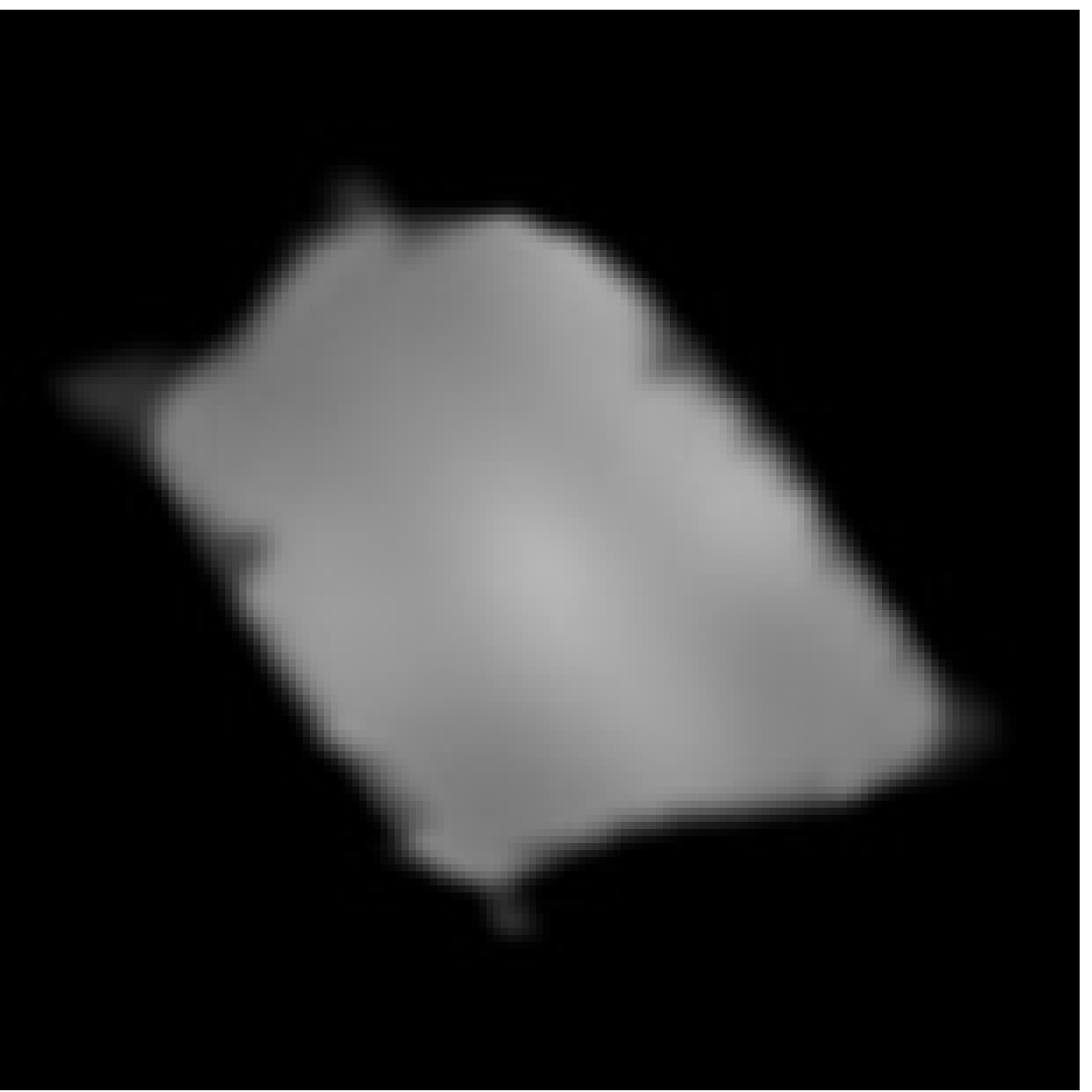} 
\includegraphics[width=0.15\textwidth]{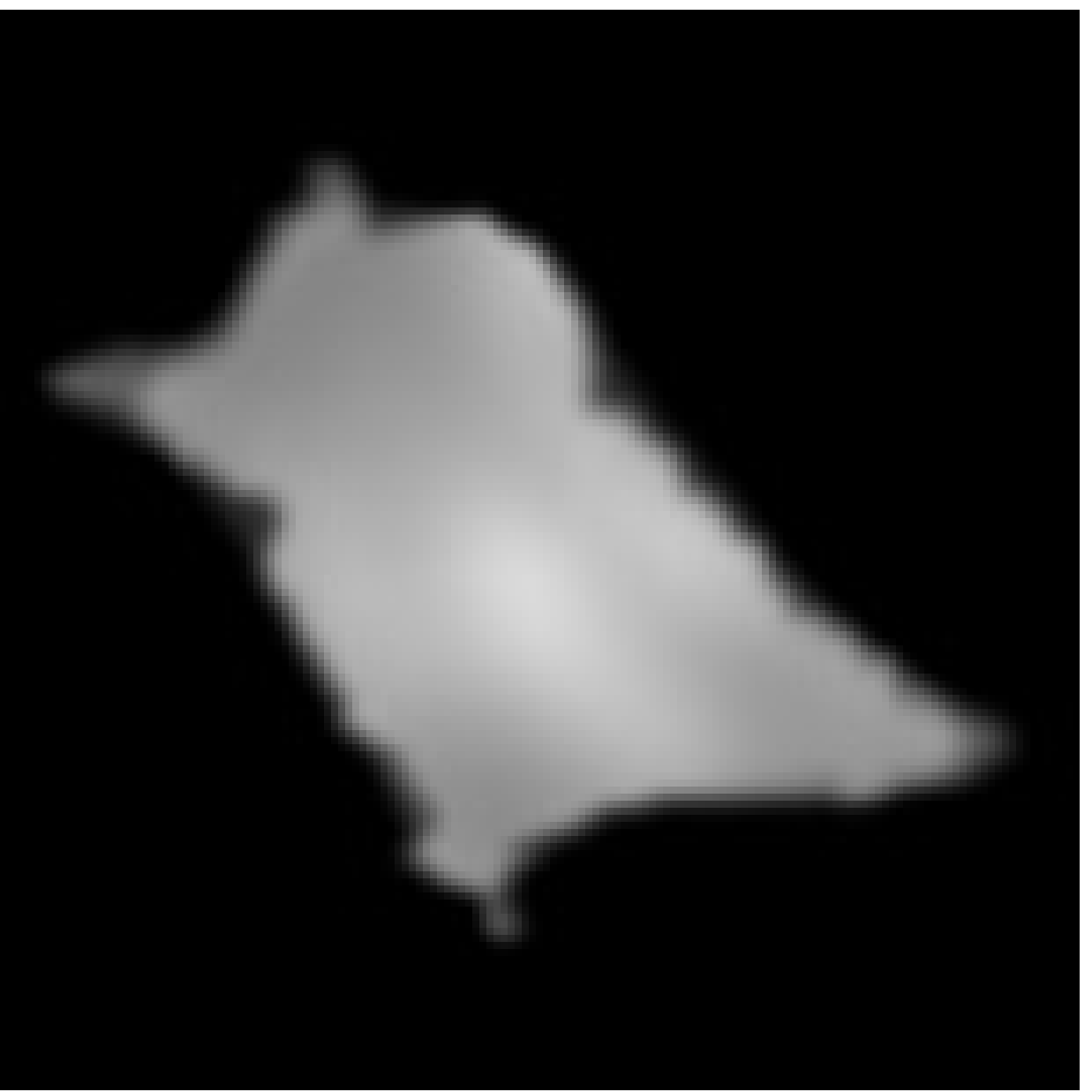} 
\includegraphics[width=0.15\textwidth]{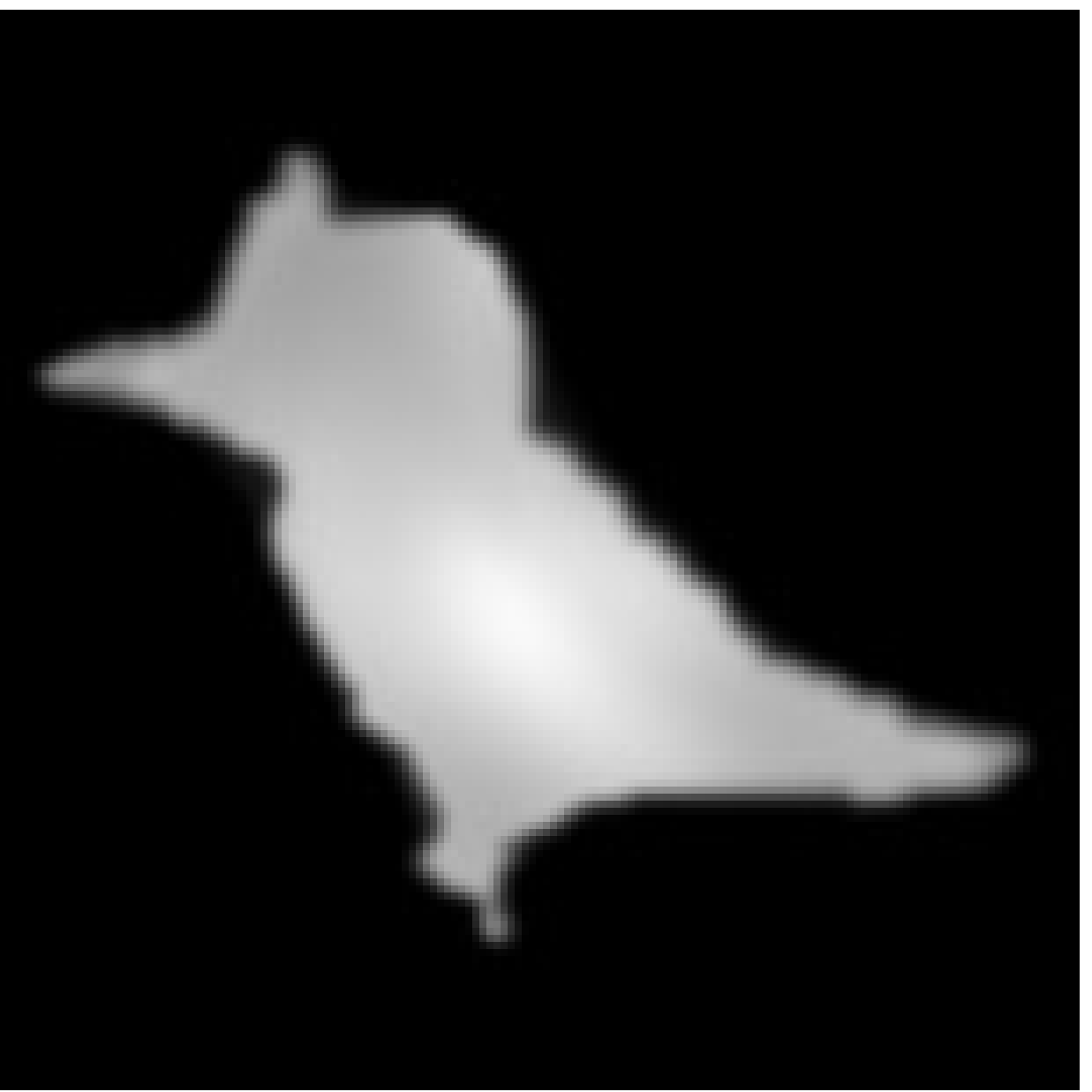} 
\includegraphics[width=0.15\textwidth]{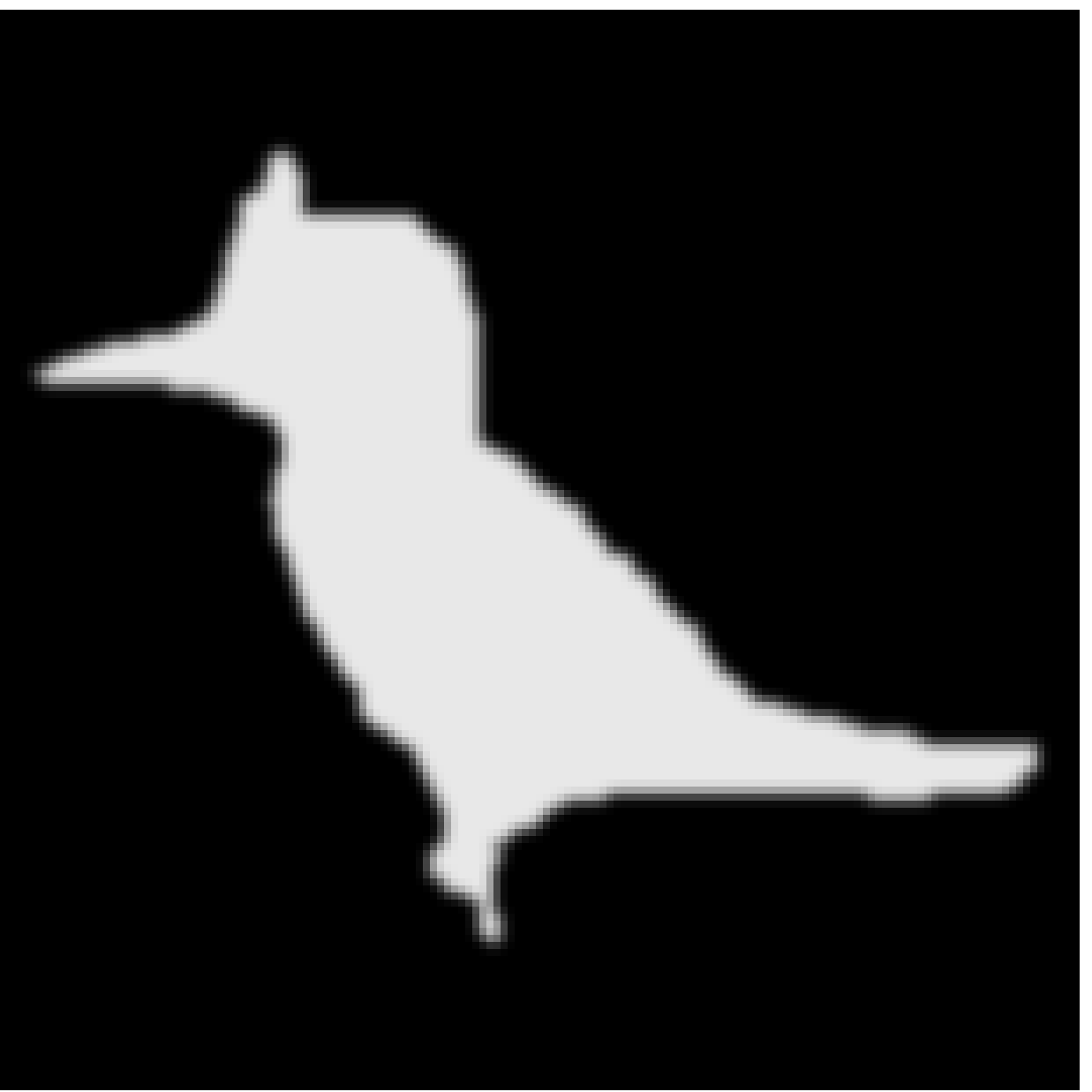} 
\\
\includegraphics[width=0.15\textwidth]{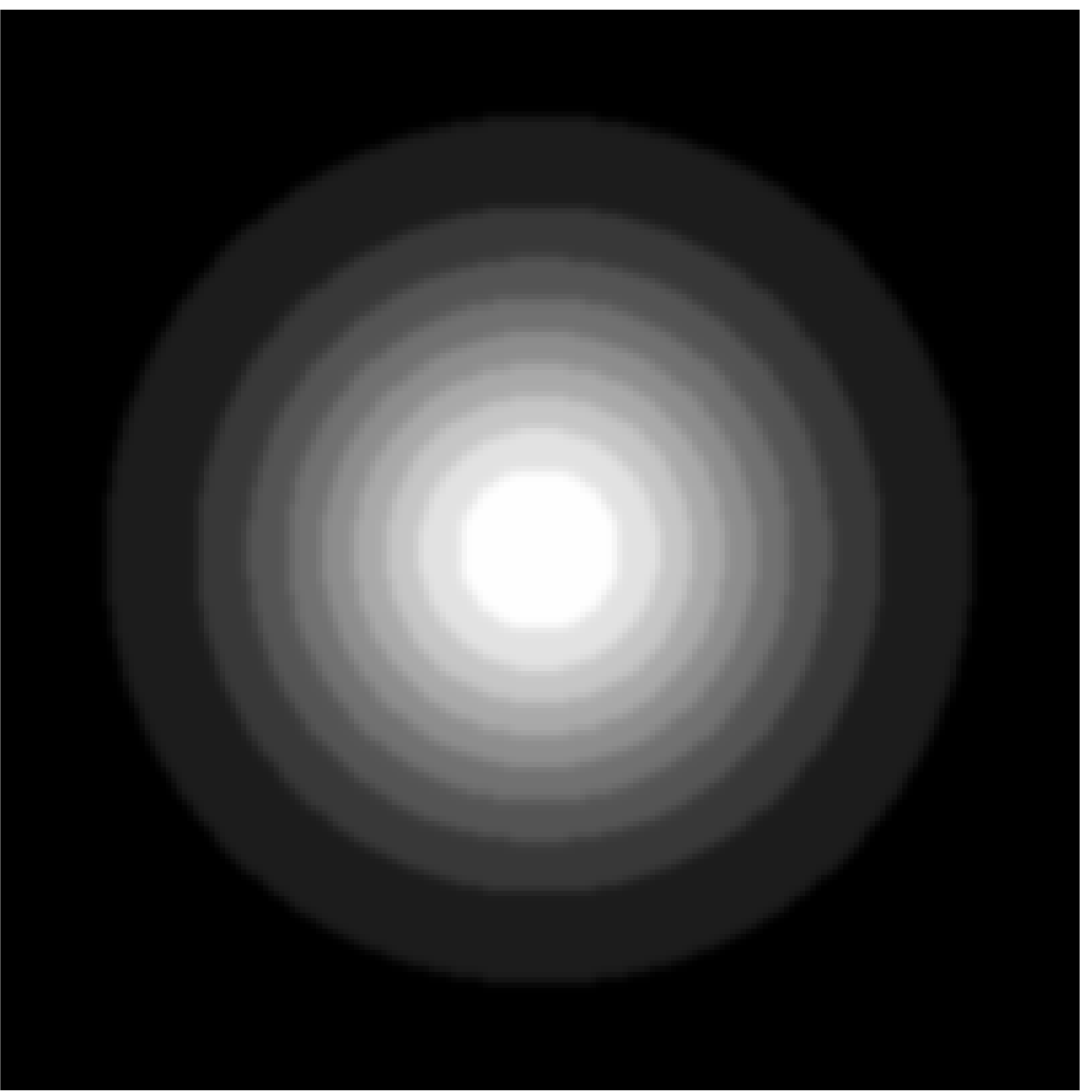} 
\includegraphics[width=0.15\textwidth]{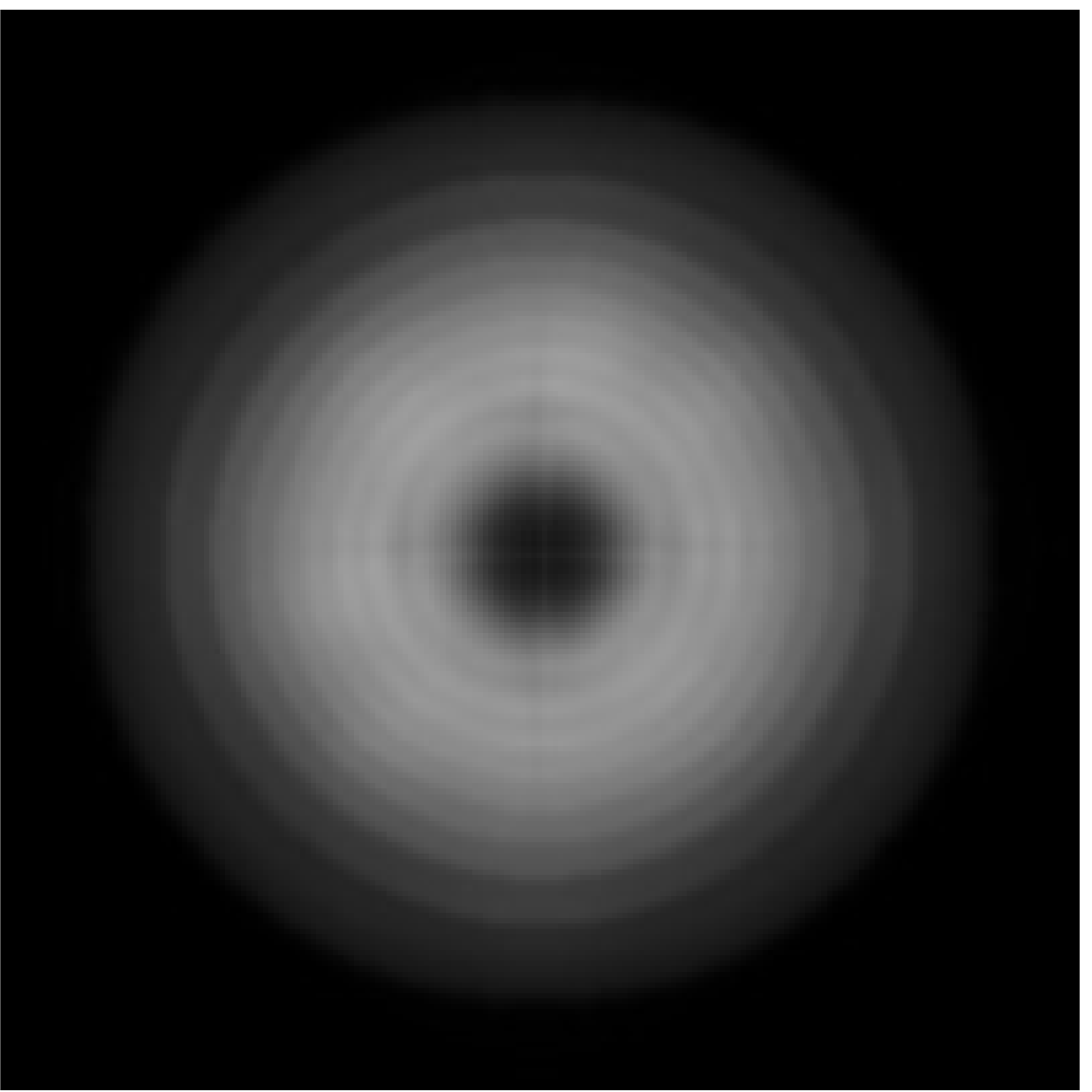} 
\includegraphics[width=0.15\textwidth]{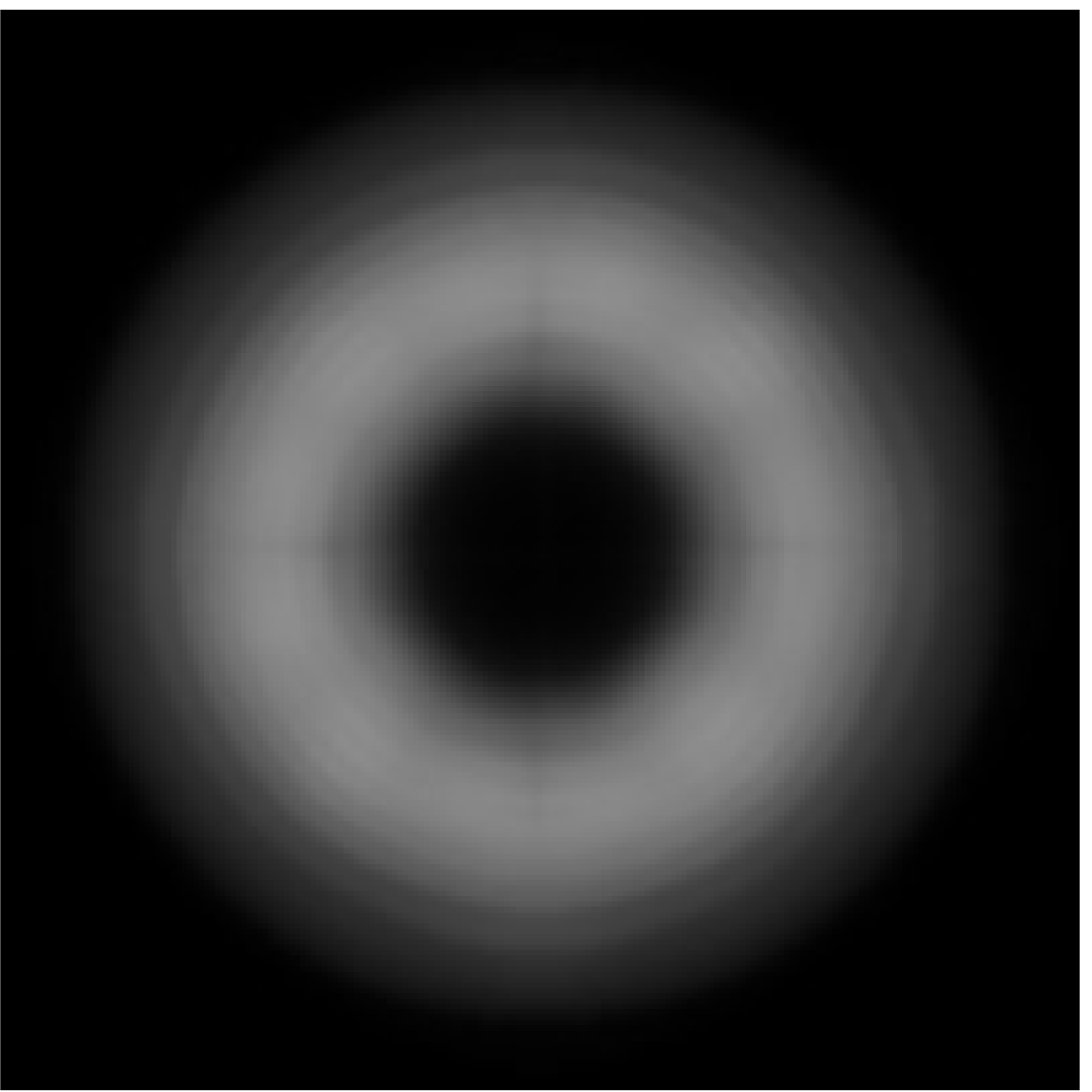} 
\includegraphics[width=0.15\textwidth]{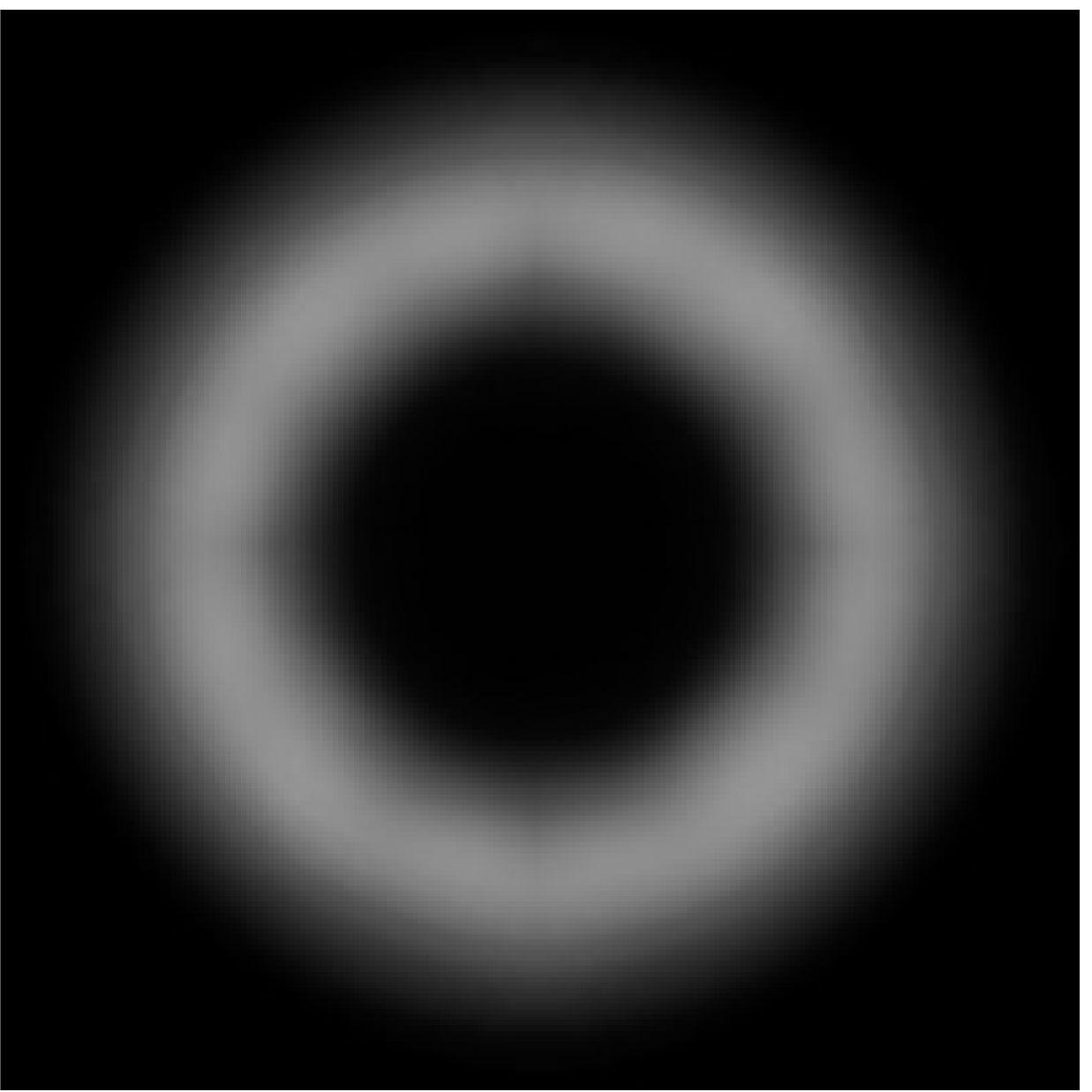} 
\includegraphics[width=0.15\textwidth]{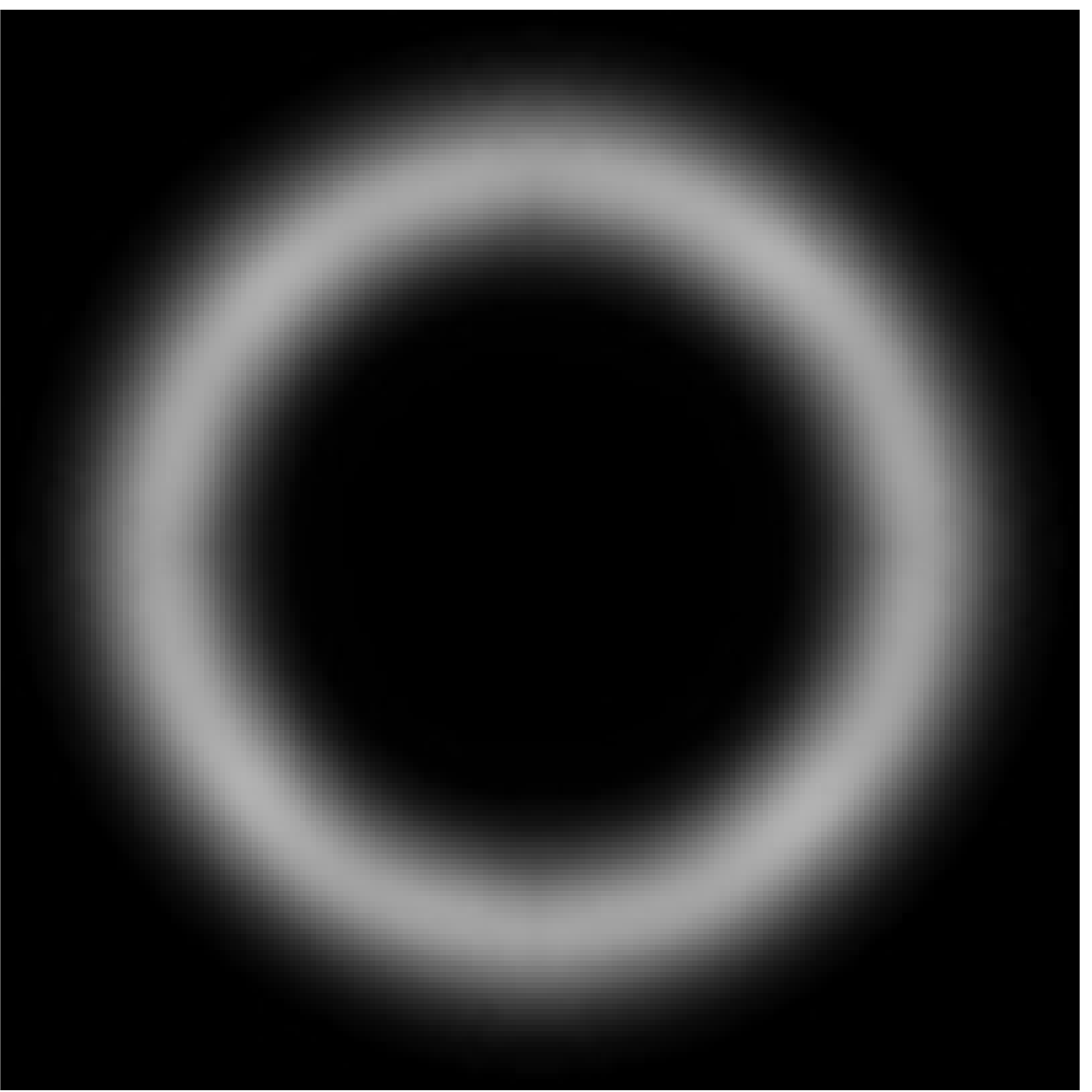} 
\includegraphics[width=0.15\textwidth]{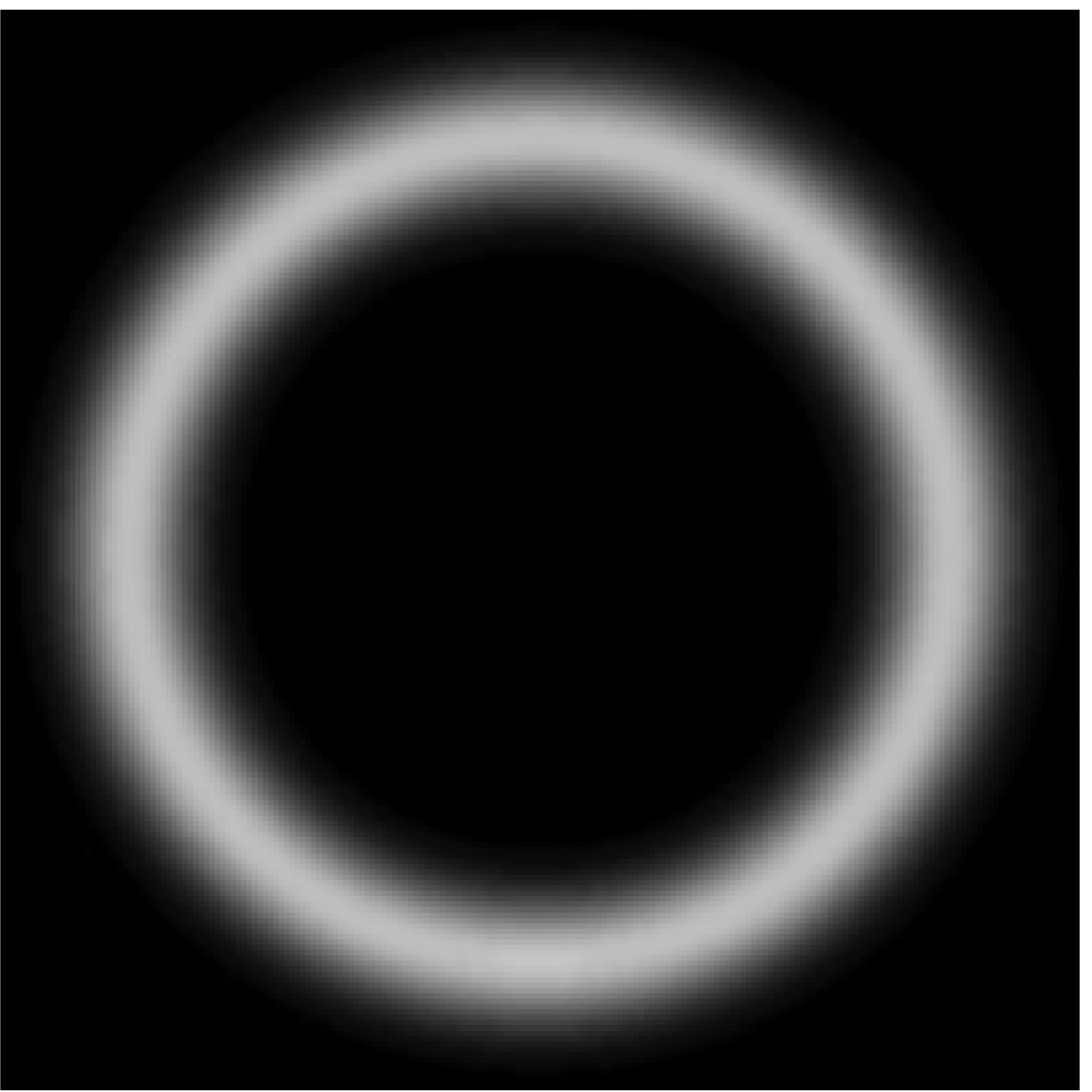} 
\caption{\label{fig:meta.dens} {}These gray-scale images show optimal metamorphoses between two density distributions with equal total mass from \cite{HoTrYo2009}. The optimization approach would compute the distance along the optimal path between  between the first and last density in each row. In the evolutionary approach, the optimal \emph{trajectories} for $n(t)$  are computed. The images between the endpoints show snapshots along the optimal path $n(t)$ in each row at intermediate points in time. In particular, the second row shows that metamorphosis allows a change in topology along its optimal path. Our interest focuses on the evolutionary equations for the process of metamorphosis. The dynamical system of metamorphosis equations obtained in registering such gray-scale image densities is given in Section \ref{examples_section} as one of the examples of the general approach. In one dimension, the metamorphosis equations for this class of images comprises a completely integrable Hamiltonian system. }
\end{figure}
\end{center}

\noindent
{\bf Metamorphosis approach:} Given $n_0$ and $n_T$, minimize
\begin{equation}\label{metamorphosis_functional}
\int_0^T\left( \ell(u(t))+ \frac{1}{2 \sigma ^2} \|\dot n (t) + {\rm div} \big( n(t) u(t)\big)\|^2_{L^2} \right) dt
\end{equation} 
over time dependent vector field $u(t)$ and scalar densities $n(t)$. As one sees in Figure \ref{fig:meta.dens} for the metamorphosis of shapes characterized as densities, the term ``metamorphosis'' introduced in \cite{TrYo2005} for this process can be understood in practice by its ordinary meaning, as ``change of shape'', such as the gradual and continuous metamorphosis of a tadpole into a frog. 
{}

The paper begins by contrasting optimal control problems with distributed optimization problems in a geometric setting. In particular, we discuss the geometric properties of Lie algebra controls acting on state space manifolds. The latter optimal control approach parallels the familiar Clebsch variational formulation of dynamical equations continuum mechanics (e.g., \cite{HoKu1983}). In fact, continuum mechanics was one of the early paradigms for image registration \cite{Yo1998}. The  Clebsch variational formulation of continuum mechanics has recently been developed and applied in the study of the dynamical aspects of optimal control problems in a geometric setting (see \cite{GBRa2011,Ho2009}). Conversely, our concern here is to continue this parallel development by studying the implications for dynamics of the geometric approach to \emph{distributed} optimization problems.

\subsection{Plan and main contributions of the paper} 

In the remainder of the paper, we compare the dynamical equations that arise from optimal control problems with those arising from distributed optimization.  This comparison provides several examples of how the two approaches differ and, in particular, how their dynamical equations differ when their variational problem is regarded as Hamilton's principle for the dynamics. Their comparison also identifies the aspects of these approaches that are fundamentally the same. Section \ref{sec: Standard OptConProbs} begins by explaining the dynamical set up for standard optimal control problems treated by the Pontryagin Maximum Principle. 
Section \ref{GXM-eg} provides several examples illustrating the consequences of applying Lie group controls acting on state manifolds by using the Clebsch framework for optimal control. These examples introduce the \emph{momentum map} for the cotangent-lifted action of the Lie group controls on the state manifold. The  cotangent-lift momentum map is a fundamental concept in the application of geometric mechanics methods in the Clebsch framework for optimal control. It turns out that the same momentum map is also the organizing principle for the distributed optimization dynamics introduced in Section 
\ref{distributed_Optimiz}. After establishing  this background for our comparison of optimization and dynamical systems methods, Section \ref{sec: overview} provides an overview of the rest of the paper. 
\medskip

Section \ref{sec_OCP} begins by reviewing  the Clebsch framework for optimal control problems introduced and studied in  
\cite{GBRa2011}. A new class of optimization problems is then introduced which is the subject of study of this paper. The stationarity conditions are obtained and the associated equations of motion are determined. Inspired by the extremum problems presented earlier, Section \ref{sec_LP_and_metamorph} presents two Lagrangian reduction procedures for Lagrangian functions defined on $T(G \times Q)$, where $G $ is a Lie group acting on the manifold $Q $.  These reduction methods are used in Section \ref{sec_Lagr_approach} to rederive the equations of motion that were found in Section \ref{sec_OCP}. Hamiltonian reduction is carried out in Section 
\ref{section_HP_metamorphosis}. As before, there are two reduction methods and, in the case of a representation, one of them leads to Lie-Poison equations with a symplectic cocycle on the dual of a larger semidirect product Lie algebra. In Section \ref{optimal_hamiltonian_section} we apply these Hamiltonian reduction methods to the optimization problems introduced earlier. Section \ref{examples_section}, by far the longest of the paper, presents a number of examples. We begin by studying examples where $G $ is represented on a vector space. The concrete examples treated are the heavy top and a class of problems using the adjoint representation. For example, we
find a modification of the pair of double bracket equations studied in \cite{BlBrCr1997}, \cite{BlCr1996}. Next, we study optimization problems associated to affine actions. Actions by group multiplication is the next topic. The concrete examples include the $N $-dimensional free rigid body, Euler's equations for an ideal incompressible homogeneous and for a barotropic fluid. The $N $-dimensional Camassa-Holm equation is presented from this optimization point of view, inspired by the construction of singular solutions. Finally, the optimization problem is used to obtain the equations of metamorphosis dynamics for use in computational anatomy. Section \ref{sec-ConclusionsOutlook} briefly summarizes the paper and gives an outlook for future work. 

\section{Review of optimal control problems}\label{sec: Standard OptConProbs}

\subsection{Definitions}
We begin by recalling the definition of optimal control problems.
\medskip

\begin{definition}{\rm (}Optimal control problems{\rm)}
\label{def_optimal_control}
A standard \textit{optimal control problem} comprises:
\begin{itemize}
\item a differentiable manifold $Q$ on which \textit{state variables} $n\in Q$ evolve in time $t$ during an interval ${I}=[0,T]$ along a curve $n: {I} \to Q$  from $n(0)=n_0$ to $n(T)=n_T$, with specified values $n_0,n_T\in Q$;
\item a vector space $U$ of \textit{control variables} ${u}\in U$ whose time dependence ${u}: I\to U$ is at our disposal to affect the evolution  $n(t)$ of the state variables;
\item a smooth map $F:Q\times U\to TQ$ such that $F(\cdot , u) : Q \rightarrow TQ $ is a vector field on $Q$ for any $u \in U $ whose associated evolution equation%
\footnote{The over-dot notation in $\dot{n}$ means time derivative. Several forms of time derivative appear in applications and the meaning should be clear from the usage.  Besides the over-dot notation, we shall use the equivalent notation $d/dt$ to mean either partial or ordinary time derivative in the abstract formulas, as needed in the context. For fluids, we shall also use $\partial_t$ for the Eulerian time derivative at fixed spatial location. Finally, the covariant time derivation on a Riemannian manifold will be denoted as $D/Dt$.}

\begin{equation}\label{statesyst-eqn}
\dot{n}=F(n, {u})
\end{equation}
relates the unknown state and control variables $(n(t), {u}(t)): I \to Q\times U$;
\item a \textit{cost functional} depending on the state and control variables
\begin{equation}\label{S}
S:=\int_0^T \ell(u(t),n(t))\,dt,
\end{equation}
subject to the prescribed initial and final conditions, at $n(0)=n_0$ and $n(T)=n_T$. 
The integrand $\ell: Q\times U\to \mathbb{R}$, called the \textit{Lagrangian}, is assumed to be  $C^1$ on $Q\times U$.
\end{itemize}
The goal of the optimal control problem is to find the evolution  $(n(t), {u}(t))$ of the state and control variables such that $S $ is minimal subject to the prescribed dynamics \eqref{statesyst-eqn} and the prescribed initial and final conditions $n(0) = n_0 $, $n(T) = n_T $.
\end{definition}
\medskip

The coupling between the control and state variables may be made explicit by using the pairing $\left\langle\,\cdot,\,\cdot\right\rangle_Q :T^*Q\times TQ\to\mathbb{R}$ and a Lagrange multiplier $\alpha\in T^*Q$ that imposes the state system as a constraint on the cost functional, 
\begin{equation}\label{action-constrained}
S_c:=\int_0^T \Big[\ell(u,n) + \left\langle \alpha,\dot{n}-F(n,u)\right\rangle_Q\Big]dt.
\end{equation}
This is a consequence of the well-known \textit{Pontryagin maximum principle} \cite{Bl2004,Ju1997}.

The variable $\alpha\in T^*Q$ is called a \textit{costate variable}. We now compute the equations associated to the variational principle $\delta S_c=0$. For simplicity, we suppose here that the state manifold $Q$ is a vector space, say $W$. In this case the cotangent space is $T^*W=W\times W^*$ and the costate variable is of the form $\alpha=(n,p)\in W \times W ^\ast$. The stationary variations of the constrained cost function $S_c$ in (\ref{action-constrained}) yield
\begin{align*}
0=\delta S_c 
=&\int_0^T  \Bigg[\left\langle \frac{\delta \ell}{\delta n} - \left(\frac{\delta F}{\delta n} 
\right)^T p-\dot{p},\delta n\right\rangle_W+ \left\langle\frac{\delta \ell}{\delta {u}} - \left(\frac{\delta F}{\delta {u}}\right)^T p, 
\delta {u}\right\rangle_U\\
&\qquad\qquad +\left\langle \delta p,\dot{n}-F(n, {u})\right\rangle_W
\Bigg]dt
+ \left\langle p,\delta n\right\rangle_Q\Big|_0^T,
\end{align*}
where $\langle\,\cdot\,,\,\cdot\,\rangle_U:U^*\times U\rightarrow\mathbb{R}$ denotes the duality pairing for the control vector space $U$.

Stationarity in the variations $\delta{u}$ gives a relation that determines the controls ${u}$ in terms of the state and costate variables, $n$ and $\alpha$, respectively, while stationarity in the variations $(\delta n, \delta \alpha)$ determines the evolution equations for the state and costate variables that minimize the cost function $S$.  Since the values of $n$ at the endpoints in time are fixed, $\delta n$ vanishes at the endpoints. We thus get the stationarity conditions
\[
\frac{\delta \ell}{\delta {u}}=\left(\frac{\delta F}{\delta {u}}\right)^T p
,\qquad 
\dot{n}=F(n, {u})
,\qquad 
\dot p=\frac{\delta \ell}{\delta n} - \left(\frac{\delta F}{\delta n} 
\right)^T p.
\]

\begin{remark}\normalfont
Although we shall confine our considerations to the Lagrangian description, we point out that the relation to the \textit{Pontryagin Maximum Principle} in the Hamiltonian description is obtained via the Legendre transformation of the integrand in the cost functional given by \eqref{action-constrained} which, for each point
$u$ in the control space $U$, defines the corresponding Hamiltonian $H_u: T^*Q\to \mathbb{R}$ by 
\begin{equation}\label{Hamiltonian-Clebsch}
H_u(\alpha_n) =  \left\langle\alpha_n,F(n, {u})\right\rangle_Q - \ell(n,u).
\end{equation}
The notation $\alpha_n$ for a covector in $T^*Q$ means that it belongs to the fiber $T^*_nQ$ of the cotangent bundle.
For more information about the Hamiltonian approach to geometric optimal control theory and the Pontryagin Maximum Principle, see \cite{Bl2004,Ju1997}.
\quad $\blacklozenge$
\end{remark}

\subsection{Examples: Lie group controls acting on state manifolds}\label{GXM-eg}

As an example that illustrates the theory developed in this paper, we consider the case of continuum mechanical systems with advected quantities; see Section 6 in \cite{HoMaRa1998a}. In this case, the state manifold $M$ is some vector subspace $V^*$ of $\mathfrak{T}(\mathcal{D})\otimes\operatorname{Den}(\mathcal{D})$, the tensor field densities on a manifold $\mathcal{D}$. We will denote by $a\in V^*$ these tensor field densities. The group $\operatorname{Diff}(\mathcal{D})$ of all diffeomorphisms of the manifold $\mathcal{D}$ acts on $V^*$ by pull back, that is,
\[
a\mapsto\eta^*a = a\circ\eta,
\quad\text{for all}\quad 
\eta\in\operatorname{Diff}(\mathcal{D}).
\]
It is thus a \textit{right} representation of $\operatorname{Diff}(\mathcal{D})$ on $\mathfrak{T}(\mathcal{D})\otimes\operatorname{Den}(\mathcal{D})$. We consider here the group $\operatorname{Diff}(\mathcal{D})$ of diffeomorphism as an infinite dimensional Lie group (either formally or in some 
Fr\'echet sense) whose Lie algebra is given by vector fields $\mathbf{v}\in\mathfrak{X}(\mathcal{D})$.
The right action of the Lie algebra $\mathfrak{X}(\mathcal{D})$ on $V^*$ is given by the Lie derivative
\[
\left.\frac{d}{dt}\right|_{t=0}\operatorname{exp}(t\mathbf{v})^*a :=\boldsymbol{\pounds}_{\mathbf{v}} a,
\]
where $t \mapsto\exp(t \mathbf{v}) $ denotes the flow of $\mathbf{v}$.

\subsection*{Example 1} We present a simple example of optimal control problem based on the geometric formulation of continuum mechanics described above.
In this example, the control space $U $ is the Lie algebra $\mathfrak{X}( \mathcal{D}) $ and thus the control variable is a vector field $u:=\mathbf{v}\in \mathfrak{X}(\mathcal{D})$. The state manifold $Q$ is the vector space $V^*$ of tensor field densities. The state variable $n:=a\in V^*$ is constrained to evolve according to the ODE
\[
\dot a=F(a,\mathbf{v}):=\boldsymbol{\pounds}_\mathbf{v}a
\]
and one wants to minimize
\[
S:=\frac{1}{2}\int_0^T\|\mathbf{v}\|^2_{\mathfrak{g}}dt,
\]
where $\|\cdot\|_\mathfrak{g}$ is an inner product norm on the Lie algebra $\mathfrak{g}=\mathfrak{X}(\mathcal{D})$. Note that we are in the setting of Definition \ref{def_optimal_control} with $M=V^*$ and $U=\mathfrak{X}(\mathcal{D})$.
This is an example of a \textit{Clebsch optimal control problem}, as studied from a geometric point of view in \cite{GBRa2011}. For this class of problems, the vector field $F$ is given by the infinitesimal generator associated to a group action on the state manifold. In the present example, this infinitesimal generator turns out to be the Lie derivative.

According to \eqref{action-constrained}, the constrained cost function in this case is
\[
S_c=\int_0^T\left(\frac{1}{2}\|\mathbf{v}\|^2_{\mathfrak{g}}+\langle p,\dot a-\boldsymbol{\pounds}_\mathbf{v}a\rangle_V\right)dt,
\]
where $p\in V$ is the costate variable. This is nothing else than the \textit{Clebsch approach} to continuum mechanics; see, e.g., \cite{HoKu1983}. The variational principle $\delta S_c=0$ gives the control
\[
\mathbf{v}=-\,(p\diamond a)^\sharp\in\mathfrak{g},
\]
where $\sharp:\mathfrak{g}^*\rightarrow\mathfrak{g}$ is the sharp operator associated to the inner product on $\mathfrak{g}$ and the bilinear operator $\diamond:V \times V ^\ast \rightarrow \mathfrak{g}^\ast$ is defined by
\begin{equation}\label{diamond_first_case}
\langle p\diamond a,\mathbf{v}\rangle:=-\langle\boldsymbol{\pounds}_\mathbf{v}a,p\rangle,\quad\text{for all}\quad p\in V, \quad a\in V^*, \quad \mathbf{v}\in \mathfrak{g}.
\end{equation}
The other stationarity conditions are
\begin{equation}\label{statecostate-eqns-Clebsch}
\left\{
\begin{array}{l}
\vspace{0.2cm}\displaystyle\dot{a}+\boldsymbol{\pounds}_{(p\diamond a)^\sharp}a=0
,\\
\dot{p}-\boldsymbol{\pounds}_{(p\diamond a)^\sharp}^\mathsf{T}p=0,
\end{array}
\right.
\end{equation}
where $\boldsymbol{\pounds}_\mathbf{v}^\mathsf{T}p\in V$ is defined by
\begin{equation}\label{LieDerT-star-def}
\left\langle a,\boldsymbol{\pounds}_\mathbf{v}^\mathsf{T}p\right\rangle=\left\langle \boldsymbol{\pounds}_\mathbf{v}a,p\right\rangle,\quad\text{for all}\quad p\in V, \quad a\in V^*, \quad \mathbf{v}\in \mathfrak{g}.
\end{equation}
The Clebsch state-costate equations (\ref{statecostate-eqns-Clebsch}) are canonically Hamiltonian
with
\[
H(a, p) = \frac{1}{2}\|(p\diamond a)^\sharp\|^2_{\mathfrak{g}}
= \frac{1}{2} \Pair{p\diamond a}{(p\diamond a)^\sharp}_{\mathfrak{g}}
.
\]
As is well known, \cite{HoKu1983}, using the cotangent-lift momentum map given by $\Pi=-\,p\diamond a$ to project the equations \eqref{statecostate-eqns-Clebsch} on $T^*M$ to $\mathfrak{g}^\ast$, yields the (left) Lie-Poisson bracket on the dual Lie algebra $\mathfrak{g}^*$. Explicitly, this Lie-Poisson bracket is given by 
\begin{equation} 
\dot{\Pi} 
= {\rm ad}^\ast_{\delta h/\delta \Pi}{\Pi}= {\rm ad}^\ast_{\Pi^\sharp}{\Pi}
\end{equation} 
where the Hamiltonian has the expression
\begin{equation}
h(\Pi) = \frac{1}{2} \pair{\Pi}{\Pi^\sharp}_{\mathfrak{g}}.
\end{equation} 

\subsection*{Example 2} This example will use the geometric setting of continuum mechanics as described before. However, the control vector space will now be given by $U:=\mathfrak{g}\times V^*\ni (\mathbf{v},\nu)$. We choose the quadratic Lagrangian
\[
\ell(\mathbf{v},\nu):=\frac{1}{2}\|\mathbf{v}\|_\mathfrak{g}^2+\frac{1}{2\sigma^2}\|\nu\|_{L^2}^2,
\]
where $\|\cdot\|_{L^2}$ denotes an $L^2$ norm on $V^*\subset \mathfrak{T}( \mathcal{D}) \otimes \operatorname{Den}( \mathcal{D})$. As before, the state manifold $Q $ is  $V^*$ and the state variable $a \in V ^\ast$ is constrained to evolve as
\[
\dot a=F(a,\mathbf{v},\nu):=\boldsymbol{\pounds}_\mathbf{v}a+\nu.
\]
Note that the advection law $\dot a=\boldsymbol{\pounds}_\mathbf{v}a$ is not imposed. Instead, the  penalty term in the Lagrangian introduces the additional term $\nu$ into the advection law.

Thus, the constrained action \eqref{action-constrained} becomes in this case
\begin{equation}\label{costfn-1}
S_c=\int_0^T\left(
\frac{1}{2}\|\mathbf{v}\|^2_\mathfrak{g} +  \frac{1}{2\sigma^2}\|\nu\|_{L^2}^2 
+ \left\langle p, \dot{a}-\boldsymbol{\pounds}_\mathbf{v} a - \nu\right\rangle_V
\right)dt,
\end{equation}
whose stationary variation results in
\begin{align*}
0=\delta S_c=&\int_0^T  \Big[
\left\langle- \boldsymbol{\pounds}_\mathbf{v}^\mathsf{T} p -\dot{p},\delta a\right\rangle_V
+\left\langle\mathbf{v}^\flat + p \diamond a, 
\delta \mathbf{v}\right\rangle_{\mathfrak{g}}
\\
&\qquad\qquad
+ \left\langle \frac{1}{\sigma^2}\nu^\flat - p, 
\delta \nu\right\rangle_V
+
\left\langle \delta p,\dot{a}-\boldsymbol{\pounds}_\mathbf{v}a - \nu\right\rangle_V
\Big]dt
+ \left\langle p, \delta a\right\rangle_V\Big|_0^T,
\end{align*}
where the flat operators $\flat :\mathfrak{g}\rightarrow\mathfrak{g}^*$ and $\flat :V^*\rightarrow V$ are associated to the inner products on $\mathfrak{g}$ and $V^*$, respectively.
Here the endpoint terms vanish because the values of $a$ at the endpoints in time are fixed. 
According to the variational formula for $\delta S_c$, the cost functional in (\ref{costfn-1}) is optimized when the controls satisfy 
\begin{equation}\label{control-eqn}
\mathbf{v} = -\,(p \diamond a)^\sharp\in\mathfrak{g}
\qquad\hbox{and}\qquad
\nu = \sigma^2p^\sharp\in V^*,
\end{equation}
in which the sharp maps are the inverses of the flat maps defined above.
For the controls $(\mathbf{v},\nu)\in\mathfrak{g}\times V^*$, the state and costate variables $(a,p)\in V^*\times V$ evolve according to the following closed system
\begin{equation}\label{statecostate-eqns-LT}
\left\{
\begin{array}{l}
\dot{a}+\boldsymbol{\pounds}_{(p\diamond a)^\sharp}a=\sigma^2p^\sharp
\,,\\[2mm]
\dot{p}-\boldsymbol{\pounds}_{(p\diamond a)^\sharp}^\mathsf{T}p=0
\,.
\end{array}
\right.
\end{equation}
These are Hamilton's canonical equations for the Hamiltonian
\begin{equation} 
H(p,a) = \frac{1}{2} \pair{(p \diamond a)}{(p \diamond a)^\sharp}_{\mathfrak{g}}
+  \frac{\sigma^2}{2} \pair{p}{p^\sharp}_{V}\,.
\end{equation} 

\begin{remark}\normalfont
Thus, the evolution of the state $a$ and costate $p$ variables occurs by the corresponding Lie derivative actions of the vector field $(p\diamond a)^\sharp\in \mathfrak{g} = \mathfrak{X}( \mathcal{D})$ calculated by applying the sharp map $\sharp$ to raise indices on the cotangent momentum map $(a,p) \in V^* \times V  = T ^* V^* \mapsto \mathbf{J}(a,p)=-\,p\diamond a\in \mathfrak{g}^*$ of the cotangent-lifted action. 
\quad $\blacklozenge$
\end{remark} 
\medskip

The evolution of the momentum $V^* \times V \rightarrow \mathfrak{g}^\ast$ itself is the last formula to be found, just as in the Clebsch approach, \cite{HoKu1983}.\\

\begin{proposition}\label{momap-evol}
Denote the momentum map of the cotangent-lifted action by 
\[
\Pi:= -\,p\diamond a
\]
and its dual vector field by 
\[
\mathbf{v}:=-\,( p\diamond a)^\sharp = \Pi^\sharp
\,.
\] 
Then the state and costate equations \eqref{statecostate-eqns-LT} imply the following Euler-Poincar\'e equation for the evolution for the momentum map:
\begin{equation}\label{momentum_map-eqns}
\dot{\Pi}
= -  \boldsymbol{\pounds}_\mathbf{v}^*\Pi - \sigma^2  p\diamond  p^\sharp,
\end{equation}
where the operator $\boldsymbol{\pounds}_\mathbf{v}^* : \mathfrak{g}^\ast \rightarrow \mathfrak{g}^\ast$ is defined by $\left\langle \boldsymbol{\pounds}_\mathbf{v}^* \Pi, \mathbf{u} \right\rangle : = \left\langle \Pi, [ \mathbf{v}, \mathbf{u}]_{JL} \right\rangle$ for any $ \mathbf{u}, \mathbf{v} \in \mathfrak{g} = \mathfrak{X}( \mathcal{D}) $, $\Pi \in \mathfrak{g}^\ast = \Omega^1( \mathcal{D}) \otimes \operatorname{Den}( \mathcal{D}) $ and $[ \mathbf{v}, \mathbf{u}]_{JL}=\boldsymbol{\pounds}_\mathbf{v}\mathbf{u}$ denotes the standard Lie bracket of vector fields.
\end{proposition}
\medskip

\noindent\textbf{Proof.} The proof proceeds by a direct calculation. In the computation below we use the standard Jacobi-Lie bracket of vector fields $[X, Y]_{JL} (f) = X(Y(f)) - Y(X(f))$ for any $f \in C ^{\infty}( \mathcal{D}) $. For a fixed Lie algebra element $Z\in \mathfrak{g} = \mathfrak{X}( \mathcal{D})$ we compute,
\begin{align*}\label{momentum_map-eqns-proof}
\left\langle\dot{\Pi},Z\right\rangle
&=-\left\langle\dot{ p}\diamond a +  p\diamond \dot{a},Z\right\rangle
\\&= \left\langle\dot{ p} ,\boldsymbol{\pounds}_Z a\right\rangle + \left\langle p ,\boldsymbol{\pounds}_Z\dot{a}\right\rangle \\
&= -\left\langle\boldsymbol{\pounds}_\mathbf{v}^\mathsf{T}  p ,\boldsymbol{\pounds}_Z a\right\rangle + \left\langle  p,\boldsymbol{\pounds}_Z\boldsymbol{\pounds}_\mathbf{v} a\right\rangle
+ \sigma^2 \left\langle p ,\boldsymbol{\pounds}_Z p^\sharp\right\rangle \\
&=\left\langle  p ,\boldsymbol{\pounds}_{[Z,\mathbf{v}]} a\right\rangle
+ \sigma^2 \left\langle  p,\boldsymbol{\pounds}_Z p^\sharp\right\rangle \\
&=- \left\langle  p\diamond a ,[Z, \mathbf{v}] \right\rangle
-\sigma^2 \left\langle  p \diamond  p^\sharp,Z\right\rangle \\
& = - \left\langle \Pi, \boldsymbol{\pounds}_\mathbf{v} Z \right\rangle 
-\sigma^2 \left\langle  p \diamond  p^\sharp,Z\right\rangle \\
&= -\left\langle \boldsymbol{\pounds}_\mathbf{v}^*\Pi ,Z \right\rangle
- \sigma^2 \left\langle  p \diamond  p^\sharp ,Z\right\rangle,
\end{align*}
which proves the Proposition \ref{momap-evol}.
$\qquad\blacksquare$
\medskip

\begin{remark}\normalfont 
(Lie algebra formulation of the equations)
Recall the the Lie algebra bracket $[\mathbf{u},\mathbf{v}]=\operatorname{ad}_\mathbf{u}\mathbf{v}$ on $\mathfrak{g}$ is minus the Lie bracket of vector fields, that is,
\[
[\mathbf{u},\mathbf{v}]= - [ \mathbf{u}, \mathbf{v}]_{JL} := - \left(\mathbf{u} \cdot \nabla \mathbf{v} - \mathbf{v} \cdot \nabla \mathbf{u}\right).
\]
We may thus identify $\boldsymbol{\pounds}_\mathbf{v}^\ast=-\operatorname{ad}^*_\mathbf{v}$ and the previous equations can be rewritten as
\begin{equation}
\label{Ex2mod-eqn}
\left\{
\begin{array}{l}
\dot{\Pi}
= {\rm ad}^*_\mathbf{v}\Pi - \sigma^2  p\diamond  p^\sharp
\,,\\[2mm]
\dot{a}=-\boldsymbol{\pounds}_{\mathbf{v}}a+\sigma^2p^\sharp
\,,\\[2mm]
\dot{p}=\boldsymbol{\pounds}_{\mathbf{v}}^\mathsf{T}p
\,.
\end{array}
\right.
\end{equation}
These are Lie-Poisson equations with a cocycle for the Hamiltonian
\begin{equation} 
h(\Pi,a, p) = \frac{1}{2} \pair{\Pi}{\Pi^\sharp}_{\mathfrak{g}}
+  \frac{\sigma^2}{2} \pair{p}{p^\sharp}_{V}
\,,
\end{equation} 
with respect to the Lie-Poisson bracket given by,
\begin{equation} 
    \begin{bmatrix}
   {\dot{\Pi}} \\[2mm]
 \  {\dot{a}} \   \\[2mm]
   {\dot{p}}
    \end{bmatrix}
    =
    \begin{bmatrix}
     {\rm ad}^\ast_\Box{\Pi}
   &    
   {a} \diamond \Box
    &
     -{p} \diamond \Box
      \\[2mm]
       -\boldsymbol{\pounds}_{\Box}{a}
       & 0 & 1
          \\[2mm]
          \boldsymbol{\pounds}_{\Box}^\mathsf{T}\,{p}
      & -1 & 0 
    \end{bmatrix}
    \begin{bmatrix}
   \partial h/\partial{\Pi} = \Pi^\sharp = \mathbf{v} \\[2mm]
   \partial h/\partial{a} = 0\\[2mm]
   \partial h/\partial{p} = \sigma^2 p^\sharp  
    \end{bmatrix}
\label{EP-gXV*XVeqns}
\end{equation}
in which the variational derivatives of the Hamiltonian are to be substituted into the corresponding places indicated by a box $(\Box)$.
This matrix is identified as the Hamiltonian operator for the Lie-Poisson bracket dual to the semidirect product Lie algebra $\mathfrak{g} \,\circledS\, (V^*\times V)$ \emph{plus} a symplectic 2-cocycle on $(a,p)\in V\times V^*$.
\quad $\blacklozenge$
\end{remark} 
\medskip

\begin{remark}\label{rem_isotropic_K} \normalfont  
This Hamiltonian matrix will block-diagonalize in the Lagrange-Poincar\'e formulation discussed in Section \ref{sec_LP_and_metamorph}. Roughly speaking, this amounts to transforming variables $\Pi\to \tilde{\Pi}:=(\Pi + p\diamond a)$ and $(a,\nu)\to (a,\dot{a})$.
\quad $\blacklozenge$
\end{remark} \medskip

\subsection*{Example 3} We now consider an example analogous to the preceding one but in finite dimensions. We let the orthogonal group $G=SO(3)$ act on $\mathbb{R}^3$ by matrix multiplication on the \textit{left} and we choose $U: = \mathfrak{so}(3)\times \mathbb{R}^3\ni ( \boldsymbol{\Omega}, \nu)$ as control space. As usual, we identify the Lie algebra $\mathfrak{so}(3)$ with $\mathbb{R}^3$. We choose the quadratic Lagrangian $\ell: \mathfrak{so}(3) \times \mathbb{R}^3 \rightarrow \mathbb{R}$ given by 
\[
\ell(\boldsymbol{\Omega},\nu):=\frac{1}{2}\mathbb{I}\boldsymbol{\Omega}\cdot\boldsymbol{\Omega}+\frac{1}{2\sigma^2}\mathbb{K}\nu\cdot\nu,
\]
for symmetric positive definite matrices $\mathbb{I}$ and $\mathbb{K}$. We impose the evolution equation
\begin{equation}\label{X_equation}\dot{\mathbf{X}}=-\,\boldsymbol{\Omega}\times \mathbf{X}+\nu
\end{equation}
for the state variable $\mathbf{X}\in \mathbb{R}^3 = : Q$. 
As before, the variational principle $\delta S_c=0$ with
\[
S_c=\int_0^T\left(\frac{1}{2}\mathbb{I}\boldsymbol{\Omega}\cdot\boldsymbol{\Omega}+\frac{1}{2\sigma^2}\mathbb{K}\nu\cdot\nu+
\mathbf{P} \cdot \left(\dot{\mathbf{X}}+\boldsymbol{\Omega}\times \mathbf{X}-\nu\right) \right)dt
\]
yields the controls
\[
\mathbb{I}\boldsymbol{\Omega}=\mathbf{P}\times \mathbf{X}\quad\text{and}\quad \mathbb{K}\nu=\sigma^2\mathbf{P},
\]
as in \eqref{control-eqn}. Note that $\boldsymbol{\Omega} = \mathbb{I}^{-1}( \mathbf{P}\times \mathbf{X}) = ( \mathbf{P}\times \mathbf{X})^ \sharp $ and $\mathbb{K} ^{-1} \mathbf{P} = \mathbf{P}^\sharp$, by the definition of the sharp maps. Then the state and costate evolution equations \eqref{statecostate-eqns-LT} take canonical Hamiltonian form with Hamiltonian function
\begin{equation}\label{RBmod-Ham}
H( \mathbf{X}, \mathbf{P})=  \frac12( \mathbf{P}\times \mathbf{X})\cdot( \mathbf{P}\times \mathbf{X})^ \sharp
+
\frac{\sigma^2}{2} \mathbf{P}\cdot\mathbf{P}^ \sharp
\,.
\end{equation}
Intriguingly, the resulting canonical Hamiltonian equations,
\begin{equation}
\label{statecostate-veceqns}
\left\{
\begin{array}{l}
\dot{\mathbf{X}}
=\displaystyle \frac{\partial H}{\partial \mathbf{P}}
= -\,(\mathbf{P}\times \mathbf{X})^\sharp \times\mathbf{X} + \sigma^2\mathbf{P}^\sharp
\,,\\[5mm]
\dot{\mathbf{P}} 
= \displaystyle -\, \frac{\partial H}{\partial \mathbf{X}}
= -\,(\mathbf{P}\times \mathbf{X})^\sharp \times \mathbf{P}
\,,
\end{array}
\right.
\end{equation}
involve the double cross product of the state and costate vectors $(\mathbf{X}, \mathbf{P})\in \mathbb{R}^3 \times\mathbb{R}^3$. The double cross products correspond to the Lie derivatives in equations (\ref{statecostate-eqns-LT}) which for this case become cross products. For more information about the roots of the Hamiltonian approach in geometric control theory, see \cite{BaWi1999}. \medskip

Upon defining the vector $\boldsymbol{\Pi}:=\mathbb{I}\boldsymbol{\Omega} =\mathbf{P}\times {\bf X}$, equations \eqref{statecostate-veceqns} imply 
\begin{equation}
\label{RBmod-eqn}
\left\{
\begin{array}{l}
\boldsymbol{\dot{\Pi}} 
= 
-\,\boldsymbol{\Omega} \times\boldsymbol{\Pi}
-
\sigma^2(\mathbb{K}^{-1}\mathbf{P}) \times\mathbf{P} 
\,,\\[2mm]
\dot{\mathbf{X}}
= -\,{\boldsymbol{\Omega}}\times\mathbf{X} + \sigma^2\mathbf{P}^\sharp
\,,\\[2mm]
\mathbf{\dot{P}} 
= -\,{\boldsymbol{\Omega}}\times \mathbf{P},
\end{array}
\right.
\end{equation}
which recovers the momentum map system (\ref{Ex2mod-eqn}) for this case. Indeed, one may  compute directly that
\begin{align*}\dot{ \boldsymbol{\Pi}} 
&= \dot{ {\bf P}} \times \mathbf{X} + {\bf P} \times \dot{ \mathbf{X}}
\\[2mm]& = \left(-\boldsymbol{\Omega} \times \mathbf{P}\right)\times\mathbf{X}+    \mathbf{P}\times \left(- \boldsymbol{\Omega} \times \mathbf{X} + \sigma^2 \mathbf{P}^\sharp\right)
\\[2mm]& =  ( \mathbf{P} \times  \boldsymbol{\Omega}) \times \mathbf{X}+\left( \boldsymbol{\Omega} \times \mathbf{X} \right) \times \mathbf{P}
+ \sigma^2 \mathbf{P}\times \mathbf{P}^\sharp 
\\[2mm]& = - ( {\bf X} \times \mathbf{P}) \times \boldsymbol{\Omega} + \sigma^2 \mathbf{P}\times ( \mathbb{K}^{-1} \mathbf{P})
\\[2mm]& = \boldsymbol{\Pi} \times \boldsymbol{\Omega} + 
\sigma^2 \mathbf{P}\times ( \mathbb{K}^{-1} \mathbf{P})
\,,
\end{align*}
from which the result follows. 
\medskip

\begin{remark}\normalfont (Lie algebra formulation) The Lie algebra bracket 
on $\mathfrak{se}(3)\simeq \mathfrak{so}(3)\,\circledS\, \mathbb{R}^3$ may be written 
on $\mathbb{R}^3\times\mathbb{R}^3$ as,
\[
{\rm ad}_{(\boldsymbol{\Omega}, \boldsymbol{\alpha} )}(\tilde{\boldsymbol{\Omega}},\tilde{\boldsymbol{\alpha}})
=
\Big[(\boldsymbol{\Pi}, \boldsymbol{\alpha})\,,\, (\tilde{\boldsymbol{\Omega}},\tilde{\boldsymbol{\alpha}}) \Big]
=
\Big(\boldsymbol{\Omega}\times \boldsymbol{\tilde{\Omega}}
\,,\,
\boldsymbol{\Omega}\times \boldsymbol{\tilde{\alpha}}
-
 \boldsymbol{\tilde{\Omega}} \times \boldsymbol{\alpha}
\Big)
\]
Its dual operation is 
\[
{\rm ad}^*_{(\boldsymbol{\Omega}, \boldsymbol{\alpha})}(\boldsymbol{\Pi},\mathbf{P})
=
\Big(
- \boldsymbol{\Omega}\times \boldsymbol{\Pi}
-
 \boldsymbol{\alpha} \times \mathbf{P},\,
 -
 \boldsymbol{\Omega} \times \mathbf{P}
\Big).
\]
In terms of the ad$^*$ operation on $\mathfrak{se}(3)^*$, the motion equations for $(\boldsymbol{\Pi}, \mathbf{P})$ in (\ref{RBmod-eqn}) can be rewritten as
\begin{align*}
\big(\boldsymbol{\dot{\Pi}}\,,\,\mathbf{\dot{P}} \big)
&=
\Big(
- \boldsymbol{\Omega}\times \boldsymbol{\Pi}
-
\sigma^2\mathbf{P}^\sharp \times \mathbf{P}
 \,,\,
 -\,
 \boldsymbol{\Omega} \times \mathbf{P}
\Big)
\\&=
\Big(\operatorname{ad}^*_{\boldsymbol{\Omega}}\boldsymbol{\Pi} + 
\sigma^2 \mathbf{P}\diamond\mathbf{P}^\sharp
\,,\,
-\,{\boldsymbol{\Omega}}\times \mathbf{P}\Big)
\\&=
\operatorname{ad}^*_{(\boldsymbol{\Omega},\, \sigma^2\mathbf{P}^\sharp)}
\big(\boldsymbol{\Pi}\,,\, \mathbf{P}\big)
\,.
\end{align*}
The result of the last calculation may be rewritten in Lie-Poisson bracket form as
\begin{equation}\label{LP-PiP}
\big(\boldsymbol{\dot{\Pi}}\,,\,\mathbf{\dot{P}} \big)
= 
\operatorname{ad}^*_{\big({\partial h}/{\partial\boldsymbol{\Pi}},{\partial h}/{\partial\mathbf{P}} \big)}
\big(\boldsymbol{\Pi}, \mathbf{P}\big)
\,,
\end{equation}
with Hamiltonian (\ref{RBmod-Ham}) rewritten in these variables as 
\begin{equation}\label{RBmod-Ham-LP}
h( \boldsymbol{\Pi}, \mathbf{P})=  \frac12\boldsymbol{\Pi}\cdot\boldsymbol{\Pi}^ \sharp
+
\frac{\sigma^2}{2} \mathbf{P}\cdot\mathbf{P}^ \sharp
\,,
\end{equation}
and using the (left) Lie-Poisson bracket defined on the dual Lie algebra  $\mathfrak{se}(3)^*$. This is the Hamiltonian and Lie-Poisson bracket for the motion of an ellipsoidal underwater vehicle in the body representation. See, e.g., \cite{Ho2008} for more discussion and references to the literature about the geometrical approach to the dynamics and control of underwater vehicles.
\quad $\blacklozenge$
\end{remark} 
\medskip

We have seen that equations (\ref{statecostate-veceqns}) for the state-costate vectors $(\mathbf{X},\mathbf{P})$ are canonically Hamiltonian and that the system (\ref{LP-PiP}) for $(\boldsymbol{\Pi},\mathbf{P})$ is Lie-Poisson on the dual of a semidirect product Lie algebra. Now, it remains to include the dynamics of the coordinate $\mathbf{X}$ into a single structure for the entire system (\ref{RBmod-eqn}) for $(\boldsymbol{\Pi},\mathbf{X},\mathbf{P})$. 
We observe that equations (\ref{RBmod-eqn}) may be put into Lie-Poisson form, as
\begin{equation} 
    \begin{bmatrix}
   \boldsymbol{\dot{\Pi}} \\
 \  \mathbf{\dot{X}}\   \\
   \mathbf{\dot{P}} 
    \end{bmatrix}
    =
    \begin{bmatrix}
    \boldsymbol{\Pi}\times
   &    
     \mathbf{X} \times
    &
     \mathbf{P} \times
      \\
    \mathbf{X} \times    & 0 & 1
          \\
    \mathbf{P} \times   & -1 & 0 
    \end{bmatrix}
    \begin{bmatrix}
   \partial h/\partial\boldsymbol{\Pi} \\
    \partial h/\partial\mathbf{X}\\
   \partial h/\partial\mathbf{P} 
    \end{bmatrix}
    =
    \begin{bmatrix}
    \boldsymbol{\Pi}\times
   &    
     \mathbf{X} \times
    &
     \mathbf{P} \times
      \\
    \mathbf{X} \times    & 0 & 1
          \\
    \mathbf{P} \times   & -1 & 0 
    \end{bmatrix}
    \begin{bmatrix}
   \boldsymbol{\Omega} \\
   0\\
\sigma^2 \mathbf{P}^\sharp 
    \end{bmatrix}
    .
\label{EP-so3XR3XR3eqns}
\end{equation}
This is the Lie-Poisson bracket dual to the semidirect product Lie algebra $\mathfrak{so}(3) \,\circledS\, (\mathbb{R}^3 \times \mathbb{R}^3) $ plus a symplectic 2-cocycle on $(\mathbf{X},\mathbf{P})\in \mathbb{R}^3 \times \mathbb{R}^3$.
\medskip

\begin{remark} \normalfont
As mentioned earlier, the Lagrange-Poincar\'e and Hamilton-Poincar\'e formulations in Sections \ref{sec_LP_and_metamorph} and \ref{section_HP_metamorphosis} will block-diagonalize this Hamiltonian matrix. 
\quad $\blacklozenge$
\end{remark} 
\medskip

\begin{remark}\normalfont (Comparison of the examples) The major difference between Example 1 and Examples 2 and 3 is the following. In Example 1, we impose the \textit{advection equation} $\dot a=\boldsymbol{\pounds}_\mathbf{v}a$ as a \textit{constraint} on the minimization problem. This is done, as usual, by introducing a new variable $\mathbf{p}$ and adding the term $\left\langle\mathbf{p},\dot{a}-\boldsymbol{\pounds}_\mathbf{v}a\right\rangle$ in the action functional. In Examples 2 and 3, the advection law is not imposed \textit{exactly}, but only up to an error term
\[
\nu:=\dot{a}-\boldsymbol{\pounds}_\mathbf{v}a,
\]
whose norm is added to the Lagrangian as a penalty, and needs to be minimized. Of course, in this case, the relation $\nu=\dot{a}-\boldsymbol{\pounds}_\mathbf{v}a$ is a constraint as seen in the term $\left\langle\mathbf{p},\dot{a}-\boldsymbol{\pounds}_\mathbf{v}a-\nu\right\rangle$. As we have seen in Proposition \ref{momap-evol}, this error term implies a modification of the equations of motion.

One of the aims of the present paper is to transform the control problem corresponding to the cost function in (\ref{costfn-1})  into an optimization problem in which the penalty term $\|\dot a-\boldsymbol{\pounds}_\mathbf{v}a\|^2$ appears. This objective motivates the introduction of the distributed optimization problem in the next section.
\quad $\blacklozenge$
\end{remark}

\subsection{Distributed optimization problems}\label{distributed_Optimiz}
\medskip
\begin{definition}{\rm (}istributed optimization 
problems{\rm )}
\label{def_DOCP}
A distributed optimization problem imposes the evolutionary state system in \eqref{statesyst-eqn} as a \emph{penalty} involving a chosen norm, rather than as a constraint. The resulting cost functional is thus taken to be of the form
\begin{equation}\label{action-distrib}
S_d:=\int_0^T \Bigg[\ell(u,n) + \frac{1}{2\sigma^2}\|\dot{n}-F(n,u)\|^2\Bigg]dt,
\end{equation}
where the norm is associated to a Riemannian metric on $Q$.
In this cost functional, the state system dynamics \eqref{statesyst-eqn} is imposed only in a distributed sense; namely, as a penalty enforced by the norm on $Q$, not pointwise on $Q$, as in \eqref{action-constrained}. We assume that $\sigma^2>0$.
\end{definition}
\medskip

We may initially regard this second approach as simply modifying the cost function in the optimal control problem (\ref{action-constrained}) by introducing a penalty based on a norm of the state system. We will show later that the solutions of the two types of optimization problems coincide in the limit $\sigma^2\to0$.

In the case where $Q$ is a vector space, denoted by $W$, and the norm is associated to an inner product, the variations of the distributed cost function $S_d$ in (\ref{action-distrib}) now yield
\begin{equation}
\delta S_d 
=\int_0^T  \left[
\left\langle\frac{\delta \ell}{\delta n} - \left(\frac{\delta F}{\delta n} 
\right)^T  p-\dot{ p},\delta n\right\rangle_W
+ \left\langle\frac{\delta \ell}{\delta {u}} - \left(\frac{\delta F}{\delta {u}} \right)^T  p,
\delta {u}\right\rangle _V\,\right]dt
+ \left\langle p,\delta n\right\rangle_W\Big|_0^T,
\end{equation}
where the momentum variable $p$ obtained from the variation with respect to the vector field $\dot{n}\in W$ is defined by
\begin{equation}
\sigma^2p := \Big(\dot{n}-F(n, {u}) \Big)^\flat \in W^*,
\label{costate-mom}
\end{equation}
and in this case the $\flat$ map (index lowering) is applied with respect to the inner product on $W$.

Let us return to Example 2 above and treat it as distributed optimization problem.

\subsection*{Example} As in \S\ref{GXM-eg}, we consider the geometric setting of continuum mechanics. Contrary to Example 1 above, we do not impose the advection equation $\dot a=\boldsymbol{\pounds}_\mathbf{v} a$ as a constraint but as a penalty. The problem is now to minimize the expression
\[
S_d:=\int_0^T \left[
\frac12 \|\mathbf{v}\|^2_\mathfrak{g} +\frac{1}{2\sigma^2}\| \dot{a}-\boldsymbol{\pounds}_\mathbf{v} a\|_{L^2}^2
\right]dt,
\]
where $\|\cdot\|_{L^2}$ is a $L^2$ norm on the space of tensor field densities. This problem is clearly equivalent to that of Example 2 in \S\ref{GXM-eg}. The variational principle $\delta S_d=0$ yields the control
\[
\mathbf{v} = -\,(p \diamond a)^\sharp\in\mathfrak{g}
\]
and the same equations as before
\begin{equation}
\left\{
\begin{array}{l}
\vspace{0.2cm}\displaystyle\dot{a}+\boldsymbol{\pounds}_{(p\diamond a)^\sharp}a=\sigma^2p^\sharp
,\\
\dot{p}-\boldsymbol{\pounds}_{(p\diamond n)^\sharp}^Tp=0,
\end{array}
\right.
\end{equation}
where we have defined the variable $p$ by
\begin{equation}\label{def_p}
p:=\frac{1}{\sigma^2}(\dot a-\boldsymbol{\pounds}_\mathbf{v}a)^\flat\in  V.
\end{equation}
It is important to observe that in this approach the variable $p$ is not really needed, since it is \textit{defined} in terms of the other variables. This is not the case for the Clebsch approach described in the Examples of \S\ref{GXM-eg} for which $p$ is an independent variable. For the Clebsch approach, the relation \eqref{def_p} is recovered as a consequence of the variational principle $\delta S_c=0$.

\subsection*{Control problems versus optimization problems}
We now make some simple comments concerning the role of the variational principles in control problems and optimization problems.

Let $\ell=\ell(u,n):U\times Q\rightarrow\mathbb{R}$ be a cost function and $F$ a vector field as in the general Definition \ref{def_optimal_control}. As we have seen, one associates to these objects the following problems.
\begin{description}
\item {(1)} The \textit{optimal control problem} consists of minimizing the integral
\[
S:=\int_0^T\ell(u,n)dt\quad\text{subject to the conditions}\quad \dot n=F(n,u)
\]
and the usual endpoint conditions. The resolution of this problem uses the Pontryagin maximum principle which, under sufficient smoothness condition, implies that a solution of this problem is necessarily a solution of the variational principle
\[
\delta S_c
=
\delta \int_0^T\Big(\ell(\xi,n)
+
\langle\alpha,\dot n-F(n,u)\rangle\Big)dt=0.
\]
Example 1 in \S\ref{GXM-eg}, for which the cost function is a kinetic energy and the vector field $F$ is given by a Lie derivative, illustrates this method.

\item {(2)} The \textit{optimization problem with penalty} described above consists of minimizing the integral
\[
S_d:=\int_0^T\left(\ell(u,n)+\frac{1}{2\sigma^2}\|\dot n-F(n,u)\|^2\right)dt
\]
subject to the usual endpoint conditions. Of course, the solutions of this problem are necessarily solutions of the variational principle
\[
\delta S_d=\delta\int_0^T\left(\ell(\xi,n)+\frac{1}{2\sigma^2}\|\dot n-\xi_Q(n)\|^2\right)dt=0.
\]
The examples in here illustrate this point.
\end{description}
\medskip

\begin{remark}\normalfont
Despite the analogy between the two variational principles $\delta S_c=0$ and $\delta S_d=0$, the origins of these principles are quite different.

In the first problem, the functional $S$ is minimized under a constraint, leading to the construction of the functional $S_c$ by introducing the costate variable $\alpha$. The well-known Pontryagin approach tells us that the solutions of the optimal control problem are necessarily critical points of $S_c$.

The variational principle of the second problem is simply the stationarity condition implied by optimization of the functional $S_d$, without other constraints, except the endpoint conditions.
\quad $\blacklozenge$
\end{remark}

\subsection{Overview}\label{sec: overview}

In \cite{GBRa2011} a general formulation for a large class of optimal control problems was given. These problems, called \textit{Clebsch optimal control problems}, are associated to the action of a Lie group $G$ on a manifold $Q$ and to a cost function $\ell: \mathfrak{g} \times Q \rightarrow \mathbb{R}$, where $\mathfrak{g}$ denotes the Lie algebra of $G$. The Clebsch optimal control problem is, by definition,
\begin{equation}\label{Clebsch_optimal_control_intro}
\min_{\xi (t)}\int_0^T\ell(\xi(t),n(t))dt\,,
\end{equation}
subject to the following conditions:
\begin{itemize}
\item[\rm (A)] Either \quad $\dot n(t)=\xi(t)_Q(n(t))$\,, 
\quad\hbox{or}\quad $\;(\rm A)'$  \quad  $\dot
n(t)=-\,\xi(t)_Q(n(t))$\,;
\item[\rm (B)] Both \quad $n(0)=n_0$ \quad \hbox{and} \quad $n(T)=n_T$\,,
\end{itemize}
where $\xi_Q$ denotes the infinitesimal generator of the $G$-action, that is,
\[
\xi_Q(n):=\left.\frac{d}{dt}\right|_{t=0}\Phi_{\operatorname{exp}(t\xi)}(n)
\,.
\]
These optimal control problems comprise abstract formulations of many
systems such as the symmetric representation of the rigid body and Euler fluid equations \cite{BlCr1996,Ho2009}, the double bracket equations on symmetric spaces \cite{BlBrCr1997}, the singular solutions of the Camassa-Holm equation \cite{CaHo1993}, control problems on Stiefel manifolds \cite{BlCrSa2006}, and others \cite{Bl2004,BlCrMaSa2008}. 

Optimal control problems on Lie groups have a long history; see
\cite{Bl2004}, \cite{Ju1997} and references therein. Some of
the earliest papers dealing with such problems are 
\cite{Br1973} and \cite{HeMa1983}.

\medskip

\subsection*{Goals of the paper}  The first goal of the present paper is to replace the constraints in the Clebsch optimal control problem with a penalty function added to the cost function and to obtain in this way a classical (unconstrained) optimization problem. The fundamental idea is to use the constraints to form a quadratic penalty function in order to get the Lagrangian
\begin{equation}
\label{Lagrangian_intro}
\int_0^T \left(\ell(u,n)+\frac{1}{2\sigma^2}\|\dot n\mp\,\xi_Q(n)\|^2\right)dt.
\end{equation}
We first determine necessary and sufficient conditions characterizing the critical points of this Lagrangian. Taking the time derivative of one of the conditions and using the others leads directly to certain equations of motion. We then show that these equations are naturally obtained by Lagrangian reduction and that they are the Lagrange-Poincar\'e equations of a Lagrangian function in the material representation that is the sum of the original Lagrangian plus the square of the norm on the velocity vector. This approach links directly to the approach used in \cite{HoTrYo2009} in the study of the metamorphosis of shapes. From a variational point of view, one replaces the Hamilton-Pontryagin variational principle in the Clebsch framework
\[
\delta\int_0^T\left(\ell(u,n)+\langle\alpha,\dot n\mp u_Q(n)\rangle\right)dt=0
\,,\]
by the principle
\[
\delta\int_0^T\left(\ell(u,n)+\frac{1}{2\sigma^2}\|\dot n\mp u_Q(n)\|^2\right)dt=0
\,,\]
in the framework of distributed optimization. 
\medskip

This paper traces how the dynamical equations change on moving from constraints (optimal control), to optimization via imposition of a cost, and then on to metamorphosis. Passing from optimal control to optimization preserves the momentum map, but this passage modifies the reconstruction relation. The evolution is no longer only for the momentum map of the reduced Lagrangian. Instead, the momentum canonically conjugate to the velocity on the configuration manifold becomes coupled to the momentum map equations (which are the Euler-Poincar\'e equations), with coupling constant $\sigma^2$.
\medskip

Another feature of the paper, directly related to the dynamics of our optimization problem, is the description of the equations of motion by Lagrangian and Hamiltonian reduction. In particular, we carry out a certain type of Lagrangian reduction adapted to the problem, that we naturally call \textit{metamorphosis reduction}, since it was directly inspired by the example of the metamorphosis approach to image dynamics \cite{HoTrYo2009}. This Lagrangian reduction leads to the expression of the associated variational principles and Hamiltonian structures. In metamorphosis, the optimization problem involves Riemannian structures induced by Lie group actions on themselves and on Lie subgroups by group homomorphisms.  This is a rich field whose possibilities are still being developed. In particular, metamorphosis and related  variants of the geometric approach to control and optimization can be expected to produce opportunities for new applications and analysis in geometric dynamics.

\section{Distributed optimization}\label{sec_OCP}

In this section we begin with a quickly review the Clebsch optimal control problem studied in \cite{GBRa2011}. Then we introduce the class of optimization problems investigated in this paper, obtained by adding to the cost function a penalty given by the norm of the constraints in the previous approach. 

\subsection{Review of Clebsch optimal control}\label{review_COCP}

\subsection*{Clebsch optimal control formulation and main results} 
We recall from \cite{GBRa2011} some facts concerning Clebsch optimal control problems.
Let $\Phi:G\times Q\rightarrow Q$ be a left (resp. right) action of a Lie group $G$ on the manifold $Q$ and let $\ell:\mathfrak{g}\times Q\rightarrow\mathbb{R}$ be a cost function. The \textit{Clebsch
optimal control problem} for the curves $\xi(t) \in \mathfrak{g}$ and $n (t) \in Q$ is
\begin{equation}\label{Clebsch_optimal_control_gen}
\min_{\xi(t)}\int_0^T\ell(\xi(t),n(t))dt
\end{equation}
subject to the following conditions:
\begin{itemize}
\item[\rm (A)] Either \quad $\dot n(t)=\xi(t)_Q(n(t))\;$, \quad or \quad$\;(\rm A)'$ \quad $\dot
n(t)=-\xi(t)_Q(n(t))$\,; 
\item[\rm (B)] Both \quad $n(0)=n_0$ \quad and \quad $n(T)=n_T$,
\end{itemize}
where $\xi_Q$ denotes the infinitesimal generator of the $G$-action associated to $\xi\in \mathfrak{g}$, that is,
\[
\xi_Q(q):=\left.\frac{d}{ds}\right|_{s=0}\Phi_{\operatorname{exp}(s \xi)}(q), \quad  \quad q \in Q.
\]

If condition (A) is assumed, then by applying the Pontryagin maximum principle, we obtain that an extremal curve $n(t)\in Q$ is necessarily the projection of a curve $\alpha(t)\in T^*Q$ that is a solution of the equations  \cite{GBRa2011}
\begin{equation}\label{optimal_control_condition}
\frac{\delta \ell}{\delta \xi}=\mathbf{J}(\alpha),\quad \dot\alpha=\xi_{T^*Q}(\alpha)+\operatorname{Ver}_\alpha\frac{\partial
\ell}{\partial n}
\,.
\end{equation}
Here $\mathbf{J}:T^*Q\rightarrow\mathfrak{g}^*$ denotes the momentum map associated to the cotangent-lifted action of $G$ on $T^*Q$. Recall that $\mathbf{J}$ is given by \cite{MaRa1999}
\[
\langle\mathbf{J}(\alpha_q),\xi\rangle=\langle\alpha_q,\xi_Q(q)\rangle.
\]
\[
\langle\mathbf{J}(\alpha_q),\xi\rangle=\langle\alpha_q,\xi_Q(q)\rangle
=-\langle\alpha_q\diamond q,\xi\rangle
\,,\quad\hbox{when}\quad
 \xi_Q(q) = \pounds_\xi q
\]
The expression $\frac{\delta\ell}{\delta\xi}\in\mathfrak{g}^*$ denotes the usual functional derivative of $\ell(\cdot,n)$ for each fixed $n\in Q$ whereas $\frac{\partial\ell}{\partial n}: = \mathbf{d} \ell( \xi, \cdot ) \in T^*_nQ$ denotes the differential of the function $\ell(\xi,\cdot): Q \rightarrow \mathbb{R}$ for each fixed $\xi\in\mathfrak{g}$. For $\alpha, \beta \in T _q^\ast Q$, the map $\operatorname{Ver}_ \alpha \beta$ denotes the \textit{vertical lift of $\beta \in T _q^\ast Q$ relative to $\alpha \in T _q^\ast Q$}, defined by
\[
\operatorname{Ver}_ \alpha \beta : = \left.\frac{d}{ds}\right|_{s=0} ( \alpha+ s \beta) \in T_{ \alpha} ( T ^\ast Q).
\]
In \eqref{optimal_control_condition}, $\xi_{T^*Q}$ denotes the infinitesimal generator of the cotangent-lifted action of $G$ on $T^*Q$. Note that the vector field $\xi_{T^*Q}(\alpha)+\operatorname{Ver}_\alpha\frac{\partial
\ell}{\partial n}$ on $T^*Q$ is the Hamiltonian vector field associated to the Hamiltonian
\[
\alpha_n\in T^*Q\mapsto \langle\alpha_n,\xi_Q(n)\rangle-\ell(\xi,n)\in\mathbb{R},
\]
in which the Lie algebra element $\xi\in\mathfrak{g}$ is regarded as a parameter.
Using these equations, we determine that the optimal control $\xi$ is the solution of the equations
\begin{equation}\label{generalized_EP}
\frac{d}{dt}\frac{\delta \ell}{\delta
\xi}=-\operatorname{ad}^*_\xi\frac{\delta \ell}{\delta
\xi}+\mathbf{J}\left(\frac{\partial \ell}{\partial
n}\right),\quad\text{resp.}\quad \frac{d}{dt}\frac{\delta
\ell}{\delta \xi}=\operatorname{ad}^*_\xi\frac{\delta \ell}{\delta
\xi}+\mathbf{J}\left(\frac{\partial \ell}{\partial n}\right).
\end{equation}
If condition $\rm (A)'$ is assumed, then \eqref{optimal_control_condition} is replaced by
\begin{equation}\label{optimal_control_condition_A'}
\frac{\delta \ell}{\delta \xi}=-\,\mathbf{J}(\alpha),\quad
\dot\alpha=-\,\xi_{T^*Q}(\alpha)+\operatorname{Ver}_\alpha\frac{\partial
\ell}{\partial n}
\end{equation}
and the optimal control $\xi$ is the solution of the equations
\begin{equation}\label{generalized_EP_inverse}
\frac{d}{dt}\frac{\delta \ell}{\delta
\xi}=\operatorname{ad}^*_\xi\frac{\delta \ell}{\delta
\xi}-\mathbf{J}\left(\frac{\partial \ell}{\partial
n}\right),\quad\text{resp.}\quad \frac{d}{dt}\frac{\delta
\ell}{\delta \xi}=-\operatorname{ad}^*_\xi\frac{\delta \ell}{\delta
\xi}-\mathbf{J}\left(\frac{\partial \ell}{\partial n}\right).
\end{equation}
We refer to \cite{GBRa2011} for proofs of these statements and further discussion.

\subsection*{Variational principle} We shall prove that \textit{equations \eqref{optimal_control_condition} or \eqref{optimal_control_condition_A'}, together with the constraint $\dot n=\pm\,\xi_Q(n)$ follow from the variational principle
\begin{equation}\label{variational_principle_clebsch}
\delta\int_0^T\big(\ell(\xi,n)+\langle\alpha,\dot n\mp\xi_Q(n)\rangle\big)dt=0\,,
\end{equation}
for curves $t\mapsto \xi(t)\in\mathfrak{g}$ and $t\mapsto \alpha(t)\in T^*_{n(t)}Q$. The variations  $\delta\xi$ are free, whereas the variations $\delta\alpha$ are such that the induced variations $\delta n$ vanish at the endpoints, that is, $\delta n(0) = \delta n(T) = 0$. }
\medskip

To see this, let $\xi_s \in \mathfrak{g}$ and $\alpha_s \in T_{n_s} ^\ast Q $ be curves whose infinitesimal variations at $s=0 $ are $\delta \xi \in \mathfrak{g}$ and $\delta\alpha \in T_n ^\ast Q$. We have 
\begin{align} \label{intermediate_three}
\delta\int_0^T\big(\ell(\xi,n)+& \langle\alpha,\dot n\mp\xi_Q(n)\rangle\big)dt
= \int_0^T\left\langle \frac{\delta\ell}{ \delta \xi}, \delta\xi \right\rangle dt +
\int_0^T\left\langle \frac{\partial\ell}{ \partial n}, \delta n \right\rangle dt  \nonumber \\
&  + 
\left.\frac{d}{ds}\right|_{s=0}\int_0^T \left\langle \alpha_s, \dot{n}_s \right\rangle dt  \mp \left.\frac{d}{ds}\right|_{s=0} \int _0^T \left\langle \mathbf{J}( \alpha_s), \xi_s \right\rangle dt.
\end{align}
A direct computation in canonical coordinates, using $\delta n(0) = \delta n (T) = 0 $ in an integration by parts, shows that
\begin{equation}\label{intermediate_one}
\left.\frac{d}{ds}\right|_{s=0}\int_0^T \left\langle \alpha_s, \dot{n}_s \right\rangle dt = \int_0^T \Omega_{\rm can}( \dot{ \alpha}, \delta\alpha) dt,
\end{equation}
where $\Omega_{can}$ denotes the canonical symplectic form on $T^*Q$.
In addition, using the definition of the momentum map $\mathbf{J}: T ^\ast Q \rightarrow \mathfrak{g}^\ast$ we have
\begin{equation}\label{intermediate_two}
\left.\frac{d}{ds}\right|_{s=0} \left\langle \mathbf{J}( \alpha_s), \xi_s \right\rangle = \left\langle T_ \alpha \mathbf{J}( \delta \alpha) , \xi \right\rangle + \left\langle\mathbf{J}( \alpha),  \delta \xi\right\rangle = \Omega_{\rm can}\left(\xi_{T^\ast Q} ( \alpha),  \delta\alpha \right) + \left\langle\mathbf{J}( \alpha),  \delta \xi\right\rangle.
\end{equation}
Using relations \eqref{intermediate_one} and \eqref{intermediate_two} in formula \eqref{intermediate_three} yields \eqref{optimal_control_condition} and \eqref{optimal_control_condition_A'}.

\subsection*{Alternative form of the stationarity conditions} Note that the \textit{equations 
\begin{equation}\label{stat_cond}
\dot\alpha=\pm\,\xi_{T^*Q}(\alpha)+\operatorname{Ver}_\alpha\frac{\partial\ell}{\partial n}
\end{equation}
imply the constraint} $\dot n=\pm\,\xi_Q(n)$. To see this, it suffices to apply the tangent map $T\pi$ to \eqref{stat_cond}, where $\pi:T^*Q\rightarrow Q$ is the projection, and recall that $\xi_{T ^\ast Q} $ and $\xi_Q $ are $\pi$-related. By introducing a Riemannian metric $g$ on $Q$, it is possible to rewrite the stationarity condition in a more explicit way, as we show in the following lemma.
\medskip

\begin{lemma}\label{equivalent_form_of_stationarity_conditions} Suppose that $Q$ is endowed with a Riemannian metric $g$ and denote by $\nabla$ and $D/Dt$ the associated Levi-Civita covariant 
derivatives.

Then the equation $\dot\alpha=\pm\,\xi_{T^*Q}(\alpha)+\operatorname{Ver}_\alpha\frac{\partial\ell}{\partial n}$ in (\ref{stat_cond}) is equivalent to the system
\begin{equation}\label{system_stationarity}
\left\{
\begin{array}{l}
\displaystyle\vspace{0.2cm}\dot n=\pm\,\xi_Q(n),\\
\displaystyle\frac{D}{Dt}\alpha=\mp \langle\alpha,\nabla \xi_Q(n)\rangle+\frac{\partial\ell}{\partial n}
\,.
\end{array}
\right.
\end{equation}
\end{lemma}

\noindent\textbf{Proof.} We begin by recalling the definition and main property of the \textit{connector} $K:TTQ\rightarrow TQ$ associated to a Riemannian manifold $(Q, g)$. A general detailed treatment for connectors associated to linear connections can be found in 
\cite{Michor2008}, Section 13.8. In infinite dimensions we need to assume that the given weak Riemannian metric has a smooth geodesic spray  $S \in \mathfrak{X}(TQ)$. In natural local charts of $TTQ$, the intrinsic map $K $ is defined by
\begin{equation}
\label{connector_def}
K_{loc}(x,e,u,v)=(x,v+\Gamma(x)(e,u)),
\end{equation}
where $\Gamma(x)$ is the Christoffel map defined by the quadratic form in the fourth component of the geodesic spray  $S(x,u)=(x,u,u,-\Gamma(x)(u,u))$.  In finite dimensions, the Christoffel map has the familiar expression $\Gamma(x)(e,u)^i=\Gamma_{jk}^i(x)e^iu^k$, where $\Gamma^i_{jk}$ are usual the Christoffel symbols associated to the metric $g$. The relation between the connector and the Levi-Civita covariant derivative is given for all $X, Y \in \mathfrak{X}(Q)$ by
\begin{equation}
\label{connector_property}
\nabla_YX=K\circ TX\circ Y.
\end{equation}
The connector $K $ induces an intrinsic map, also denoted by $K : T T ^\ast Q \rightarrow T^*Q$ defined in natural local charts by
\begin{equation}
\label{connector_def_cotangent}
K_{loc}(x,\beta,u,\gamma)=(x,\gamma-\beta(\Gamma(x)(u,\cdot))).
\end{equation}
The associated covariant derivative 
\begin{equation}
\label{connector_property_cotangent}
\nabla_X\alpha:=K\circ T\alpha\circ X
\end{equation}
on $T^*Q$ recovers the Levi-Civita connection on one-forms $\alpha\in \Omega^1(Q)$. Although the same notation is used for the connector on $TQ$ and on $T^*Q$, it will be clear from the context which one is meant.

The proof of Lemma \ref{equivalent_form_of_stationarity_conditions} begins by recalling the vector bundle isomorphism $TT^*Q\rightarrow T^*Q\oplus TQ\oplus T^*Q$ given by
\begin{equation*}
X\mapsto \left(\sigma_{T^*Q}(X),T\pi(X),K(X)\right),
\end{equation*}
where $\sigma_{T^*Q}:TT^*Q\rightarrow T^*Q$ is the projection. Therefore, to prove the equivalence it suffices to apply the maps $T\pi$ and $K$ to the equation $\dot\alpha=\pm\,\xi_{T^*Q}(\alpha)+\operatorname{Ver}_\alpha\frac{\partial\ell}{\partial n}$. As we have seen before, applying $T\pi$ yields the first equation in the system (\ref{system_stationarity}). The definition \eqref{connector_def_cotangent} of $K $ and \eqref{connector_property_cotangent} immediately imply the equalities
\[
K(\dot\alpha)=\frac{D}{Dt}\alpha\quad\text{and}\quad K\left(\operatorname{Ver}_\alpha\frac{\partial\ell}{\partial n}\right)=\frac{\partial\ell}{\partial n}\,.
\]
Thus, to finish the proof, it suffices to compute $K\left(\xi_{T^*Q}(\alpha)\right)$. Given $v_n \in T_nQ$, $\alpha_n\in T^\ast_nQ$, and $\xi\in\mathfrak{g}$, we have
\[
\left\langle T^*\Phi^{-1}_{\operatorname{exp}(s\xi)}(\alpha_n),T\Phi_{\operatorname{exp}(s\xi)}(v_n)\right\rangle=\left\langle\alpha_n,v_n\right\rangle.
\]
Taking the $s$-derivative at $s=0$ yields
\begin{equation}\label{relation_connectors}
\left\langle K\left(\xi_{T^*Q}(\alpha_n)\right),v_n\right\rangle+\left\langle\alpha_n,K\left(\xi_{TQ}(v_n)\right)\right\rangle=0.
\end{equation}
Noting the equalities $K(\xi_{TQ}(v_n))=K(T\xi_Q(v_n))=\nabla_{v_n}\xi_Q(n)$, we obtain the formula
\[
K\left(\xi_{T^*Q}(\alpha_n)\right)=-\left\langle\alpha_n,\nabla\xi_Q(n)\right\rangle,
\]
which proves Lemma \ref{equivalent_form_of_stationarity_conditions}, that the stationarity conditions \eqref{stat_cond} and \eqref{system_stationarity} are equivalent for a Riemannian manifold. $\qquad\blacksquare$

\medskip

System \eqref{system_stationarity} may also be obtained directly from the variational principle $\delta S_c=0$, $S_c = \int_0^T\big(\ell(\xi,n)+\langle\alpha,\dot n\mp\,\xi_Q(n)\rangle\big)dt$, by using a Riemannian metric on $Q$. However, we have chosen to derive the stationarity conditions \eqref{optimal_control_condition} or \eqref{optimal_control_condition_A'} together with the constraint $\dot n=\pm\,\xi_Q(n)$ for the functional $S_c$
without introducing a Riemannian metric; see \eqref{intermediate_three}--\eqref{intermediate_two} above.

\subsection*{Lagrangian and Hamiltonian approach} Equations \eqref{generalized_EP} and \eqref{generalized_EP_inverse} can be obtained via Euler-Poincar\'e reduction for the $G$-invariant function $\mathcal{L}:TG\times Q\rightarrow\mathbb{R}$ induced by $\ell$. More precisely, upon fixing $q\in Q$ and defining the Lagrangian $\mathcal{L}_{q}(u_g):=\mathcal{L}(u_g,q)$ on $TG$, one finds that the equations \eqref{generalized_EP} and \eqref{generalized_EP_inverse} are equivalent to the Euler-Lagrange equations for $\mathcal{L}_{q}$ by invoking a generalization of the Euler-Poincar\'e reduction theorem. We refer to \cite{GBTr2010} for a proof of this assertion  and for applications to systems with broken symmetry. If $Q$ is a representation space of $G$, one recovers the Euler-Poincar\'e reduction theorem for semidirect products; see \cite{HoMaRa1998a,HoMaRa1998b}. 

If the Legendre transform $\xi \in \mathfrak{g} \mapsto \frac{\delta \ell}{ \delta \xi} \in \mathfrak{g}^\ast$ is a diffeomorphism, we can form the associated Hamiltonian $h:\mathfrak{g}^*\times Q \rightarrow \mathbb{R}$ defined by 
\[
h(\mu,n):=\langle\mu,\xi\rangle-\ell(\xi,n), 
\quad\hbox{where} \quad
\frac{\delta\ell}{\delta\xi}=\mu
\,.
\]
In this case, the Lagrangian $\mathcal{L}$ is hyperregular on $TG$, the variable $q\in Q$ being considered as a parameter, and we can form the Hamiltonian $\mathcal{H}:T^*G\times Q\rightarrow\mathbb{R}$. More precisely, fixing $q\in Q$, we define
\[
\mathcal{H}_q:=E_q\circ\mathbb{F}\mathcal{L}_q^{-1},
\]
where $E_q$ is the energy associated to the Lagrangian $\mathcal{L}_q:TG\rightarrow\mathbb{R}$ and $\mathbb{F}\mathcal{L}_q:TQ \rightarrow T ^\ast Q$ is the classical Legendre transform of $\mathcal{L}_q$. The function $\mathcal{H}:T^*G\times Q\rightarrow\mathbb{R}$ is then defined by $\mathcal{H}(\alpha_g,q):=\mathcal{H}_q(\alpha_g)$. Equations \eqref{generalized_EP} and \eqref{generalized_EP_inverse} can be written in Hamiltonian form as
\begin{equation}
\left\{
\begin{array}{l}
\vspace{0.2cm}\displaystyle\dot\mu=\mp\,\operatorname{ad}^*_{\frac{\delta h}{\delta
\mu}}\mu-\mathbf{J}\left(\frac{\partial h}{\partial n}\right),\\
\displaystyle\dot n=\left(\frac{\delta h}{\delta\mu}\right)_Q(n),
\end{array}
\right.
\end{equation}
and
\begin{equation}
\left\{
\begin{array}{l}
\vspace{0.2cm}\displaystyle\dot\mu=\pm\operatorname{ad}^*_{\frac{\delta h}{\delta
\mu}}\mu+\mathbf{J}\left(\frac{\partial h}{\partial
n}\right),\\
\displaystyle\dot n=-\left(\frac{\delta h}{\delta\mu}\right)_Q(n),
\end{array}
\right.
\end{equation}
respectively. They are obtained by Poisson reduction of Hamilton's equations for $\mathcal{H}$ on $T^*G\times Q$, where $Q$ is endowed with the zero Poisson structure.

In terms of $h$, the equations \eqref{optimal_control_condition} or \eqref{optimal_control_condition_A'} read
\begin{equation}\label{optimal_control_condition_h}
\mu=\mathbf{J}(\alpha),\quad \dot\alpha=\left(\frac{\delta h}{\delta\mu}\right)_{T^*Q}(\alpha)-\operatorname{Ver}_\alpha\frac{\partial
h}{\partial n},
\end{equation}
and
\begin{equation}\label{optimal_control_condition_A'_h}
\mu=-\mathbf{J}(\alpha),\quad
\dot\alpha=-\left(\frac{\delta h}{\delta\mu}\right)_{T^*Q}(\alpha)-\operatorname{Ver}_\alpha\frac{\partial
h}{\partial n}.
\end{equation}

As in Lemma \ref{equivalent_form_of_stationarity_conditions}, by introducing a Riemannian metric $g$ on $Q$, these equations can be rewritten as
\begin{equation}\label{stationary_h_riemannian}
\mu=\pm\,\mathbf{J}(\alpha),\quad \dot n=\pm \left(\frac{\delta h}{\delta\mu}\right)_Q(n),\quad\frac{D}{Dt}\alpha=\mp\left\langle\alpha,\nabla\left(\frac{\delta h}{\delta\mu}\right)_Q(n)\right\rangle-\frac{\partial
h}{\partial n}\,.
\end{equation}

\subsection{Optimization using penalties}

As before, we consider a left (resp. right) action $\Phi:G\times Q\rightarrow Q$ and a cost function $\ell:\mathfrak{g}\times Q\rightarrow\mathbb{R}$. We suppose that the manifold $Q$ is endowed with a Riemannian metric $g$. The basic idea is to treat the condition $\rm (A)$ or $\rm (A)'$ as a penalty rather than a constraint. Therefore, in the case of condition $\rm (A)$ above, we consider the minimization problem
\begin{equation}\label{minimization_A}
\min_{\xi,n}\int_0^T\left(\ell(\xi,n)+\frac{1}{2\sigma^2}\|\dot n-\xi_Q(n)\|^2\right)dt,
\end{equation}
and if condition $\rm (A)'$ holds, we consider 
\begin{equation}\label{minimization_A'}
\min_{\xi,n}\int_0^T\left(\ell(\xi,n)+\frac{1}{2\sigma^2}\|\dot n+\xi_Q(n)\|^2\right)dt.
\end{equation}
These two problems are subject to the condition
\[
n(0)=n_0\quad\text{and}\quad  n(T)=n_T,
\]
for given $n_0, n_T\in Q$. Here the norm is taken with respect to the Riemannian metric $g$ on $Q$ and $\sigma\neq 0$.

\subsection*{Stationarity conditions} In order to find the critical curves, we consider the variational principle
\begin{equation}\label{unconstrained_variational_principle}
\delta \int_0^T\left(\ell(\xi,n)+\frac{1}{2\sigma^2}\|\dot n\mp\,\xi_Q(n)\|^2\right)dt=0
\end{equation}
for the two curves $(\xi,n):[0,T]\mapsto \mathfrak{g}\times Q$, where $n$ has fixed endpoints. That is, the variation $\delta \xi$ is free and the variation $\delta n$ vanishes at the endpoints.

We will treat condition $\rm (A)$ and $\rm (A)'$ simultaneously. In all the expressions below, the upper sign refers to condition $\rm (A)$ and the lower sign refers to condition $\rm (A)'$. The $\xi$-variation  yields the condition
\begin{equation}\label{condition_1}
\frac{\delta \ell}{\delta \xi}=\pm \frac{1}{\sigma^2}\mathbf{J}(\nu_n^\flat), \quad \text{where} \quad \nu_n : = \dot n\mp\,\xi_Q(n),
\end{equation}
and $\nu_n^\flat: = g(n)( \nu_n, \cdot ) \in T_n^\ast Q$ .
We now compute the variations of $n$, where we denote by $\nabla$ and $D/Dt$ the covariant derivatives associated to the Levi-Civita connection of the metric $g$. For $\delta n=\left.\frac{d}{ds}\right|_{s=0}n_s$, we have
\begin{align*}
&\int_0^T\left(\left\langle\frac{\partial \ell}{\partial n},\delta n\right\rangle+\frac{1}{\sigma^2}\left\langle\nu_n^\flat,\left.\frac{D}{Ds}\right|_{s=0}\dot n\mp \left.\frac{D}{Ds}\right|_{s=0}\xi_Q(n_s)\right\rangle\right)dt\\
&\qquad\qquad =\int_0^T\left(\left\langle\frac{\partial \ell}{\partial n},\delta n\right\rangle-\frac{1}{\sigma^2}\left\langle\frac{D}{Dt}\nu_n^\flat,\delta n\right\rangle \mp\,\frac{1}{\sigma^2}\left\langle\nu_n^\flat, \nabla_{\delta n}\xi_Q(n)\right\rangle\right)dt.
\end{align*}
Upon exchanging the order of derivatives, $\frac{D}{Dt} \frac{d}{ds} = \frac{D}{Ds} \frac{d}{dt}$ (which is allowed because the Levi-Civita connection has no torsion) one finds the equation
\begin{equation}\label{condition_2}
\frac{D}{Dt}\nu_n^\flat=\mp g(\nu_n,\nabla\xi_Q)+\sigma^2\frac{\partial \ell}{\partial n}\,.
\end{equation}
Consequently, $( \xi, n) $ is a solution of \eqref{unconstrained_variational_principle} if and only if \eqref{condition_1} and \eqref{condition_2} hold.
In what follows, equations \eqref{condition_1} and \eqref{condition_2} will be called the \textit{stationarity conditions}.

Note that here, in contrast to the argument in \S\ref{review_COCP}, specific use of the Riemannian metric is made in computing the stationarity equations from the condition $\delta S_d=0$, where
\begin{equation}
\label{def_s_d}
S_d: = \int_0^T\left(\ell(\xi,n)+\frac{1}{2\sigma^2}\|\dot n\mp\,\xi_Q(n)\|^2\right)dt.
\end{equation}
This is natural, because a Riemannian metric is provided by the penalty term in the problem statement. Using the notation
\[
\pi:=\frac{1}{\sigma^2}\nu_n^\flat=\frac{1}{\sigma^2}
\left(\dot n\mp \xi_Q(n)\right)^\flat\in T^*Q\,,
\]
enables the stationarity conditions \eqref{condition_1} and \eqref{condition_2} to be written as
\begin{equation}\label{stationarity_cond_pi}
\frac{\delta\ell}{\delta\xi}=\pm\,\mathbf{J}(\pi),\quad \dot n=\pm\,\xi_Q(n)+\sigma^2\pi^\sharp,\quad \frac{D}{Dt}\pi=\mp \langle\pi,\nabla\xi_Q\rangle+\frac{\partial \ell}{\partial n}\,.
\end{equation}
These equations should be compared with the other stationarity conditions (\ref{optimal_control_condition}) and (\ref{system_stationarity}),
\begin{equation}\label{stationarity_cond_alpha}
\frac{\delta\ell}{\delta\xi}=\pm\,\mathbf{J}(\alpha),\quad \dot n=\pm\,\xi_Q(n),\quad \frac{D}{Dt}\alpha=\mp \langle\alpha,\nabla\xi_Q\rangle+\frac{\partial \ell}{\partial n}\,,
\end{equation}
associated to the Clebsch optimal control problem. These two sets of stationarity conditions are analogous. However, the corresponding variables $\alpha$ and $\pi$ have different origins. Namely, the costate variable $\alpha$ was introduced as the Lagrange multiplier in formulating the constrained Clebsch variational principle (\ref{variational_principle_clebsch}), whereas the variable $\pi$ arises as a canonical momentum, dual to the penalty variable $\nu_n$ in the unconstrained variational principle (\ref{unconstrained_variational_principle}).

Recall from Lemma \ref{equivalent_form_of_stationarity_conditions} that the last two stationarity conditions of the system (\ref{stationarity_cond_alpha}) are equivalent to
\[
\dot\alpha=\pm\,\xi_{T^*Q}(\alpha)+\operatorname{Ver}_\alpha\frac{\partial \ell}{\partial n}\,.
\]
An analogous result concerning the stationarity conditions of the distributed optimal control problem is given by the following lemma. Let $\sharp: = \flat ^{-1}: T ^\ast Q \rightarrow T Q $.
\medskip

\begin{lemma}
\label{equivalent_form_of_stationarity_conditions_distributed} 
The system of two equations
\begin{equation}
\left\{
\begin{array}{l}
\displaystyle\vspace{0.2cm}\dot n=\pm\,\xi_Q(n)+\sigma^2\pi^\sharp\,,\\
\displaystyle\frac{D}{Dt}\pi=\mp \langle\pi,\nabla \xi_Q(n)\rangle+\frac{\partial\ell}{\partial n}
\,,
\end{array}
\right.
\end{equation}
is equivalent to the single equation
\[
\dot\pi=\pm\,\xi_{T^*Q}(\pi)+\operatorname{Ver}_\pi\frac{\partial\ell}{\partial n}+\sigma^2\mathcal{S}(\pi),
\]
where $\mathcal{S}\in\mathfrak{X}(T^*Q)$ is the Hamiltonian vector field associated to the kinetic energy of the Riemannian metric.
\end{lemma}
\medskip

\noindent\textbf{Proof.} It suffices to observe that the vector field $\mathcal{S}$ verifies the properties
\[
K(\mathcal{S}(\alpha))=0\quad\text{and}\quad T\pi(\mathcal{S}(\alpha))=\alpha^\sharp,
\]
for all $\alpha\in T^*Q$. Then the proof is similar to that of Lemma \ref{equivalent_form_of_stationarity_conditions}.
$\qquad\blacksquare$

\medskip

\begin{remark}\label{stationarity_cond_h}\normalfont 
In terms of the Hamiltonian $h$ associated to $\ell$, the stationarity conditions \eqref{condition_1} and \eqref{condition_2} read
\[
\dot n=\pm\left(\frac{\delta h}{\delta \kappa}\right)_Q(n)+\sigma^2\pi^\sharp,\quad \frac{D}{Dt}\pi=\mp \left\langle\pi,\nabla \left(\frac{\delta h}{\delta \kappa}\right)_Q(n)\right\rangle-\frac{\partial h}{\partial n}\,,
\]
or, equivalently,
\[
\dot\pi=\pm\left(\frac{\delta h}{\delta \kappa}\right)_{T^*Q}(\pi)-\operatorname{Ver}_\pi\frac{\partial h}{\partial n}+\sigma^2\mathcal{S}(\pi).
\]
These equations should be compared to their analogues in \eqref{optimal_control_condition_h} -- \eqref{stationary_h_riemannian}.
\quad $\blacklozenge$
\end{remark}

\subsection*{Equations of motion associated to the stationarity conditions} We now compute the differential equation associated to condition \eqref{condition_1}, that is, the analogue of equations \eqref{generalized_EP}, \eqref{generalized_EP_inverse}. The formulation will involve the following $\mathfrak{g}^*$-valued $(1,1)$ tensor field.
\medskip

\begin{definition}\label{def_F_nabla} Consider a Lie group $G$ acting on a Riemannian manifold $(Q, g)$. We define the $\mathfrak{g}^*$-valued $(1,1)$ tensor field $\mathcal{F}^\nabla:T^*Q\times TQ\rightarrow\mathfrak{g}^*$ associated to the Levi-Civita connection $\nabla$ by
\begin{equation}\label{definition_F_nabla}
\left\langle\mathcal{F}^\nabla(\alpha_q,u_q),\eta\right\rangle:=\left\langle\alpha_q,\nabla_{u_q}\eta_Q(q)\right\rangle,
\end{equation}
for all $u_q\in T_qQ$, $\alpha_q\in T^*_qQ$, and $\eta\in\mathfrak{g}$.
\end{definition}
\medskip

The main properties of the tensor field $\mathcal{F}^\nabla$ are given in the following lemmas.
\medskip

\begin{lemma}\label{property_F_connector} 
For all $\alpha_q\in T^*_qQ$, $u_q\in T_qQ$, and $\xi\in\mathfrak{g}$, 
\[
\left\langle \mathcal{F}^\nabla(\alpha_q,u_q), \xi \right\rangle =\left\langle\alpha_q,K(\xi_{TQ}(u_q))\right\rangle=-\left\langle K(\xi_{T^*Q}(\alpha_q)),u_q\right\rangle,
\]
where $K$ denotes the connectors of the covariant derivatives on $TQ$ and $T^*Q$, respectively {\rm (}see formulas \eqref{connector_def}-\eqref{connector_property_cotangent}{\rm )}.
\end{lemma}
\medskip

\noindent\textbf{Proof.} It suffices to use formula \eqref{relation_connectors} in the proof of Lemma \ref{equivalent_form_of_stationarity_conditions}.
$\qquad\blacksquare$

\medskip

The following important property of $\mathcal{F}^\nabla$ is valid when $G$ acts by isometries.
\medskip

\begin{lemma}\label{antisymmetry} If $G$ acts by isometries, then $\mathcal{F}^\nabla$ is antisymmetric, that is
\[
\mathcal{F}^\nabla(\alpha_q,u_q)=-\mathcal{F}^\nabla(u_q^\flat,\alpha_q^\sharp),
\]
for all $u_q\in T_qQ$, $\alpha_q\in T^*_qQ$.
\end{lemma}
\medskip

\noindent\textbf{Proof.} Since $G $ acts by isometries, $\boldsymbol{\pounds}_{ \xi_Q}g = 0$ which implies  $(\nabla\xi_Q)^T=-\nabla\xi_Q$. $\qquad\blacksquare$

\medskip

We also need the following preparatory lemma, valid for any action.
\medskip

\begin{lemma}\label{first_lemma} Let $\mathbf{J}:T^*Q\rightarrow\mathfrak{g}$, $\langle\mathbf{J}(\alpha_q),\xi\rangle=\langle\alpha_q,\xi_Q(q)\rangle$ be the momentum map of the cotangent-lifted action of $G$ on $T^*Q$ and let $g$ be a Riemannian metric on $Q$. Then for a curve $\alpha(t)\in T^*_{q(t)}Q$ we have
\[
\frac{d}{dt}\mathbf{J}(\alpha(t))=\mathbf{J}\left(\frac{D}{Dt}\alpha(t)\right)+\mathcal{F}^\nabla(\alpha(t),\dot q(t))),
\]
where $D/Dt$ and $\nabla$ denote the Levi-Civita covariant derivatives associated to $g$.
\end{lemma}
\medskip

\noindent\textbf{Proof.} For all $\eta\in\mathfrak{g}$, we have
\begin{align*}
\frac{d}{dt}\langle\mathbf{J}(\alpha(t)),\eta\rangle&=\frac{d}{dt}\langle\alpha(t),\eta_Q(q(t))\rangle=\left\langle\frac{D}{Dt}\alpha(t),\eta_Q(q(t))\right\rangle+\left\langle\alpha(t),\frac{D}{Dt}\eta_Q(q(t))\right\rangle\\
&=\left\langle\mathbf{J}\left(\frac{D}{Dt}\alpha(t)\right),\eta\right\rangle
+
\Big\langle\alpha(t),\nabla_{\dot q(t)}\eta_Q( q (t)) \Big\rangle.
\end{align*}
Using the definition of $\mathcal{F}^\nabla$ implies the required formula.$\qquad\blacksquare$

\medskip

Note that this proof of Lemma \ref{first_lemma} did not assume that the metric is $G$-invariant and that the formula is valid for left and right actions.

Lemma \ref{first_lemma}, and equations \eqref{condition_1}, \eqref{condition_2} enable one to compute the motion equations associated to the minimization problems \eqref{minimization_A}, \eqref{minimization_A'} as follows:
\begin{align*}
\frac{d}{dt}\frac{\delta \ell}{\delta \xi}&=\pm\,\frac{d}{dt}\frac{1}{\sigma^2}\mathbf{J}(\nu_n^\flat)=\pm\,\frac{1}{\sigma^2}\mathbf{J}\left(\frac{D}{Dt}\nu_n^\flat\right)\pm\,\frac{1}{\sigma^2}\mathcal{F}^\nabla(\nu_n^\flat,\dot n)\\
&=\pm \mathbf{J}\left(\frac{\partial \ell}{\partial n}\right)-\frac{1}{\sigma^2}\mathbf{J}\left(\langle\nu_n^\flat,\nabla\xi_Q\rangle\right)\pm\,\frac{1}{\sigma^2}\mathcal{F}^\nabla(\nu_n^\flat,\nu_n)+\frac{1}{\sigma^2}\mathcal{F}^\nabla(\nu_n^\flat,\xi_Q(n))\\
&=\pm \mathbf{J}\left(\frac{\partial \ell}{\partial n}\right)\pm\,\frac{1}{\sigma^2}\mathcal{F}^\nabla(\nu_n^\flat,\nu_n)+ (\mp)\frac{1}{\sigma^2}\operatorname{ad}^*_\xi\mathbf{J}(\nu_n^\flat)\\
&=\pm \mathbf{J}\left(\frac{\partial \ell}{\partial n}\right)\pm\,\frac{1}{\sigma^2}\mathcal{F}^\nabla(\nu_n^\flat,\nu_n)\pm (\mp)\operatorname{ad}^*_\xi\frac{\delta\ell}{\delta\xi}\,,
\end{align*}
where in $(\mp)$ one chooses $-$ (resp. $+$) when $G$ acts on $Q$ by a left (resp. right) action; so in the last term there are four choices of sign. Thus, when the penalty is given by $\|\dot n-\xi_Q(n)\|^2$ (condition (A)), the critical curves of the variational principle \eqref{unconstrained_variational_principle} are solutions of
\begin{equation}\label{unconstrained_A}
\left\{
\begin{array}{l}
\vspace{0.2cm}\displaystyle \frac{d}{dt}\frac{\delta \ell}{\delta \xi}=\mp\,\operatorname{ad}^*_\xi\frac{\delta\ell}{\delta\xi}+ \mathbf{J}\left(\frac{\partial \ell}{\partial n}\right)+\frac{1}{\sigma^2}\mathcal{F}^\nabla(\nu_n^\flat,\nu_n)\,,\\
\displaystyle \frac{D}{Dt}\nu_n^\flat=-\langle\nu_n^\flat,\nabla\xi_Q\rangle+\sigma^2\frac{\partial \ell}{\partial n}, \quad \nu_n : = \dot n-\xi_Q(n)
\,.
\end{array}\right.
\end{equation}
When the penalty $\|\dot n+\xi_Q(n)\|^2$ (condition $(\rm A)'$) is chosen instead, one finds,
\begin{equation}\label{unconstrained_A'}
\left\{
\begin{array}{l}
\vspace{0.2cm}\displaystyle
 \frac{d}{dt} \frac{\delta \ell}{\delta \xi}
=
\pm\,\operatorname{ad}^*_\xi\frac{\delta\ell}{\delta\xi}
- 
\mathbf{J}\left(\frac{\partial \ell}{\partial n}\right)
-
\frac{1}{\sigma^2}\mathcal{F}^\nabla(\nu_n^\flat,\nu_n)
\,,\\
\displaystyle \frac{D}{Dt}\nu_n^\flat
=
\langle\nu_n^\flat,\nabla\xi_Q\rangle+\sigma^2\frac{\partial \ell}{\partial n}
\,, \qquad 
\nu_n : = \dot n +\xi_Q(n)\,.
\end{array}\right.
\end{equation}

\begin{remark}\normalfont The motion equations 
\eqref{unconstrained_A} and \eqref{unconstrained_A'} should be compared to the analogous motion equation \eqref{generalized_EP} and \eqref{generalized_EP_inverse}, respectively, obtained by the Clebsch optimal control approach. Note that the term $\mathcal{F}^\nabla(\nu_n^\flat,\nu_n)$ is an additional force term that is due to the presence of the quantity $\nu_n$. The variable $\nu_n=\dot n\pm\,\xi_Q(n)$ measures the inexact matching
and evolves according to the second equation $\frac{D}{Dt}\nu_n^\flat=\pm g(\nu_n,\nabla\xi_Q)+\sigma^2\frac{\partial \ell}{\partial n}$. We shall return to the discussion of inexact matching for images in Section \ref{optimage-reg}.
\quad $\blacklozenge$
\end{remark} 
\medskip

Thanks to Lemma \ref{antisymmetry} we obtain the following important result, when $G$ acts by isometries.
\medskip

\begin{theorem}\label{actions_by_isometries} Let $G$ be a Lie group acting on a manifold $Q$ and let $\ell:\mathfrak{g}\times Q\rightarrow\mathbb{R}$ be a cost function. We consider the two associated Clebsch optimal control  and distributed optimization problems. Suppose that the Riemannian metric used in the penalty term is $G$-invariant. Then the two problems yield the same equations of motion.
\end{theorem}
\medskip

\noindent\textbf{Proof.} It suffices to use Lemma \ref{antisymmetry}, and to compare equations \eqref{unconstrained_A'}, \eqref{unconstrained_A} with equations \eqref{generalized_EP}, \eqref{generalized_EP_inverse}.$\qquad\blacksquare$
\medskip
 
For completeness we rewrite below the equations \eqref{unconstrained_A} and \eqref{unconstrained_A'} in the particular case where $G$ acts by isometries. Using $\mathcal{F}^\nabla(\nu_n^\flat,\nu_n)=0$ and $\nabla\xi_Q^T=-\nabla\xi_Q$, for this case yields 
\begin{equation}\label{unconstrained_A_iso}
\left\{
\begin{array}{l}
\vspace{0.2cm}\displaystyle \frac{d}{dt}\frac{\delta \ell}{\delta \xi}=\mp\,\operatorname{ad}^*_\xi\frac{\delta\ell}{\delta\xi}+ \mathbf{J}\left(\frac{\partial \ell}{\partial n}\right)\\
\displaystyle \frac{D}{Dt}\nu_n=\nabla_{\nu_n}\xi_Q+\sigma^2\frac{\partial \ell}{\partial n}^\sharp, \quad \nu_n : = \dot n-\xi_Q(n)
\end{array}\right.
\end{equation}
and
\begin{equation}\label{unconstrained_A'_iso}
\left\{
\begin{array}{l}
\vspace{0.2cm}\displaystyle \frac{d}{dt}\frac{\delta \ell}{\delta \xi}=\pm\,\operatorname{ad}^*_\xi\frac{\delta\ell}{\delta\xi}- \mathbf{J}\left(\frac{\partial \ell}{\partial n}\right)\\
\displaystyle \frac{D}{Dt}\nu_n=-\nabla_{\nu_n}\xi_Q+\sigma^2\frac{\partial \ell}{\partial n}^\sharp, \quad \nu_n : = \dot n+\xi_Q(n).
\end{array}\right.
\end{equation}

\begin{remark}{\rm 
The remainder of the present paper will investigate these equations as dynamical systems, rather than as optimal control problems. See \cite{HoRaTrYo2004}, in which a similar approach is taken.}
\quad $\blacklozenge$
\end{remark}

\section{Lagrange-Poincar\'e and metamorphosis reduction}\label{sec_LP_and_metamorph}

In this section, we present two Lagrangian reduction approaches that will be useful in understanding the geometry of the equations \eqref{unconstrained_A'}, \eqref{unconstrained_A} associated to the minimization problem \eqref{minimization_A'}, \eqref{minimization_A}.

Let $G$ act on the left (resp. right) on $Q$. Let $L:T(G\times Q)\rightarrow \mathbb{R}$ be a left (resp. right)-invariant Lagrangian under the action of $G$ given by
\[
(u_g,u_q)\mapsto (hu_g,hu_q)\quad\text{resp.}\quad (u_g,u_q)\mapsto (u_gh,u_qh).
\]
Two reduction processes are discussed. The first uses Lagrangian reduction (see \cite{CeMaRa2001}) and the second is a formulation of the reduction used for metamorphosis in 
\cite{HoTrYo2009}.
\medskip

\begin{theorem}{\rm (}Lagrange-Poincar\'e reduction{\rm )} Let $g\in G$ and $q\in Q$ be two curves and define the curves $n:=g^{-1}q\in Q$ and $\xi:=g^{-1}\dot g\in \mathfrak{g}$ $($resp. $n:=qg^{-1}\in Q$ and $\xi:=\dot gg^{-1}\in \mathfrak{g}$$)$.

Then $(g,q)$ is a solution of the Euler-Lagrange equations for $L$ if and only if $(n,\xi)$ is a solution of the Lagrange-Poincar\'e equations
\begin{equation}\label{LP}\!\!\!\!
\left\{
\begin{array}{l}
\vspace{0.2cm}\displaystyle\frac{d}{dt}\frac{\delta \ell_{LP}}{\delta\xi}=\operatorname{ad}^*_\xi\frac{\delta \ell_{LP}}{\delta\xi},\\
\displaystyle\frac{D}{Dt}\frac{\partial \ell_{LP}}{\partial \dot n}-\frac{\partial \ell_{LP}}{\partial n}=0,\quad 
\frac{d}{dt}n = \dot{n},
\end{array}\right.\quad\text{resp.}\quad \left\{
\begin{array}{l}
\vspace{0.2cm}\displaystyle\frac{d}{dt}\frac{\delta \ell_{LP}}{\delta\xi}=-\operatorname{ad}^*_\xi\frac{\delta \ell_{LP}}{\delta\xi},\\
\displaystyle\frac{D}{Dt}\frac{\partial \ell_{LP}}{\partial \dot n}-\frac{\partial \ell_{LP}}{\partial n}=0, \quad 
\frac{d}{dt}n = \dot{n},
\end{array}\right.
\end{equation}
where the Lagrange-Poincar\'e Lagrangian $\ell_{LP}=\ell_{LP}(n,\dot n,\xi):TQ\times \mathfrak{g}\rightarrow\mathbb{R}$ is induced from $L$ by the quotient map
\begin{equation}\label{LP_quotient_map_A'}
T(G\times Q)\rightarrow TQ\times\mathfrak{g},\quad (u_g,u_q)\mapsto (n,\dot n,\xi):=(\nu_n-\xi_Q(n),\xi)
\end{equation}
for $n:=g^{-1}q$, $\nu_n:=g^{-1}u_q$, $\xi:=g^{-1}u_g$ $($resp. $n:=qg^{-1}$, $\nu_n:=u_qg^{-1}$, $\xi:=u_gg^{-1}$$)$.

These equations are equivalent to the variational principle
\[
\delta\int_{0}^{T}\ell_{LP}(n,\dot n,\xi)dt=0,
\]
for arbitrary variations $\delta n$ and constrained variations $\delta\xi=\dot\eta+[\xi,\eta]$ $($resp. $\delta\xi=\dot\eta-[\xi,\eta]$$)$.

In the Lagrange-Poincar\'e equations, $D/Dt$ and $\partial \ell_{LP}/\partial n$ denote the covariant derivative and the partial derivative associated to a fixed torsion free connection $\nabla$ on $Q$.
\end{theorem}
\medskip

\noindent\textbf{Proof.} We treat the case of a left action and apply the results of \cite{CeMaRa2001}. The projection associated to the $G$-action reads $\pi:G\times Q\rightarrow Q$, $\pi(q,g)=g^{-1}q$. Thus, by taking the tangent map, we find $T\pi(u_g,u_q)=\left(g^{-1}u_g-(g^{-1}u_g)_Q(g^{-1}q)\right)$. The adjoint bundle $\operatorname{Ad}(G\times Q)$ can be identified with the trivial vector bundle $Q\times \mathfrak{g}$ via the identification $[(g,q),\xi]\simeq (g^{-1}q,\operatorname{Ad}_{g^{-1}}\xi)$. Using the principal connection $\mathcal{A}(u_g,u_q):=u_gg^{-1}$, the diffeomorphism $(T(G \times Q))/G \cong TQ\times \mathfrak{g}$ is given by $[u _g, u _q] \mapsto 
\left(g^{-1}u_q-(g^{-1}u_g)_Q(g^{-1}q), g ^{-1} u _g\right)$. Thus, 
the Lagrange-Poincar\'e reduction map has the required expression \eqref{LP_quotient_map_A'}. Since the chosen principal connection is flat, we obtain the Lagrange-Poincar\'e equations \eqref{LP}.$\qquad\blacksquare$

\medskip
\,
\medskip

For the same $G$-invariant Lagrangian $L:T(G\times Q)\rightarrow\mathbb{R}$ as before, we define another reduced Lagrangian $\ell_M=\ell_M(\nu_n,\xi):TQ\times\mathfrak{g}\rightarrow\mathbb{R}$ associated to the quotient map
\[
T(G\times Q)\rightarrow TQ\times\mathfrak{g},\quad (u_g,u_q)\mapsto (\nu_n,\xi):=(g^{-1}u_q,g^{-1}u_g),\quad\text{resp.}\quad (\nu_n,\xi):=(u_qg^{-1},u_gg^{-1}).
\]
This reduced Lagrangian differs from the Lagrange-Poincar\'e Lagrangian $\ell_{LP}$ defined above, but one can pass from the one to the other by the vector bundle isomorphism
\begin{equation}\label{vector_bundle_isom}
 TQ\times\mathfrak{g}\rightarrow  TQ\times\mathfrak{g},\quad (\nu_n,\xi)\mapsto (\nu_n-\xi_Q(n),\xi),
\end{equation}
that is, we have
\[
\ell_{LP}(n,\dot n,\xi)=\ell_{M}(n, \dot n+\xi_Q(n),\xi),
\]
for both the left and right cases.
The reduction associated to this quotient map will be called \textit{metamorphosis reduction}, since it is the abstract framework underlying the metamorphosis dynamics described in \cite{HoTrYo2009}.
\medskip

\begin{theorem}{\rm (}Metamorphosis reduction{\rm )} 
\label{met_thm}
Let $g\in G$ and $q\in Q$ be two curves and define the curves $\nu_n:=g^{-1}\dot q\in TQ$ and $\xi:=g^{-1}\dot g\in \mathfrak{g}$ $($resp. $\nu_n:=\dot qg^{-1}\in TQ$ and $\xi:=\dot gg^{-1}\in \mathfrak{g}$$)$.

Then $(g,q)$ is a solution of the Euler-Lagrange equations associated to $L$ if and only if $(\nu,\xi)$ is solution of the equations
\begin{equation}\label{M_left}
\left\{
\begin{array}{l}
\vspace{0.6mm}\displaystyle
\frac{d}{dt}\frac{\delta \ell_M}{\delta\xi}
=\pm\,\operatorname{ad}^*_\xi\frac{\delta \ell_M}{\delta\xi}-\mathbf{J}\left(\frac{\partial \ell_M}{\partial n}\right)-\mathcal{F}^\nabla\left(\frac{\partial \ell_M}{\partial\nu_n},\nu_n\right)
,\\[5mm]
\displaystyle
\frac{D}{Dt}\frac{\partial \ell_M}{\partial \nu_n}=\left\langle\frac{\partial \ell_M}{\partial\nu_n},\nabla\xi_Q\right\rangle+\frac{\partial \ell_M}{\partial n}
\,,\qquad 
\frac{d}{dt}n=\nu_n-\xi_Q(n)
\,,
\end{array}\right.
\end{equation}
where $+$ $($resp. $-$$)$ occurs when $G$ acts on $Q$ by a left $($resp. right$)$ action, and $\mathcal{F}^\nabla$ is the $\mathfrak{g}^*$-valued $(1,1)$ tensor field defined in \eqref{definition_F_nabla}. In \eqref{M_left}, $\partial \ell_M/ \partial n $ and $\partial \ell_M/ \partial \nu_n$ denote the horizontal and fiber derivatives, respectively.

These equations are equivalent to the variational principle
\[
\delta\int_0^T\ell_{M}(\nu,\xi)dt=0,
\]
with variations $\delta\xi=\dot\eta+[\xi,\eta]$ $($resp. $\delta\xi=\dot\eta-[\xi,\eta]$$)$ and $\delta\nu=\frac{D}{Dt}\omega+\nabla_\omega\xi_Q-\nabla_\nu\eta_Q$.
\end{theorem}
\medskip

The proof will use the following lemma.
\medskip

\begin{lemma}\label{second_lemma} Consider the two reduced Lagrangians $\ell_{LP}$ and $\ell_M$. Then we have the relations
\begin{equation}\label{LagPoin-Meta-Lag}
\frac{\delta \ell_{LP}}{\delta\xi}=\frac{\delta \ell_M}{\delta\xi}+\mathbf{J}\left(\frac{\partial \ell_M}{\partial\nu_n}\right),\quad \frac{\partial \ell_{LP}}{\partial n}=\frac{\partial \ell_M}{\partial n}+\left\langle\frac{\partial \ell_M}{\partial\nu_n},\nabla\xi_Q\right\rangle,\quad \frac{\partial \ell_{LP}}{\partial \dot n}=\frac{\partial \ell_M}{\partial \nu_n}.
\end{equation}
\end{lemma} 

\noindent\textbf{Proof.} Using the relation $\ell_{LP}(n,\dot n,\xi)=\ell_{M}(n, \dot n+\xi_Q(n),\xi)$, we easily obtain the first and third expression. For the second we recall that partial derivatives $\frac{\partial \ell_{LP}}{\partial n}, \frac{\partial \ell_M}{\partial n}$ are defined with the help of a connection $\nabla$ on $Q$. Let $c(t)\in T_{m(t)}Q$ be a smooth horizontal curve covering a curve $m(t)\in Q$ and such that $c(0)=\dot n$, $\dot m(0)=u_n\in T_nQ$. By using the decomposition of $TTQ$ into its vertical and horizontal part, we have
\begin{align*}
\left\langle\frac{\partial\ell_{LP}}{\partial n}(n,\dot n,\xi),u_n\right\rangle&=\left.\frac{d}{dt}\right|_{t=0}\ell_{LP}(c(t),\xi)
=\left.\frac{d}{dt}\right|_{t=0}\ell_{M}(c(t)+\xi_Q(m(t)),\xi) \\
&= \mathbf{d}_{TQ}\ell_M(n, \dot{n}, \xi) \left(\left.\frac{d}{dt}\right|_{t=0}c(t)+\left.\frac{d}{dt}\right|_{t=0}\xi_Q(m(t)) \right)\\
&=\left\langle\frac{\partial\ell_M}{\partial n}(\dot n+\xi_Q(n),\xi),T\pi\left(\left.\frac{d}{dt}\right|_{t=0}c(t)+\left.\frac{d}{dt}\right|_{t=0}\xi_Q(m(t))\right)\right\rangle\\
&\qquad +\left\langle\frac{\partial\ell_M}{\partial\nu_n}(\dot n+\xi_Q(n),\xi),K\left(\left.\frac{d}{dt}\right|_{t=0}c(t)+\left.\frac{d}{dt}\right|_{t=0}\xi_Q(m(t))\right)\right\rangle\\
&=\left\langle\frac{\partial\ell_M}{\partial n}(\dot n+\xi_Q(n),\xi),u_n\right\rangle+\left\langle\frac{\partial\ell_M}{\partial\nu_n}(\dot n+\xi_Q(n),\xi),\nabla_{u_n}\xi_Q\right\rangle,
\end{align*}
where $K:TTQ\rightarrow TQ$ denotes the connector map associated to $\nabla$. Here $\mathbf{d}_{TQ} $ is the exterior derivative on $TQ$ and the fourth equality is a general formula valid for linear connections that links the total derivative to the horizontal and vertical derivatives. $\qquad\blacksquare$

\medskip

\noindent\textbf{Proof of Theorem \ref{met_thm}.} We treat simultaneously the case of a left and right action. Using the second equation in \eqref{LP} and Lemma \ref{second_lemma}, we directly obtain the equations
\[
\frac{D}{Dt}\frac{\partial \ell_M}{\partial \nu_n}-\frac{\partial \ell_M}{\partial n}=\left\langle\frac{\partial \ell_M}{\partial\nu_n},\nabla\xi_Q\right\rangle,\quad \frac{d}{dt} n=\nu_n-\xi_Q(n).
\]
By Lemma \ref{first_lemma}, for any $\eta\in\mathfrak{g}$, we have
\begin{align*}
\left\langle\frac{d}{dt}\mathbf{J}\left(\frac{\partial \ell_M}{\partial\nu_n}\right),\eta\right\rangle
&=\left\langle\mathbf{J}\left(\frac{D}{Dt}\frac{\partial \ell_M}{\partial\nu_n}\right),\eta\right\rangle+\left\langle\frac{\partial \ell_M}{\partial\nu_n},\nabla_{\dot n}\eta_Q(n)\right\rangle\\
&=\left\langle\mathbf{J}\left(\frac{\partial \ell_M}{\partial n}\right),\eta\right\rangle+\left\langle\frac{\partial \ell_M}{\partial\nu_n},\nabla_{\eta_Q}\xi_Q(n)\right\rangle\\
&\qquad\qquad +\left\langle\frac{\partial \ell_M}{\partial\nu_n},\nabla_{\nu_n}\eta_Q(n)\right\rangle-\left\langle\frac{\partial \ell_M}{\partial\nu_n},\nabla_{\xi_Q}\eta_Q(n)\right\rangle\\
&=\left\langle\mathbf{J}\left(\frac{\partial \ell_M}{\partial n}\right),\eta\right\rangle+\left\langle\mathcal{F}^\nabla\left(\frac{\partial \ell_M}{\partial\nu_n},\nu_n\right),\eta\right\rangle\mp\left\langle\mathbf{J}\left(\frac{\partial \ell_M}{\partial\nu_n}\right),[\eta,\xi]\right\rangle,
\end{align*}
where we use the equalities $\nabla_{\eta_Q}\xi_Q-\nabla_{\xi_Q}\eta_Q=[\eta_Q,\xi_Q]=\mp[\eta,\xi]_Q$. We thus obtain
\[
\frac{d}{dt}\mathbf{J}\left(\frac{\partial \ell_M}{\partial\nu_n}\right)=\mathbf{J}\left(\frac{\partial \ell_M}{\partial n}\right)+\mathcal{F}^\nabla\left(\frac{\partial \ell_M}{\partial\nu_n},\nu_n\right)\pm\,\operatorname{ad}^*_\xi\mathbf{J}\left(\frac{\partial \ell_M}{\partial\nu_n}\right).
\]
Inserting the formula $\frac{\delta \ell_{LP}}{\delta\xi}=\frac{\delta \ell_M}{\delta\xi}+\mathbf{J}\left(\frac{\partial \ell_M}{\partial\nu_n}\right)$ in the first equation of \eqref{LP} and using the previous expression for $\frac{d}{dt}\mathbf{J}\left(\frac{\partial \ell_M}{\partial\nu_n}\right)$, we get the required equation
\[
\frac{d}{dt}\frac{\delta \ell_M}{\delta\xi}=\pm\,\operatorname{ad}^*_\xi\frac{\delta \ell_M}{\delta\xi}-\mathbf{J}\left(\frac{\partial \ell_M}{\partial n}\right)-\mathcal{F}^\nabla\left(\frac{\partial \ell_M}{\partial\nu_n},\nu_n\right).\qquad\blacksquare
\]

\subsection*{Left (right) reduction and right (left) action} In some applications, we need to consider left-invariant (resp. right-invariant) Lagrangians whereas $G$ acts on $Q$ by a right (resp. left) action. We quickly present here the situation, by giving the main formulas in this case.
Let $G$ act on the left (resp. right) on $Q$. We consider here the case of a right (resp. left) invariant Lagrangian $L:T(G\times Q)\rightarrow \mathbb{R}$ under the action of $G$ given by
\[
(u_g,u_q)\mapsto (u_gh,h^{-1}u_q)\quad\text{resp.}\quad (u_g,u_q)\mapsto (hu_g,u_qh^{-1}).
\]
The Lagrange-Poincar\'e Lagrangian $\ell_{LP}:TQ\times \mathfrak{g}\rightarrow\mathbb{R}$ is now induced by the quotient map
\begin{equation}\label{LP_quotient_map_A}
T(G\times Q)\rightarrow\mathbb{R},\quad (u_g,u_q)\mapsto (n,\dot n,\xi):=(\nu_n+\xi_Q(n),\xi),
\end{equation}
for $n:=gq$, $\nu_n:=g u_q$, $\xi:=u_gg^{-1}$ (resp. $n:=qg$, $\nu_n:=u_qg$, $\xi:=g^{-1}u_g$$)$. The Lagrange-Poincar\'e equations are now given by

\begin{equation}\label{LP_A}
\!\!\!\!
\left\{
\begin{array}{l}
\vspace{0.6mm}\displaystyle\frac{d}{dt}\frac{\delta \ell_{LP}}{\delta\xi}=-\operatorname{ad}^*_\xi\frac{\delta \ell_{LP}}{\delta\xi},\\[4mm]
\vspace{0.2mm}\displaystyle\frac{D}{Dt}\frac{\partial \ell_{LP}}{\partial \dot n}-\frac{\partial \ell_{LP}}{\partial n}=0, \quad \frac{d}{dt}n = \dot{n},
\end{array}\right.\quad\text{resp.}\quad \left\{
\begin{array}{l}
\vspace{0.6mm}\displaystyle\frac{d}{dt}\frac{\delta \ell_{LP}}{\delta\xi}=\operatorname{ad}^*_\xi\frac{\delta \ell_{LP}}{\delta\xi},\\[4mm]
\vspace{0.2mm}\displaystyle\frac{D}{Dt}\frac{\partial \ell_{LP}}{\partial \dot n}-\frac{\partial \ell_{LP}}{\partial n}=0, \quad \frac{d}{dt}n = \dot{n}.
\end{array}\right.
\end{equation}
Note the change in the sign when compared to \eqref{LP}. Note that we now have the relation $\ell_{LP}(n,\dot n,\xi)=\ell_{M}(\dot n-\xi_Q(n),\xi)$. Therefore, the conclusions of Lemma \ref{second_lemma} should be replaced by
\[
\frac{\delta \ell_{LP}}{\delta\xi}=\frac{\delta \ell_M}{\delta\xi}-\mathbf{J}\left(\frac{\partial \ell_M}{\partial\nu_n}\right),\quad \frac{\partial \ell_{LP}}{\partial n}=\frac{\partial \ell_M}{\partial n}-\left\langle\frac{\partial \ell_M}{\partial\nu_n},\nabla\xi_Q\right\rangle,\quad \frac{\partial \ell_{LP}}{\partial \dot n}=\frac{\partial \ell_M}{\partial \nu_n}.
\]
Thus, equations \eqref{M_left} are replaced by
\begin{equation}\label{M_left_right}
\left\{
\begin{array}{l}
\vspace{0.6mm}\displaystyle\frac{d}{dt}\frac{\delta \ell_M}{\delta\xi}=\mp\,\operatorname{ad}^*_\xi\frac{\delta \ell_M}{\delta\xi}+\mathbf{J}\left(\frac{\partial \ell_M}{\partial n}\right)+\mathcal{F}^\nabla\left(\frac{\partial \ell_M}{\partial\nu_n},\nu_n\right)
,\\[4mm]
\vspace{0.2mm}\displaystyle\frac{D}{Dt}\frac{\partial \ell_M}{\partial \nu_n}=-\left\langle\frac{\partial \ell_M}{\partial\nu_n},\nabla\xi_Q\right\rangle+\frac{\partial \ell_M}{\partial n},\quad \frac{d}{dt}n=\nu_n+\xi_Q(n),
\end{array}\right.
\end{equation}
where $-$ (resp. $+$) occurs when $G$ act on $Q$ by a left (resp. right) action.
\medskip

\subsection*{Alternative form of the equations} For completeness, we give here an alternative form for the second and third equations in systems \eqref{M_left}, \eqref{M_left_right}. This alternative form is analogous to that given in Lemmas \ref{equivalent_form_of_stationarity_conditions}, \ref{equivalent_form_of_stationarity_conditions_distributed}, and reads
\begin{equation}\label{alternative_form_Lagrangian_side}
\frac{d}{dt}\frac{\partial \ell_M}{\partial \nu_n}=\pm\,\xi_{T^*Q}\left(\frac{\partial \ell_M}{\partial \nu_n}\right)+\operatorname{Ver}_{\frac{\partial \ell_M}{\partial \nu_n}}\frac{\partial \ell_M}{\partial n}+\operatorname{Hor}_{\frac{\partial \ell_M}{\partial \nu_n}}\nu_n,
\end{equation}
where, for $\alpha_n\in T^*_nQ$, 
$\operatorname{Hor}_{\alpha_n}:T_nQ\rightarrow T_{\alpha_n}T^*Q$ denotes the horizontal lift associated to the Levi-Civita connection on $T^*Q$. Note that we have the formula $\operatorname{Hor}_{\nu_n^\flat}\nu_n=\mathcal{S}(\nu_n)$, where as before, $\mathcal{S}\in \mathfrak{X}(T^*Q)$ is the Hamiltonian vector field associated to the kinetic energy of the Riemannian metric.

\section{Optimization, the Lagrangian approach}
\label{sec_Lagr_approach}

In this section, we show how to obtain the motion equations associated to the distributed optimization problem by using Lagrangian reduction. More precisely, we will use the metamorphosis reduction, starting from the unreduced Lagrangian associated to $\ell$, augmented by the square of the norm of the velocity vector.

Let $G$ act on the left (resp. right) on $Q$ and consider a cost function 
$\ell:=\ell(\xi,n)$ on $\mathfrak{g}\times Q$. Let $\mathcal{L}:TG\times 
Q\rightarrow\mathbb{R}$ be the associated $G$-invariant Lagrangian on 
$TG\times Q$. The definition of $\mathcal{L}$ depends on the condition 
((A) or ${\rm (A)'}$) we want to impose.
\begin{itemize}
\item If $\rm (A)$ holds, we suppose that $\mathcal{L}$ is invariant under the right (resp. left) action
\begin{equation}\label{actions_A}
(u_g,q)\mapsto (u_gh,h^{-1}q),\quad\text{resp.}\quad (u_g,q)\mapsto (hu_g,qh^{-1}),
\end{equation}
i.e., we define $\mathcal{L}(u_g,q):=\ell(u_gg^{-1},gq)$, resp. $\mathcal{L}(u_g,q):=\ell(g^{-1}u_g,qg)$.
\item 
If $\rm (A)'$ holds, we suppose that $\mathcal{L}$ is invariant under the left (resp. right) action
\begin{equation}\label{actions_A'}
(u_g,q)\mapsto (hu_g,hq),\quad\text{resp.}\quad (u_g,q)\mapsto (u_gh,qh),
\end{equation}
i.e., we define $\mathcal{L}(u_g,q):=\ell(g^{-1}u_g,g^{-1}q)$, resp. $\mathcal{L}(u_g,q):=\ell(u_gg^{-1},qg^{-1})$.
\end{itemize}

\subsection*{Definition of the unreduced Lagrangian} The $G$-invariant Lagrangian $\mathcal{L}:TG\times Q\rightarrow\mathbb{R}$ produces the function $\ell$ by reduction. We now want to modify $\mathcal{L}$ in order to obtain, by reduction, the integrand
\begin{equation}\label{integrand}
\ell(\xi,n)+\frac{1}{2\sigma^2}\|\dot n\pm\,\xi_Q(n)\|^2
\end{equation}
of the distributed optimization problem. This will be done by constructing, from $\mathcal{L}$, a $G$-invariant Lagrangian $L$ defined on the tangent bundle $T(G\times Q)$. Of course, the definition of $L$ depends on the condition ((A) or ${\rm (A)'}$) we want to impose.
\begin{itemize}
\item If $\rm (A)$ holds, we define $L:T(G\times Q)\rightarrow\mathbb{R}$ by
\begin{equation}\label{def_L_A}
L(u_g,u_q):=\mathcal{L}(u_g,q)+\frac{1}{2\sigma^2}\|gu_q\|^2,\quad\text{resp.}\quad L(u_g,u_q):=\mathcal{L}(u_g,q)+\frac{1}{2\sigma^2}\|u_qg\|^2.
\end{equation}
\item If $\rm (A)'$ holds, we define $L:T(G\times Q)\rightarrow\mathbb{R}$ by
\begin{equation}\label{def_L_A'}
L(u_g,u_q):=\mathcal{L}(u_g,q)+\frac{1}{2\sigma^2}\|g^{-1}u_q\|^2,\quad\text{resp.}\quad L(u_g,u_q):=\mathcal{L}(u_g,q)+\frac{1}{2\sigma^2}\|u_qg^{-1}\|^2.
\end{equation}
\end{itemize}
Of course, the norm appearing in the second term of the Lagrangian is the same as the norm used in the integrand \eqref{integrand} of the distributed optimization problem. It is associated to a Riemannian metric on the manifold $Q$. The presence of the group action in the second term is needed in order to make the Lagrangian $G$-invariant.

In the particular case where the Riemannian metric is $G$-invariant, the Lagrangian $L$ is simply given by
\[
L(u_g,u_q):=\mathcal{L}(u_g,q)+\frac{1}{2\sigma^2}\|u_q\|^2
\]
and the associated Euler-Lagrange equations for $L $ read
\[
\frac{D}{Dt}\dot q=\sigma^2\frac{\partial\mathcal{L}}{\partial q},\qquad \frac{d}{dt}\frac{\partial\mathcal{L}}{\partial \dot g}-\frac{\partial\mathcal{L}}{\partial g}=0,
\]
where $D/Dt$ denotes the covariant derivative associated to the Riemannian metric on $Q$.
\medskip

\subsection*{Lagrangian reduction} Using the quotient maps \eqref{LP_quotient_map_A} and \eqref{LP_quotient_map_A'} associated to Lagrange-Poincar\'e reduction, we can compute the reduced Lagrangian associated to $L$. When the $G$-invariance \eqref{actions_A} (condition (A)) holds, we get
\[
\ell_{LP}(n,\dot n,\xi)=\ell(\xi,n)+\frac{1}{2\sigma^2}\|\dot n-\xi_Q(n)\|^2,
\]
and when the $G$-invariance \eqref{actions_A'} (condition (A)') holds, we get
\[
\ell_{LP}(n,\dot n,\xi)=\ell(\xi,n)+\frac{1}{2\sigma^2}\|\dot n+\xi_Q(n)\|^2.
\]
We have thus obtained the integrand of the distributed optimization problem by Lagrange-Poincar\'e reduction. However, in order to compute the associated equations of motion, it will be more appropriate to use metamorphosis reduction. For this approach, the reduced Lagrangian is readily seen to be
\[
\ell_{M}(\nu_n,\xi)=\ell(\xi,n)+\frac{1}{2\sigma^2}\|\nu_n\|^2,
\]
in all cases.

We now compute the reduced equations of motions. Since the functional derivatives of $\ell_M$ are
\[
\frac{\delta \ell_M}{\delta\xi}=\frac{\delta \ell}{\delta\xi},\qquad\frac{\partial \ell_M}{\partial \nu_n}=\frac{1}{\sigma^2}\nu_n^\flat,\quad\text{and}\qquad \frac{\partial \ell_M}{\partial n}=\frac{\partial\ell}{\partial n},
\]
the reduced equations \eqref{M_left_right} (associated to condition $\rm (A)$) and \eqref{M_left} (associated to condition $\rm (A)'$) become, respectively
\begin{equation}
\left\{
\begin{array}{l}
\vspace{0.6mm}\displaystyle\frac{d}{dt}\frac{\delta \ell}{\delta\xi}=\mp\,\operatorname{ad}^*_\xi\frac{\delta \ell}{\delta\xi}+\mathbf{J}\left(\frac{\partial \ell}{\partial n}\right)+\frac{1}{\sigma^2}\mathcal{F}^\nabla\left(\nu_n^\flat,\nu_n\right),\\[4mm]
\vspace{0.2mm}\displaystyle\frac{D}{Dt}\nu_n^\flat=-\langle\nu_n^\flat,\nabla\xi_Q\rangle+\sigma^2\frac{\partial \ell}{\partial n},\quad \dot n=\nu_n+\xi_Q(n),
\end{array}\right.
\end{equation}
and
\begin{equation}
\left\{
\begin{array}{l}
\vspace{0.6mm}\displaystyle\frac{d}{dt}\frac{\delta \ell}{\delta\xi}=\pm\,\operatorname{ad}^*_\xi\frac{\delta \ell}{\delta\xi}-\mathbf{J}\left(\frac{\partial \ell}{\partial n}\right)-\frac{1}{\sigma^2}\mathcal{F}^\nabla\left(\nu_n^\flat,\nu_n\right),\\[4mm]
\vspace{0.2mm}\displaystyle\frac{D}{Dt}\nu_n^\flat=\langle\nu_n^\flat,\nabla\xi_Q\rangle+\sigma^2\frac{\partial \ell}{\partial n},\quad \dot n=\nu_n-\xi_Q(n).
\end{array}\right.
\end{equation}
These are exactly the equation \eqref{unconstrained_A} and \eqref{unconstrained_A'} that are verified by the extremals of the distributed optimization problem, obtained here by metamorphosis reduction.
\medskip

\begin{remark}\normalfont The fact that metamorphosis reduction recovers the motion equations verified by the extremals of the distributed optimization problem is natural in the following sense. The extremals are given by the \textit{unconstrained variational principle}
\[
0=\delta S_d=\delta\int_0^T\left(\ell(\xi,n)+\frac{1}{2\sigma^2}\|\dot n\pm\,\xi_Q(n)\|^2\right)dt;
\]
this gives the stationarity conditions \eqref{condition_1}, \eqref{condition_2}.
These imply (but are not equivalent to) the metamorphosis equations \eqref{unconstrained_A}, \eqref{unconstrained_A'} obtained form the same action $S_d$ under \textit{constrained variations}. 
\quad $\blacklozenge$
\end{remark}

\section{Hamilton-Poincar\'e and metamorphosis reduction}
\label{section_HP_metamorphosis}

In this section, we present the Hamiltonian side of the two Lagrangian reduction approaches described in Section \ref{sec_LP_and_metamorph}.

As before, we let $G$ act on the left (resp. right) on $Q$. We consider a left (resp. right)-invariant Hamiltonian $H:T^*(G\times Q)\rightarrow \mathbb{R}$ under the action of $G$ given by
\[
(\alpha_g,\alpha_q)\mapsto (h\alpha_g,h\alpha_q)\quad\text{resp.}\quad (\alpha_g,\alpha_q)\mapsto (\alpha_gh,\alpha_qh).
\]
As before, there are two reduction processes. The first uses Hamilton-Poincar\'e reduction (see \cite{CeMaPeRa2003}) and the second is the Hamiltonian version of the metamorphosis reduction described in Section \ref{sec_LP_and_metamorph}.
\medskip

\begin{theorem}{\rm (}Hamilton-Poincar\'e reduction{\rm )} 
Let $\alpha_g\in T^*G$ and $\alpha_q\in T^*Q$ be two curves 
and define the curves $\pi_n:=g^{-1}\alpha_q\in T^*Q$ and 
$\mu:=g^{-1}\alpha_g+\mathbf{J}(g^{-1}\alpha_q)\in 
\mathfrak{g}^*$ $($resp. $\pi_n:=\alpha_qg^{-1}\in T^*Q$ and 
$\mu:=\alpha_gg^{-1}+\mathbf{J}(\alpha_qg^{-1})\in 
\mathfrak{g}^*$$)$.

Then $(\alpha_g,\alpha_q)$ is a solution of the canonical Hamilton equations for $H$ on $T ^\ast G \times T ^\ast Q $ if and only if $(\pi_n,\mu)$ is a solution of the \textit{Hamilton-Poincar\'e equations}
\begin{equation}\label{HP}
\left\{
\begin{array}{l}
\vspace{0.2cm}\displaystyle\frac{d}{dt}\mu=\operatorname{ad}^*_{\frac{\delta h_{HP}}{\delta\mu}}\mu
\,,\\[4mm]
\displaystyle\frac{d}{dt}n=\frac{\partial h_{HP}}{\partial\pi},\quad \frac{d}{dt}\pi
=-\frac{\partial h_{HP}}{\partial n}
\,,
\end{array}\right.\quad\text{resp.}\quad \left\{
\begin{array}{l}
\vspace{0.2cm}\displaystyle\frac{d}{dt}\mu=-\operatorname{ad}^*_{\frac{\delta h_{HP}}{\delta\mu}}\mu
\,,\\[4mm]
\displaystyle\frac{d}{dt}n=\frac{\partial h_{HP}}{\partial\pi},\quad \frac{d}{dt}\pi
=-\frac{\partial h_{HP}}{\partial n}\,,
\end{array}\right.
\end{equation}
where the Hamilton-Poincar\'e Hamiltonian $h_{HP}=h_{HP}(\pi_n,\mu):T^*Q\times \mathfrak{g}^*\rightarrow\mathbb{R}$ is induced from $H$ by the quotient map
\begin{equation}\label{HP_quotient_map_A'}
T^*(G\times Q)\rightarrow T^*Q\times\mathfrak{g}^*,\quad (\alpha_g,\alpha_q)\mapsto (\pi_n,\mu):=(\pi_n,\kappa+\mathbf{J}(\pi_n))
\end{equation}
for $n:=g^{-1}q$, $\pi_n:=g^{-1}\alpha_q$, $\kappa:=g^{-1}\alpha_g$, $($resp. $n:=qg^{-1}$, $\pi_n:=\alpha_qg^{-1}$, $\kappa:=\alpha_gg^{-1}$$)$.

In the Hamilton-Poincar\'e equations \eqref{HP}, the second equation is written in Darboux coordinates. One can write it intrinsically as
\[
\frac{d}{dt}\pi_n=X_{h_{HP}}(\pi_n),
\]
where $X_{h_{HP}}$ is the Hamiltonian vector field of $h_{HP}$ viewed as a function on $T^*Q$, the variable 
$\mu\in\mathfrak{g}^*$ being considered as a parameter.
\end{theorem}
\medskip

\noindent\textbf{Proof.} We treat the case of a left action and apply the results in \cite{CeMaPeRa2003}. The coadjoint bundle $\operatorname{Ad}^*(G\times Q)$ can be identified with the trivial vector bundle $Q\times \mathfrak{g}^*$ via the identification $[(g,q),\mu]\simeq (g^{-1}q,\operatorname{Ad}^*_g\mu)$. Using the principal connection $\mathcal{A}(u_g,u_q):=u_gg^{-1}$, the diffeomorphism $(T^*(G \times Q))/G \cong T^*Q\times \mathfrak{g}^*$ is given by $[\alpha_g, \alpha_q] \mapsto 
\left(g^{-1}\alpha_q,g^{-1}\alpha_g+\mathbf{J}(g^{-1}\alpha_q)\right)$. Indeed, the horizontal-lift associated to $\mathcal{A}$ reads $\operatorname{Hor}_{(g,q)}:T_nQ\rightarrow T_gG\times T_qQ$, $\operatorname{Hor}_{(g,q)}v_n=(0_g,gv_n)$, its dual map is $\left[\operatorname{Hor}_{(g,q)}\right]^*(\alpha_g,\alpha_q)=g^{-1}\alpha_q$, and the momentum map $\mathbb{J}:T^*(G\times Q)\rightarrow\mathfrak{g}^*$ is $\mathbb{J}(\alpha_g,\alpha_q)=\alpha_gg^{-1}+\mathbf{J}(\alpha_q)$. Thus, 
the Hamilton-Poincar\'e reduction map has the required expression \eqref{HP_quotient_map_A'}. Since the chosen principal connection is flat, we obtain the Hamilton-Poincar\'e equations \eqref{HP}.$\qquad\blacksquare$
\medskip

It is convenient to write the equations of motion \eqref{HP} in matrix form, namely
\begin{equation}
\label{matrix_HP}
\frac{d}{dt}
\begin{bmatrix}
\mu \\[4mm] \pi_n
\end{bmatrix} =
\begin{bmatrix}
\pm \operatorname{ad}^\ast_{ \Box } \mu& 0 \\[4mm]
0 & \Omega_{can} ^ \sharp( \pi_n)
\end{bmatrix}
\begin{bmatrix}
\displaystyle
\frac{ \delta h_{HP}}{ \delta \mu} \\[4mm]
\displaystyle
\mathbf{d}_{T^*Q}h_{HP}
\end{bmatrix}
\end{equation}
where $\mathbf{d}_{T^*Q} $ is the exterior derivative on $T ^\ast Q$.

\bigskip

For the same $G$-invariant Hamiltonian $H:T^*(G\times Q)\rightarrow\mathbb{R}$ as before, we define another reduced Hamiltonian $h_M=h_M(\pi_n,\kappa):T^*Q\times\mathfrak{g}^*\rightarrow\mathbb{R}$ associated to the quotient map
\[
T^*(G\times Q)\rightarrow T^*Q\times\mathfrak{g}^*,
\]
\[\quad (\alpha_g,\alpha_q)\mapsto (\pi_n,\kappa):=(g^{-1}\alpha_q,g^{-1}\alpha_g),\quad\text{resp.}\quad (\pi_n,\kappa):=(\alpha_qg^{-1},\alpha_gg^{-1}).
\]
This reduced Hamiltonian differs from the Hamilton-Poincar\'e Hamiltonian $h_{HP}$ defined above, but one can pass from the one to the other by the vector bundle isomorphism
\[
T^*Q\times\mathfrak{g}^*\rightarrow  T^*Q\times\mathfrak{g}^*,\quad (\pi_n,\mu)\mapsto (\pi_n,\mu-\mathbf{J}(\pi_n)),
\]
that is, we have
\[
h_{HP}(\pi_n,\kappa+\mathbf{J}(\pi_n))=h_{M}(\pi_n,\kappa),
\]
for both the left and right cases. Of course, the previous isomorphism is the dual map to \eqref{vector_bundle_isom}.

As on the Lagrangian side, we fix a Riemannian metric $g$ on $Q$. This allows us to write the reduced Hamilton equation a bit more explicitly.
Note, however, that it is possible to write the reduced Hamilton equations without the help of a metric; see \eqref{reduced_Ham_without_metric} below.
\medskip

\begin{theorem}{\rm (}Metamorphosis reduction{\rm )} Let $\alpha_g\in T^*G$ and $\alpha_q\in T^*Q$ be two curves and define the curves $\pi_n:=g^{-1}\alpha_q\in T^*Q$ and $\kappa:=g^{-1}\alpha_g\in \mathfrak{g}^*$ $($resp. $\pi_n:=\alpha_qg^{-1}\in T^*Q$ and $\kappa:=\alpha_gg^{-1}\in \mathfrak{g}^*$$)$.

Then $(\alpha_g,\alpha_q)$ is a solution of the canonical Hamilton equations for $H$ on $T ^\ast G \times T ^\ast Q $ if and only if $(\pi_n,\kappa)$ is a solution of the equations
\begin{equation}\label{M_Hamiltonian}
\left\{
\begin{array}{l}
\vspace{0.2cm}\displaystyle\frac{d}{dt}\kappa=\pm\,\operatorname{ad}^*_{\frac{\delta h_M}{\delta\kappa}}\kappa+\mathbf{J}\left(\frac{\partial h_M}{\partial n}\right)-\mathcal{F}^{\nabla}\left(\pi_n,\frac{\partial h_M}{\partial \pi_n}\right)
,\\[4mm]
\displaystyle\frac{d}{dt}\pi_n=-\left(\frac{\delta h_M}{\delta\kappa}\right)_{T^*Q}(\pi_n)+X_{h_M}(\pi_n)
\,,
\end{array}\right.
\end{equation}
where $X_{h_M}$ is the Hamiltonian vector field associated to $h_M$ viewed as a function of $\pi_n$.
\end{theorem}
\medskip

\noindent\textbf{Proof.} We treat simultaneously the case of left and right actions and apply Poisson reduction. The reduced Poisson structure on $T ^\ast Q \times \mathfrak{g}^\ast $ associated to the quotient map $(\alpha_g,\alpha_q)\mapsto (g^{-1}\alpha_g,g^{-1}\alpha_q)$, resp. $(\alpha_g,\alpha_q)\mapsto (\alpha_gg^{-1},\alpha_qg^{-1})$ is given for any $f, g \in C ^{\infty}(T ^\ast Q \times \mathfrak{g}^\ast)$ by
\[
\{f, g \}_{T ^\ast Q \times \mathfrak{g}^\ast} =
\mp\left\langle\mu,\left[\frac{\delta
f}{\delta\mu},\frac{\delta
g}{\delta\mu}\right]\right\rangle-\left\langle
\boldsymbol{\mathcal{J}}\!\left(\mathbf{d}f(\pi_n)\right),\frac{\delta
g}{\delta\mu}\right\rangle+\left\langle \boldsymbol{\mathcal{J}}\!\left(\mathbf{d}g(\pi_n)\right),\frac{\delta
f}{\delta\mu}\right\rangle+\left\{f,g\right\}_{T^*Q},
\]
where $\boldsymbol{\mathcal{J}}: T^*(T^*Q) \rightarrow  \mathfrak{g}^\ast$ is the cotangent bundle momentum map and the last term is the canonical Poisson bracket on $T^*Q$; see Proposition 10.3.1 in \cite{MaMiOrPeRa2007}. Consequently, the reduced Hamilton's equation are
\begin{equation}\label{reduced_Ham_without_metric}
\left\{
\begin{array}{l}
\vspace{0.2cm}\displaystyle\frac{d}{dt}\kappa=\pm\,\operatorname{ad}^*_{\frac{\delta h_M}{\delta\kappa}}\kappa+\boldsymbol{\mathcal{J}}\!\left(\mathbf{d}h_M(\pi_n)\right)
\,,\\
\displaystyle\frac{d}{dt}\pi_n=-\left(\frac{\delta h_M}{\delta\kappa}\right)_{T^*Q}(\pi_n)+X_{h_M}(\pi_n)
\,.
\end{array}\right.
\end{equation}
Now it suffices to decompose the derivative $\mathbf{d}h_M$ into the vertical (fiber) and horizontal partial derivatives and use Lemma \ref{property_F_connector} to write
\begin{equation}
\label{big_J_formula}
\boldsymbol{\mathcal{J}}\!\left(\mathbf{d}h_M(\pi_n)\right)=\mathbf{J}\left(\frac{\partial h_M}{\partial n}\right)-\mathcal{F}^{\nabla}\left(\pi_n,\frac{\partial h_M}{\partial \pi_n}\right).
\end{equation}
This proves the result.$\qquad\blacksquare$
\medskip

Equations \eqref{M_Hamiltonian} can be conveniently written in matrix form
\begin{equation}
\label{M_Hamiltonian_matrix}
\frac{d}{dt}
\begin{bmatrix}
\kappa \\[4mm] \pi_n
\end{bmatrix} =
\begin{bmatrix}
\pm \operatorname{ad}^\ast_{ \Box } \kappa& \boldsymbol{ \mathcal{J}}
 \\[4mm]
- \left( \Box \right)_{T ^\ast Q}( \pi_n) & \Omega_{can} ^ \sharp( \pi_n)
\end{bmatrix}
\begin{bmatrix}
\displaystyle
\frac{ \delta h_M}{ \delta \kappa} \\[4mm]
\displaystyle
\mathbf{d}_{T^*Q}h_M
\end{bmatrix}
\end{equation}
where in the (1,2) entry one uses formula \eqref{big_J_formula}. We shall see in Section \ref{representation} that if $Q$ is a representation space of $G $, this formula gives rise to a Lie-Poisson equation on a semidirect product with a cocycle.

\medskip

\subsection*{Left (right) reduction and right (left) action} We quickly present here the equations arising when the Hamiltonian $H:T^*(G\times Q)\rightarrow \mathbb{R}$ is invariant under the action of $G$ given by
\[
(\alpha_g,\alpha_q)\mapsto (\alpha_gh,h^{-1}\alpha_q)
\qquad\text{resp.}\qquad 
(\alpha_g,\alpha_q)\mapsto (h\alpha_g,\alpha_qh^{-1}).
\]
The Hamilton-Poincar\'e Hamiltonian $h_{HP}:T^*Q\times \mathfrak{g}^*\rightarrow\mathbb{R}$ is now induced by the quotient map
\begin{equation}\label{HP_quotient_map_A}
T^*(G\times Q)\rightarrow\mathbb{R},\quad (\alpha_g,\alpha_q)\mapsto (\pi_n,\mu):=(\pi_n,\kappa-\mathbf{J}(\pi_n)),
\end{equation}
for $n:=gq$, $\pi_n:=g \alpha_q$, $\kappa:=\alpha_gg^{-1}$ (resp. $n:=qg$, $\pi_n:=\alpha_qg$, $\kappa:=g^{-1}u_g$$)$. The resulting Hamilton-Poincar\'e equations are given by
\begin{equation}\label{HP_condition_A}
\left\{
\begin{array}{l}
\vspace{0.2cm}\displaystyle\frac{d}{dt}\mu
=-\operatorname{ad}^*_{\frac{\delta h_{HP}}{\delta\mu}}\mu
\,,\\
\displaystyle
\frac{d}{dt}n
=\frac{\partial h_{HP}}{\partial\pi},\quad \frac{d}{dt}\pi
=-\,\frac{\partial h_{HP}}{\partial n}
\,,
\end{array}\right.
\quad\text{resp.}\quad 
\left\{
\begin{array}{l}
\vspace{0.2cm}\displaystyle
\frac{d}{dt}\mu=\operatorname{ad}^*_{\frac{\delta h_{HP}}{\delta\mu}}\mu
\,,\\
\displaystyle\frac{d}{dt}n
=\frac{\partial h_{HP}}{\partial\pi},\quad \frac{d}{dt}\pi
=-\,\frac{\partial h_{HP}}{\partial n}
\,.
\end{array}
\right.
\end{equation}
Note the change in the sign when compared to \eqref{HP}. Note that we now have the relation $h_{HP}(\pi_n,\kappa-\mathbf{J}(\pi_n))=h_{M}(\pi_n,\kappa)$.

Likewise, equations \eqref{M_Hamiltonian} are replaced by
\begin{equation}\label{M_Hamiltonian_A}
\left\{
\begin{array}{l}
\vspace{0.2cm}\displaystyle\frac{d}{dt}\kappa=\mp\,\operatorname{ad}^*_{\frac{\delta h_M}{\delta\kappa}}\kappa-\mathbf{J}\left(\frac{\partial h_M}{\partial n}\right)+\mathcal{F}^\nabla\left(\pi_n,\frac{\partial h_M}{\partial \pi_n}\right),\\
\displaystyle\frac{d}{dt}\pi_n=\left(\frac{\delta h_M}{\delta\kappa}\right)_{T^*Q}(\pi_n)+X_{h_M}(\pi_n),
\end{array}\right.
\end{equation}
where $-$ (resp. $+$) occurs when $G$ acts on $Q$ by a left (resp. right) action. As in  \eqref{matrix_HP} and \eqref{M_Hamiltonian_matrix},  equations (\ref{HP_condition_A}) and (\ref{M_Hamiltonian_A}) may be re-expressed in matrix form as
\begin{align}
\label{HP_A}
\frac{d}{dt}
\begin{bmatrix}
\mu \\[4mm] \pi_n
\end{bmatrix} &=
\begin{bmatrix}
\mp \operatorname{ad}^\ast_{ \Box } \mu& 0 \\[4mm]
0 & \Omega_{can} ^ \sharp( \pi_n)
\end{bmatrix}
\begin{bmatrix}
\displaystyle
\frac{ \delta h_{HP}}{ \delta \mu} \\[4mm]
\displaystyle
\mathbf{d}_{T^*Q}h_{HP}
\end{bmatrix}
,\\ \nonumber \\
\label{M_Hamiltonian_matrix_A}
\frac{d}{dt}
\begin{bmatrix}
\kappa \\[4mm] \pi_n
\end{bmatrix} &=
\begin{bmatrix}
\mp \operatorname{ad}^\ast_{ \Box } \kappa& -\boldsymbol{ \mathcal{J}}
 \\[4mm]
\left( \Box \right)_{T ^\ast Q}( \pi_n) & \Omega_{can} ^ \sharp( \pi_n)
\end{bmatrix}
\begin{bmatrix}
\displaystyle
\frac{ \delta h_M}{ \delta \kappa} \\[4mm]
\displaystyle
\mathbf{d}_{T^*Q}h_M
\end{bmatrix}.
\end{align}
\medskip

\begin{remark}\normalfont (Link with the untangling map) Recall that the vector bundle isomorphism
\[
(\xi,\nu_n)\in \mathfrak{g}\times TN\mapsto (\xi,\nu_n\pm \xi_Q(n))=(\xi,n,\dot n)\in \mathfrak{g}\times TN
\]
allows one to pass from the metamorphosis reduced equation to the Lagrange-Poincar\'e equations. Its dual map
\[
(\mu,\pi_n)\in \mathfrak{g}^*\times T^*N\mapsto (\mu\pm\,\mathbf{J}(\pi_n),\pi_n)=(\kappa,\pi_n)\in \mathfrak{g}^*\times T^*N,
\]
naturally passes from the Hamilton-Poincar\'e description to the metamorphosis approach. The inverse of this map is known as the \textit{untangling map} in applications (\cite{Holm1986}) since it transforms the Hamiltonian structure of the metamorphosis equation into the direct sum of the Lie-Poisson bracket on $\mathfrak{g}^*$ and the canonical Poisson bracket on $T^*N$; see \eqref{HP}-\eqref{M_Hamiltonian_matrix} and \eqref{HP_A}-\eqref{M_Hamiltonian_matrix_A}. Recent theoretical developments and new  applications of the untangling map appear in \cite{GBTr2010}. 
\quad $\blacklozenge$
\end{remark}

\subsection*{Legendre transformation and alternative formulation} When the Hamiltonian $H$ comes from a Lagrangian $L$ by Legendre transformation, then we have the following relations between the reduced objects:
\[
h_M(\pi_n,\kappa)=\langle\pi_n,\nu_n\rangle+\langle\kappa,\xi\rangle-\ell_M(\nu_n,\xi),\quad \kappa=\frac{\delta \ell_M}{\delta\xi},\quad \pi_n=\frac{\partial\ell_M}{\partial\nu_n}.
\]
The partial derivatives with respect to $n$ are related by the formula
\[
\frac{\partial h_M}{\partial n}=-\,\frac{\partial \ell_M}{\partial n}.
\]
In this case, the reduced equations on the Lagrangian  and Hamiltonian side (\eqref{M_left}, \eqref{M_left_right} and \eqref{M_Hamiltonian}, \eqref{M_Hamiltonian_A}) are readily seen to be equivalent. To see this, it suffices to use the formula
\[
X_h(\alpha_n)=\operatorname{Hor}_{\alpha_n}\frac{\partial h}{\partial\alpha_n}-\operatorname{Ver}_{\alpha_n}\frac{\partial h}{\partial n}
\]
for the Hamiltonian vector field, together with the alternative formulation \eqref{alternative_form_Lagrangian_side} for the reduced equations on the Lagrangian side. This also shows that the second equation of the systems \eqref{M_Hamiltonian} and \eqref{M_Hamiltonian_A} can be equivalently written as
\[
\frac{D}{Dt}\pi_n=\mp\left\langle \pi_n,\nabla\left(\frac{\delta h_M}{\delta\kappa}\right)_Q(n)\right\rangle-\frac{\partial h_M}{\partial n},\qquad\frac{d}{dt}n=\pm\left(\frac{\delta h_M}{\delta\kappa}\right)_Q(n)+\frac{\partial h_M}{\partial\pi_n}.
\]

\section{Optimization, the Hamiltonian approach}
\label{optimal_hamiltonian_section}

Suppose we are given a left (resp. right) action of $G$ on $Q$ and a cost function $\ell=\ell(\xi,q)$ on $\mathfrak{g}\times Q$. Let the map $\xi\mapsto \frac{\delta\ell}{\delta\xi}$ be a diffeomorphism and  consider the associated Hamiltonian $h:\mathfrak{g}^*\times Q \rightarrow \mathbb{R}$ defined by
\[
h(\mu,q):=\langle\mu,\xi\rangle-\ell(\xi,q),\quad \frac{\delta\ell}{\delta\xi}=\mu.
\]
As in Section \ref{sec_Lagr_approach} for $\ell$, the function $h$ induces a $G$-invariant function $\mathcal{H}:T^*G\times Q\rightarrow\mathbb{R}$. Of course, $\mathcal{H}$ can be obtained from $\mathcal{L}$ by a Legendre transformation, the variable $q$ being considered as a parameter. Recall that given a Riemannian metric $g$ on $Q$, we associated to $\mathcal{L}$ a $G$-invariant Lagrangian on $T(G\times Q)$ by adding to $\mathcal{L}$ a $G$-invariant expression involving the norm of the vector in $TQ$; see \eqref{def_L_A}, \eqref{def_L_A'}. For example, in the case of condition ${\rm (A)'}$ and if $G$
acts on the left we have defined
\[
L(u_g,u_q):=\mathcal{L}(u_g,q)+\frac{1}{2\sigma^2}\|g^{-1}u_q\|^2.
\]
Taking the Legendre transformation of this hyperregular Lagrangian yields the $G$-invariant Hamiltonian $H$ on $T^*(G\times Q)$ given by
\[
H(\alpha_g,\alpha_q)=\mathcal{H}(\alpha_g,q)+\frac{\sigma^2}{2}\|g^{-1}\alpha_q\|^2.
\]
The reduced Hamiltonian associated to metamorphosis reduction reads
\[
h_M(\pi_n,\kappa)=h(\kappa,n)+\frac{\sigma^2}{2}\|\pi_n\|^2.
\]
When condition $\rm (A)'$ is assumed, the reduced Hamilton-Poincar\'e Hamiltonian reads
\[
h_{HP}(\pi_n,\mu)=h_M(\pi_n,\mu-\mathbf{J}(\pi_n))=h(\mu-\mathbf{J}(\pi_n),n)+\frac{\sigma^2}{2}\|\pi_n\|^2.
\]
In the case of condition $\rm (A)$, we have
\[
h_{HP}(\pi_n,\mu)=h_M(\pi_n,\mu+\mathbf{J}(\pi_n))=h(\mu+\mathbf{J}(\pi_n),n)+\frac{\sigma^2}{2}\|\pi_n\|^2.
\]
Using the relations
\[
\frac{\partial h_M}{\partial\pi_n}=\sigma^2\pi_n^\sharp\quad\text{and}\quad \frac{\partial h_M}{\partial n}=\frac{\partial h}{\partial n},
\]
the reduced equations \eqref{M_Hamiltonian} and \eqref{M_Hamiltonian_A} become, respectively
\begin{equation}
\left\{
\begin{array}{l}
\vspace{0.2cm}\displaystyle
\frac{d}{dt}\kappa=\pm\,\operatorname{ad}^*_{\frac{\delta h}{\delta\kappa}}\kappa+\mathbf{J}\left(\frac{\partial h}{\partial n}\right)
,\\
\displaystyle
\frac{d}{dt}\pi_n=-\left(\frac{\delta h}{\delta\kappa}\right)_{T^*Q}(\pi_n)-\operatorname{Ver}_{\pi_n}\frac{\partial h}{\partial n}+\sigma^2\mathcal{S}(\pi_n)
\,,
\end{array}\right.
\end{equation}
and
\begin{equation}
\left\{
\begin{array}{l}
\vspace{0.2cm}\displaystyle
\frac{d}{dt}\kappa
=\mp\,\operatorname{ad}^*_{\frac{\delta h}{\delta\kappa}}\kappa
-\mathbf{J}\left(\frac{\partial h}{\partial n}\right)
,\\
\displaystyle\frac{d}{dt}\pi_n=\left(\frac{\delta h}{\delta\kappa}\right)_{T^*Q}(\pi_n)-\operatorname{Ver}_{\pi_n}\frac{\partial h}{\partial n}+\sigma^2\mathcal{S}(\pi_n),
\end{array}\right.
\end{equation}
where $\mathcal{S}\in\mathfrak{X}(T^*Q)$ is the Hamiltonian vector field associated to the kinetic energy
\[
\frac{1}{2}\|\pi_n\|^2=\frac{1}{2}g(\pi_n^\sharp,\pi_n^\sharp)
\,.
\]
These equations recover the motion equations associated to the distributed optimization, in Hamiltonian form, cf. Remark \ref{stationarity_cond_h}.\\

\section{Examples}
\label{examples_section}

In this section we apply the general theory to various group actions.

\subsection{Action by representation and advected quantities}\label{representation}

Let $G$ be a Lie group acting by left (resp. right) representation on the dual vector space $Q=V^*$. Given a Lie algebra element $\xi$, we denote by $\xi_{V^*}(a)=\xi a$ (resp. $\xi_{V^*}(a)= a\xi$) the associated infinitesimal generator.
Using the diamond operator $\diamond:V\times V^*\rightarrow\mathfrak{g}^*$ defined for $p \in V$ and $a \in V ^\ast$ by $\langle p\diamond a,\xi\rangle:=-\langle \xi a,p\rangle$, (resp. $\langle p\diamond a,\xi\rangle:=-\langle a\xi,p\rangle$), for any $\xi\in \mathfrak{g}$,  the cotangent bundle momentum map is $\mathbf{J}(a,p)=-\,p\diamond a$.  

\subsection*{Metamorphosis reduction and Lie-Poisson formulation with cocycles} Assume that $V $ is a left representation space of $G $ and that reduction has been performed on the left. The other cases have similar formulations. In view of the identities above, equations \eqref{M_left} become
\[
\left\{
\begin{array}{l}
\vspace{0.2cm}\displaystyle\frac{d}{dt}\frac{\delta \ell_M}{\delta\xi}=\operatorname{ad}^*_\xi\frac{\delta \ell_M}{\delta\xi}+\frac{\delta \ell_M}{\delta a}\diamond a+\frac{\delta\ell_M}{\delta\nu} \diamond \nu\\
\displaystyle\frac{d}{dt}\frac{\delta\ell_M}{\delta\nu} = - \xi \frac{\delta\ell_M}{\delta\nu}+\frac{\delta\ell_M}{\delta a},\quad \frac{d}{dt} a = - \xi a + \nu,
\end{array}
\right.
\]
where $\ell_M = \ell_M( \xi, a , \nu): \mathfrak{g}\times V ^\ast \times V ^\ast \rightarrow \mathbb{R}$ is the reduced Lagrangian. Performing the Legendre transformation $h_M( \kappa, a , \pi) : = \left\langle\kappa, \xi \right\rangle + \left\langle \pi, \nu \right\rangle- \ell_M( \xi, a ,\nu) $ where
\[
\frac{\delta \ell_M}{ \delta \xi} = \kappa, \quad \frac{\delta \ell_M}{ \delta \nu} = \pi,
\]
one finds the corresponding Hamilton equations for $h_M = h_M( \kappa, a , \pi): \mathfrak{g}^\ast \times V ^\ast\times V \rightarrow \mathbb{R}$ as
\begin{equation}
\label{M_Hamiltonian_rep}
\left\{
\begin{array}{l}
\vspace{0.2cm}\displaystyle\frac{d}{dt}\kappa=\operatorname{ad}^*_{ \frac{\delta h_M}{ \delta \kappa}} \kappa -\frac{\delta h_M}{\delta a}\diamond a+ \pi \diamond \frac{\delta h_M}{\delta\pi} \,,\\
\displaystyle\frac{d}{dt}\pi = - \frac{\delta h_M}{ \delta \kappa} \pi -\frac{\delta h_M}{\delta a},\quad \frac{d}{dt} a = - \frac{\delta h_M}{ \delta \kappa} a + \frac{\delta h_M}{ \delta \pi}\,.
\end{array}
\right.
\end{equation}
These equations recover \eqref{M_Hamiltonian} for the case of a left $G $-representation.

Note that the inverse Legendre transformation (assuming it is a diffeomorphism) is given by $\delta h_M/ \delta \kappa =  \xi$, $\delta h_M/ \delta \pi=  \nu$ and that $\delta h_M/ \delta a  = - \delta \ell_M/ \delta a$.

The proof of the following theorem is a direct verification.
\medskip

\begin{theorem}
The equations of motion \eqref{M_Hamiltonian_rep} are Lie-Poisson on $( \mathfrak{g} \,\circledS\,(V \times V ^\ast))^*$ with the cocycle $C :( V \times V ^\ast) \times ( V \times V ^\ast) \rightarrow \mathbb{R}$ given by the canonical symplectic structure $\Omega_{can} $ on $T ^\ast V  = V \times V^\ast $, where the $\mathfrak{g}$-left representation on $V \times V ^\ast$ is given by $(\xi, v) \mapsto \xi v$, $( \xi, \nu) \mapsto \xi \nu$, for $\xi \in \mathfrak{g}$, $v \in V $, $\nu\in V ^\ast$. Thus these equations can be written in matrix form as
{\small
\begin{equation} \label{matrix_representation}
\frac{\partial}{\partial t}
\left[ \begin{array}{c} 
\kappa  \\ a \\ \pi 
\end{array}\right]
= 
\left[ \begin{array}{ccc} 
\quad {\rm ad}^\ast_\Box\,\kappa \quad  & 
-\Box \diamond a  \quad     &\pi \diamond \Box      
\\ 
-\Box\, a  &  0 & 1
\\
-\Box\, \pi&  -1 & 0
\end{array} \right]
\left[ \begin{array}{c} 
\delta h_M/\delta \kappa \\ 
\delta h_M/\delta a\\
\delta h_M/\delta \pi
\end{array}\right].
\end{equation}
}

\end{theorem}

\subsection*{The Clebsch optimal control approach} Given a cost function $\ell:\mathfrak{g}\times V^*\rightarrow\mathbb{R}$, the Clebsch optimal control problem with condition $\rm (A)'$ yields (for left representation) the stationarity conditions
\begin{equation}\label{stat_cond_repres_clebsch_A'}
\frac{\delta\ell}{\delta\xi}=-\,\mathbf{J}(a,p)=p\diamond a
,\qquad \dot a=-\,\xi a, \qquad\dot p=-\,\xi p+\frac{\partial\ell}{\partial a}
\,.
\end{equation}
For a representation on the right, one replaces $\xi a$, $\xi p$ by $a\xi$, $p\xi$. These equations imply the Euler-Poincar\'e equations of motion
\[
\frac{d}{dt}\frac{\delta \ell}{\delta\xi}=\pm\,\operatorname{ad}^*_\xi\frac{\delta \ell}{\delta\xi}+\frac{\partial \ell}{\partial a}\diamond a
\,.
\]
When condition $\rm (A)$ is assumed, we get the stationarity conditions
\begin{equation}\label{stat_cond_repres_clebsch_A}
\frac{\delta\ell}{\delta\xi}=\mathbf{J}(a,p)=-p\diamond a,
\qquad \dot a=\xi a, \qquad\dot p=\xi p+\frac{\partial\ell}{\partial a}
\end{equation}
and the motion equations
\[
\frac{d}{dt}\frac{\delta \ell}{\delta\xi}=\mp\,\operatorname{ad}^*_\xi\frac{\delta \ell}{\delta\xi}-\frac{\partial \ell}{\partial a}\diamond a.
\]
These are the \textit{Euler-Poincar\'e equations for semidirect products}, useful for the study of physical systems with advected quantities; see \cite{HoMaRa1998a,HoMaRa1998b}.

Note that when the Lagrangian $\ell$ is given by the kinetic energy associated to an inner product on $\mathfrak{g}$, the control is given by $\xi=\pm (p\diamond a)^\sharp$, where $\sharp: \mathfrak{g}^\ast \rightarrow \mathfrak{g}$ is associated to the inner product on $\mathfrak{g}$. We get the equations
\[
\dot a+(p\diamond a)^\sharp  a =0,\quad \dot p +(p\diamond a)^\sharp p=0.
\]
This is the abstract formulation of the \textit{double bracket equations}; see \S\ref{adj_repr} below.
\medskip

\subsection*{The distributed optimization approach}

In order to state the optimization problem with penalty, we endow $V^*$ with a inner product. The corresponding functional is thus 
\[
\ell(\xi,a)+\frac{1}{2\sigma^2}\|\dot a\pm \xi a\|^2.
\]
The Levi-Civita covariant derivative $\nabla$ is the ordinary derivative; therefore we have $\nabla_b\xi_{V^*}(a)=\xi b$ (resp. $\nabla_b\xi_{V^*}(a)=b\xi$), for all $a,b\in V^*$. We thus obtain the expression $\mathcal{F}^\nabla((a,v),(a,b))=-v\diamond b$. If condition $(\rm A)'$ is assumed, the motion equations \eqref{unconstrained_A'} read
\begin{equation}\label{A'_representation}
\left\{
\begin{array}{l}
\vspace{0.2cm}\displaystyle\frac{d}{dt}\frac{\delta \ell}{\delta\xi}=\operatorname{ad}^*_\xi\frac{\delta \ell}{\delta\xi}+\frac{\delta \ell}{\delta a}\diamond a+\frac{1}{\sigma^2}\nu^\flat\diamond\nu\\
\displaystyle\frac{d}{dt}\nu^\flat-\sigma^2\frac{\delta\ell}{\delta a}=-\xi\nu^\flat,\quad \nu=\dot a+\xi a,
\end{array}
\right.\quad\text{resp.}\quad
\left\{
\begin{array}{l}
\vspace{0.2cm}\displaystyle\frac{d}{dt}\frac{\delta \ell}{\delta\xi}=-\operatorname{ad}^*_\xi\frac{\delta \ell}{\delta\xi}+\frac{\delta \ell}{\delta a}\diamond a+\frac{1}{\sigma^2}\nu^\flat\diamond\nu\\
\displaystyle\frac{d}{dt}\nu^\flat-\sigma^2\frac{\delta\ell}{\delta a}=-\nu^\flat\xi,\quad \nu=\dot a+a\xi,
\end{array}
\right.
\end{equation}
where $\flat:V^*\rightarrow V$ is the flat isomorphism associated to the inner product on $V^*$. When condition (A) is assumed, we have (see \eqref{unconstrained_A})
\begin{equation}\label{A_representation}
\left\{
\begin{array}{l}
\vspace{0.2cm}\displaystyle\frac{d}{dt}\frac{\delta \ell}{\delta\xi}=-\operatorname{ad}^*_\xi\frac{\delta \ell}{\delta\xi}-\frac{\delta \ell}{\delta a}\diamond a-\frac{1}{\sigma^2}\nu^\flat\diamond\nu\\
\displaystyle\frac{d}{dt}\nu^\flat-\sigma^2\frac{\delta\ell}{\delta a}=\xi\nu^\flat,\quad \nu=\dot a-\xi a,
\end{array}
\right.\quad\text{resp.}\quad
\left\{
\begin{array}{l}
\vspace{0.2cm}\displaystyle\frac{d}{dt}\frac{\delta \ell}{\delta\xi}=\operatorname{ad}^*_\xi\frac{\delta \ell}{\delta\xi}-\frac{\delta \ell}{\delta a}\diamond a-\frac{1}{\sigma^2}\nu^\flat\diamond\nu\\
\displaystyle\frac{d}{dt}\nu^\flat-\sigma^2\frac{\delta\ell}{\delta a}=\nu^\flat\xi,\quad \nu=\dot a-a\xi,
\end{array}
\right.
\end{equation}
As we have seen in the general theory, these motion equations arise by metamorphosis reduction associated to the Lagrangian $\mathcal{L}(u_g,a)+\frac{1}{2\sigma^2}\|\dot a\|^2$. They can be obtained by the stationarity conditions \eqref{condition_1}, \eqref{condition_2}. In our case, for a  left representation they read
\[
\frac{\delta\ell}{\delta \xi}=\pm\,\frac{1}{\sigma^2}\nu^\flat\diamond a,\quad \dot a=\mp \xi a +\nu,\quad \dot \nu ^\flat =\mp \xi\nu^\flat+\sigma^2\frac{\partial\ell}{\partial a}.
\]
As usual, to compare these conditions with the stationarity conditions \eqref{stat_cond_repres_clebsch_A'}, \eqref{stat_cond_repres_clebsch_A} given by the Clebsch optimal control approach, we define
\[
p:=\frac{1}{\sigma^2}\nu^\flat\in V,
\]
and we get
\begin{equation}\label{stationarity_conditions_p_representation}
\frac{\delta\ell}{\delta \xi}=\pm\, p\diamond a
\,,\qquad 
\dot a=\mp\, \xi a +\sigma^2p^\sharp
\,,\qquad 
\dot p =\mp\, \xi p+\frac{\partial\ell}{\partial a}
\,.
\end{equation}
For a right representation one simply replaces $\xi a$, $\xi p$ by $a\xi$, $p\xi$.

When the Lagrangian $\ell$ is given by the kinetic energy associated to the inner product on $\mathfrak{g}$, the control is given by $\xi=\pm (p\diamond a)^\sharp$, and we get the equations
\[
\dot a+(p\diamond a)^\sharp  a =\sigma^2p^\sharp,
\qquad \dot p +(p\diamond a)^\sharp p=0.
\]
This is the abstract formulation of the \textit{double bracket equations}, modified by the extra term $\sigma^2p^\sharp$;  see \S\ref{adj_repr} below. Note that in the formula above, there are two different sharp operators, $\sharp: \mathfrak{g}^*\rightarrow\mathfrak{g}$ and $\sharp: V\rightarrow V^*$, associated to the inner products on $\mathfrak{g}$ and $V^*$, respectively.

Note that, consistently with Theorem \ref{actions_by_isometries}, if the inner product is $G$-invariant, then $\nu^\flat\diamond\nu=0$. This has already been noticed in the Remark \ref{rem_isotropic_K} of the introduction, for the case of an isotropic inner product.

\subsubsection{Heavy top}\label{subsec-HT}
Consider the evolution equations  for a state system in the frame of a rotating body
\begin{equation}\label{state-eqns-HT}
\dot{\mathbf{X}} = \mathbf{X} \times  \boldsymbol{\Omega} 
\end{equation}
for vector state and control variables $\mathbf{X},\boldsymbol{\Omega}\in\mathbb{R}^3$ related to the rotation matrix $O\in SO(3)$ by $\mathbf{X}=O^{-1}\mathbf{\hat{z}}$  and $\boldsymbol{\Omega\times} =O^{-1}\dot{O}\in\mathfrak{so}(3)$. These vectors are, respectively,  the vertical spatial axis as seen from the rotating body and  the body angular velocity vector. 

We choose to optimize a cost functional consisting of the difference between the rotational kinetic energy and the gravitational potential energy, subject to a penalty imposed by the state system (\ref{state-eqns-HT}). This cost function is
\begin{align}
S_d &=\int_0^T \left(\ell(\boldsymbol{\Omega}, \mathbf{X})+\frac{1}{2\sigma^2}\big\|\dot{\mathbf{X}}
+
\boldsymbol{\Omega}\times \mathbf{X}\big\|^2\right){\rm d} t
\\
&= \int_0^T \left(
 \frac{1}{2}  \mathbb{I}  \boldsymbol{\Omega} \cdot \boldsymbol{\Omega}
-\,
mg\boldsymbol{\chi} \cdot \mathbf{X}
+
\frac{1}{2\sigma^2}\big\|\dot{\mathbf{X}}
+
\boldsymbol{\Omega}\times \mathbf{X}\big\|^2
\right){\rm d} t, 
\label{HT-action}
\end{align}
where $m $ is the total mass of the body, $g $ is the value of the gravitational acceleration, ${\mathbb I}$ is the real positive definite symmetric matrix  of moments of inertia in the body,  $\boldsymbol{\chi}$ is the center of mass vector in the body, and $\sigma $ is a real constant. 
The variation with respect to $\dot{\mathbf{X}}$ defines the Legendre transform relation (costate variable)
\begin{eqnarray}
\sigma^2\mathbf{P}^\sharp
:= \dot{\mathbf{X}}
+ \boldsymbol{\Omega}\times \mathbf{X}.
\label{mom-HT}
\end{eqnarray}
The variation of the cost functional is given by
\begin{eqnarray}
\delta S 
= \int_0^T \bigg[
\Big( \mathbb{I}  \boldsymbol{\Omega} 
+ 
\mathbf{X}\times \mathbf{P}
\Big) \cdot \boldsymbol{\delta\Omega}
-
\Big(
\dot{\mathbf{P}}
+
\boldsymbol{\Omega}\times \mathbf{P}
+
mg\boldsymbol{\chi} \Big) \cdot \delta\mathbf{X}
\bigg]{\rm d} t
+
\Big[\mathbf{P}\cdot\delta \mathbf{X}\Big]_0^T.
\label{HT-action-var}
\end{eqnarray}
The general system  \eqref{stationarity_conditions_p_representation} takes in this case the following \emph{double cross} form,
involving the double cross product of vectors $(\mathbf{X}, \mathbf{P})\in \mathbb{R}^3 \times\mathbb{R}^3$,
cf. equations (\ref{statecostate-veceqns}),
\begin{equation}\label{statecostate-veceqns-HT}
\left\{
\begin{array}{l}
\vspace{0.2cm}\displaystyle\dot{\mathbf{X}} - (\mathbf{X}\times \mathbf{P})^\sharp\times \mathbf{X} 
=\sigma^2\mathbf{P}^\sharp
\,,\\[2mm]
\dot{\mathbf{P}} - (\mathbf{X}\times \mathbf{P})^\sharp \times \mathbf{P} 
=-mg\boldsymbol{\chi}
\,,
\end{array}\right.
\end{equation}
with
\[
\boldsymbol{\Omega}=\mathbb{I}^{-1}(\mathbf{P}\times \mathbf{X})=(\mathbf{P}\times \mathbf{X})^\sharp. 
\]
These equations correspond to the three equations in the general system \eqref{stationarity_conditions_p_representation}, with the upper sign chosen.
After denoting the angular momentum vector $\boldsymbol{\Pi} \in \mathbb{R}^3$ by 
\begin{equation}
\boldsymbol{\Pi}:=\mathbb{I}\boldsymbol{\Omega}=\mathbf{P}\times \mathbf{X}
\,,
\label{PiOmega-defHT-HT}
\end{equation}
substitution of equations (\ref{PiOmega-defHT-HT}) into (\ref{statecostate-veceqns-HT}) yields
\begin{equation}
\dot{\boldsymbol{\Pi}}
=
\boldsymbol{\Pi}\times \boldsymbol{\Pi}^\sharp-
mg\boldsymbol{\chi} \times \mathbf{X}+\sigma^2\mathbf{P}\times\mathbf{P}^\sharp
\quad\text{and}\quad
\dot{\mathbf{X}}
+
\boldsymbol{\Pi}^\sharp\times \mathbf{X}
=
\sigma^2\mathbf{P}^\sharp
\,,
\label{chi-eqns-aux}
\end{equation}
which can be written in matrix form as

\begin{equation} 
    \begin{bmatrix}
   \boldsymbol{\dot{\Pi}} \\
 \   \mathbf{\dot{X}} \   \\
  \mathbf{\dot{P}}
    \end{bmatrix}
    =
    \begin{bmatrix}
    \boldsymbol{\Pi}\times
   &    
     \mathbf{X} \times
    &
     \mathbf{P} \times
      \\
    \mathbf{X} \times    & 0 & 1
          \\
    \mathbf{P} \times   & -1 & 0 
    \end{bmatrix}
    \begin{bmatrix}
   \partial h_M/\partial\boldsymbol{\Pi} \\
   \partial h_M/\partial\mathbf{X} \\
   \partial h_M/\partial\mathbf{P} 
    \end{bmatrix}
    =
 \begin{bmatrix}
    \boldsymbol{\Pi}\times
   &    
     \mathbf{X} \times
    &
     \mathbf{P} \times
      \\
    \mathbf{X} \times    & 0 & 1
          \\
    \mathbf{P} \times   & -1 & 0 
    \end{bmatrix}    %
    \begin{bmatrix}
   \boldsymbol{\Pi}^\sharp \\
   mg\boldsymbol{\chi}\\
   \sigma^2 \mathbf{P}^\sharp 
   \end{bmatrix}.
\label{EP-so3XR3XR3eqnsHT}
\end{equation}
where
\begin{equation}\label{HTmod-Ham}
h_M( \boldsymbol{\Pi}, \mathbf{X}, \mathbf{P})
=  \frac{1}{2}\boldsymbol{\Pi}\cdot\boldsymbol{\Pi}^ \sharp
+
mg\boldsymbol{\chi}\cdot \mathbf{X} 
+
\frac{\sigma ^2}{2} \mathbf{P}\cdot\mathbf{P}^ \sharp,
\end{equation}
which suggests that one might regard the system (\ref{EP-so3XR3XR3eqnsHT}) physically as a model of the motion of an ellipsoidal underwater vehicle, influenced by an external gravitational torque.  
These are equations \eqref{matrix_representation} in this particular case.
\medskip

\begin{remark} \normalfont
The analogous extremal problem for compressible fluids is given by
\[
\min_{u,\rho}\int_0^T\left(\ell(u,\rho)+\frac{1}{2\sigma^2}\|\dot\rho+\operatorname{div}(\rho u)\|^2\right)dt,
\]
where $u $ is the Eulerian velocity and $\rho$ is the fluid density in spatial representation. The advection law $\dot{ \rho} + \operatorname{div}( \rho u) = 0$ (exact matching) is no longer imposed. Instead its expression 
$\|\dot{ \rho} + \operatorname{div}( \rho u)\|^2_{L^2}$ (inexact matching) 
is used as a penalty. Since this can be treated in a more general case, we defer this discussion to  \S\ref{compressible_section}.
\quad $\blacklozenge$
\end{remark}

\subsubsection{Adjoint representations}\label{adj_repr}

We let $G$ act on on the right on its Lie algebra $\mathfrak{g}$ by the adjoint representation. The infinitesimal generator is thus $\xi_\mathfrak{g}(x)=[x,\xi]$, the diamond operator is $\diamond:\mathfrak{g}^*\times\mathfrak{g}\rightarrow\mathfrak{g}^*, p\diamond x =-\operatorname{ad}^*_xp$ and the momentum map is $\mathbf{J}(x,p)=\operatorname{ad}^*_xp$.

\subsection*{The Clebsch optimal control approach} The Clebsch optimal control (with condition (A), that is, $\dot x=[x,\xi]$) associated to a cost function $\ell=\ell(\xi,x)$ yields the (generalized) Euler-Poincar\'e equations
\[
\frac{d}{dt}\frac{\delta\ell}{\delta\xi}=\operatorname{ad}^*_\xi\frac{\delta\ell}{\delta\xi}+\operatorname{ad}^*_x\frac{\partial \ell}{\partial x}\,.
\]
We suppose that $\mathfrak{g}$ is endowed with a bi-invariant inner product $\gamma$. This allows us to identify the dual Lie algebra with itself and to write $\operatorname{ad}^*_xp=-[x,p]$. In this case, the motion equations are
\[
\frac{d}{dt}\frac{\delta\ell}{\delta\xi}=\left[\frac{\delta\ell}{\delta\xi},\xi\right]+\left[\frac{\partial \ell}{\partial x},x\right].
\]
These equations are obtained from the stationarity conditions
\[
\frac{\delta \ell}{\delta \xi}=[p,x],\quad \dot x=[x,\xi],\quad \dot p=[p,\xi]+\frac{\partial \ell}{\partial x}.
\]
If the Legendre transform associated to $\ell$ is a diffeomorphism, we can write $\xi=\frac{\delta h}{\delta [p,x]}$ and the equations take the form
\[
\dot x=\left[x,\frac{\delta h}{\delta [p,x]}\right],\quad \dot p=\left[p,\frac{\delta h}{\delta [p,x]}\right]+\frac{\partial\ell}{\partial x}.
\]
In the particular case where $\ell$ is given by the kinetic energy of a bi-invariant inner product, one obtains the control $\xi=[p,x]$ and the \textit{double bracket equations}
\[
\dot x=[x,[p,x]],\quad \dot p=[p,[p,x]].
\]
More generally, the Lagrangian $\ell(\xi,x)=\frac{1}{2}\|\xi\|^2-V(x)$ implies the motion equation $\dot\xi=\left[x,\frac{\delta V}{\delta x}\right]$; see \cite{BlCr1996}. An interesting example is provided by the potential $V(x)=-\frac{1}{2}\|\,[x,n]\,\|^2$; see \cite{Br1994}. 
For more discussion of the history, theoretical developments and other examples of double bracket equations, see, e.g.,  \cite{GBRa2011}.

\subsection*{The distributed optimization approach} The penalty functional is defined on $\mathfrak{g}\times T\mathfrak{g}$ and reads 
\[
\ell(\xi,x)+\frac{1}{\sigma^2}\|\dot x-[x,\xi]\|^2
\,,
\]
where the norm is associated to a inner product on the Lie algebra $\mathfrak{g}$. The associated equations of motion read
\[
\frac{d}{dt}\frac{\delta\ell}{\delta\xi}=\operatorname{ad}^*_\xi\frac{\delta\ell}{\delta\xi}+\operatorname{ad}^*_x\frac{\partial \ell}{\partial x}+\frac{1}{\sigma^2}\operatorname{ad}^*_{\nu}\nu^\flat.
\]

As above, we now suppose that $\mathfrak{g}$ is endowed with a bi-invariant inner product $\gamma$ and we use it to identify the dual Lie algebra $\mathfrak{g}^*$ with $\mathfrak{g}$. In this case, the above equations become
\[
\frac{d}{dt}\frac{\delta \ell}{\delta \xi}=\left[\frac{\delta\ell}{\delta\xi},\xi\right]+\left[\frac{\partial \ell}{\partial x},x\right]
.\]
These equations are obtained from the stationarity conditions
\[
\frac{\delta \ell}{\delta \xi}=\frac{1}{\sigma^2}[\nu,x],\quad \dot x=[x,\xi]+\nu,\quad \dot \nu=[\nu,\xi]+\sigma^2\frac{\partial \ell}{\partial x}.
\]
As usual, we define the variable $p:=\frac{1}{\sigma^2}\nu$ in order to rewrite these conditions as
\[
\frac{\delta \ell}{\delta \xi}=[p,x],\quad \dot x=[x,\xi]+\sigma^2 p,\quad \dot p=[p,\xi]+\frac{\partial \ell}{\partial x}.
\]

As before, if the Legendre transform associated to $\ell$ is a diffeomorphism, we get
\[
\dot x=\left[x,\frac{\delta h}{\delta [p,x]}\right]+\sigma^2 p,\quad \dot p=\left[p,\frac{\delta h}{\delta [p,x]}\right]+\frac{\partial\ell}{\partial x}.
\]
If the Lagrangian $\ell$ is given by the kinetic energy of a bi-invariant inner product, we get the control $\xi=[p,x]$. Now the double bracket equations are modified by an extra term:
\[
\dot x=[x,[p,x]]+\sigma^2 p,\quad \dot p=[p,[p,x]].
\]
Further investigation of this class of equations will be pursued in future research. 

\subsection{Action by affine representation}\label{affine_representation}

We now consider the more general case where $G$ acts on $V^*$ by a left affine representation, $a\mapsto ga+c(g)$, where $c:G\rightarrow V^*$ is a group one-cocycle. In this case, the infinitesimal generator is 
\[\xi_{V^*}(a)=\xi a+\mathbf{d}c(\xi)\]
and the cotangent bundle momentum map is 
\[\mathbf{J}(a,v)=-v\diamond a+\mathbf{d}c^T(v).\]
Affine representations play an important role for a comprehensive approach to the Hamiltonian dynamics of complex fluids. We quickly give below the main equations arising in that case, in order to understand the influence of the cocycle.

\subsection*{The Clebsch optimal control approach for affine action} The Clebsch optimal control problem (with condition $\rm (A)')$ yields the affine Euler-Poincar\'e equations
\[
\frac{d}{dt}\frac{\delta \ell}{\delta\xi}=\operatorname{ad}^*_\xi\frac{\delta \ell}{\delta\xi}+\frac{\delta \ell}{\delta a}\diamond a-\mathbf{d}c^T\left(\frac{\delta\ell}{\delta a}\right).
\]
These equations appear naturally in the study of spin systems and complex fluids; see \cite{GBRa2008b}.

\subsection*{The distributed optimization approach for affine action}
The penalty function in the case of an affine representation is $\|\dot a+\xi a+\mathbf{d}c(\xi)\|^2$. The presence of the cocycle $c$ does not modify the tensor field $\mathcal{F}^\nabla$ and one finds the motion equations
\begin{equation}
\left\{
\begin{array}{l}
\vspace{0.2cm}\displaystyle\frac{d}{dt}\frac{\delta \ell}{\delta\xi}=\operatorname{ad}^*_\xi\frac{\delta \ell}{\delta\xi}+\frac{\delta \ell}{\delta a}\diamond a-\mathbf{d}c^T\left(\frac{\delta\ell}{\delta a}\right)+\frac{1}{\sigma^2}\nu^\flat\diamond\nu
,\\
\displaystyle\frac{d}{dt}\nu^\flat-\sigma^2\frac{\delta\ell}{\delta a}=-\xi\nu^\flat
,\qquad 
\nu=\dot a+\xi a+\mathbf{d}c(\xi)
.
\end{array}
\right.
\end{equation}
As before, these equations can be obtained either by metamorphosis reduction, or by the stationarity conditions
\[
\frac{\delta \ell}{\delta\xi}=\frac{1}{\sigma^2}\left(\nu^\flat\diamond a-\mathbf{d}c^T(\nu^\flat)\right),\quad \dot a=-\xi a-\mathbf{d}c(\xi)+\nu,\quad
\dot\nu^\flat=-\xi\nu^\flat+\sigma^2\frac{\delta\ell}{\delta a}.
\]
Defining the variable $p:=\frac{1}{\sigma^2}\nu^\flat$, we can write
\[
\frac{\delta \ell}{\delta\xi}=p\diamond a-\mathbf{d}c^T(p),
\qquad 
\dot a=-\xi a-\mathbf{d}c(\xi)+\sigma^2p^\sharp,
\qquad
\dot p=-\xi p+\frac{\delta\ell}{\delta a}.
\]
When the affine term is not present, one recovers \eqref{stationarity_conditions_p_representation}.
If the Lagrangian $\ell$ is given by the kinetic energy associated to an inner product, then the control is given by \[\xi=(p\diamond a-\mathbf{d}c^T(p))^\sharp,\] 
and we get the equations
\begin{eqnarray*}
\dot a+(p\diamond a-\mathbf{d}c^T(p))^\sharp  a+\mathbf{d}c\left((p\diamond a-\mathbf{d}c^T(p))^\sharp\right) &=& \sigma^2p^\sharp,
\\
\dot p +(p\diamond a-\mathbf{d}c^T(p))^\sharp p&=& 0.
\end{eqnarray*}

\subsection{Actions by multiplication on Lie groups}\label{group_multiplication}

We now specialize to the case where $Q=H$ is a Lie group, containing $G$ as a subgroup. We will then apply the results to the rigid body and ideal fluids.

Suppose that $G$ acts on $H$ by left (resp. right) multiplication. Given a Lie algebra element $\xi\in\mathfrak{g}$, the infinitesimal generator is
\[
\xi_H(h)=TR_h\xi=:\xi h,\quad\text{resp.}\quad \xi_H(h)=TL_h\xi =:h\xi,
\]
and the cotangent bundle momentum map $\mathbf{J}:T^*H\rightarrow\mathfrak{g}^*$ is
\[
\mathbf{J}(\alpha_h)=i^*(T^*R_h\alpha_h)=i^*(\alpha_hh^{-1})\quad\text{resp}\quad \mathbf{J}(\alpha_h)=i^*(T^*L_h\alpha_h)=i^*(h^{-1}\alpha_h),
\]
where $i^*:\mathfrak{h}^*\rightarrow \mathfrak{g}^*$ is the dual map to the Lie algebra inclusion $i:\mathfrak{g}\rightarrow\mathfrak{h}$.

\subsection*{The Clebsch optimal control approach} Given a cost function $\ell=\ell(\xi,h)$, and assuming the constraint $\rm (A)'$, that is, $\dot h=-\xi h$ (resp. $\dot h=-h\xi$), the Clebsch optimal control problem yields the (generalized) Euler-Poincar\'e equations
\[
\frac{d}{dt}\frac{\delta\ell}{\delta\xi}=\operatorname{ad}^*_\xi\frac{\delta\ell}{\delta\xi}-i^*\left(\frac{\partial\ell}{\partial h}h^{-1}\right),\quad\text{resp.}\quad \frac{d}{dt}\frac{\delta\ell}{\delta\xi}=-\operatorname{ad}^*_\xi\frac{\delta\ell}{\delta\xi}-i^*\left(h^{-1}\frac{\partial\ell}{\partial h}\right).
\]
If the constraint $\rm (A)$ is assumed, that is $\dot h=\xi h$ (resp. $\dot h=h\xi$), the equations are
\[
\frac{d}{dt}\frac{\delta\ell}{\delta\xi}=-\operatorname{ad}^*_\xi\frac{\delta\ell}{\delta\xi}+i^*\left(\frac{\partial\ell}{\partial h}h^{-1}\right),\quad\text{resp.}\quad \frac{d}{dt}\frac{\delta\ell}{\delta\xi}=+\operatorname{ad}^*_\xi\frac{\delta\ell}{\delta\xi}+i^*\left(h^{-1}\frac{\partial\ell}{\partial h}\right).
\]
These equations are obtained by inserting the expression of the momentum map in equations \eqref{generalized_EP} and \eqref{generalized_EP_inverse}.

\subsection*{The distributed optimization approach} The penalty functional is defined on $\mathfrak{g}\times TH$ and reads
\[
\ell(\xi,h)+\frac{1}{2\sigma^2}\|\dot h\pm\,\xi h\|^2, \quad\text{resp.}\quad \ell(\xi,h)+\frac{1}{2\sigma^2}\|\dot h\pm h\xi\|^2,
\]
relative to a Riemannian metric on $H$. We will restrict to the case of an \textit{$H$-invariant metric}. More precisely, given an inner product $\gamma$ on $\mathfrak{h}$, we consider the associated left (resp. right)-invariant Riemannian metric 
$\gamma_h$ on $H$, that is, we have $\gamma_h(u_h,v_h):=
\gamma(h^{-1}u_h,h^{-1}v_h)$, resp. $\gamma_h(u_h,v_h):=
\gamma(u_hh^{-1},v_hh^{-1})$.

Since $G$ acts by isometries, the motion equation are given by \eqref{unconstrained_A_iso} and \eqref{unconstrained_A'_iso}. To compute these equations in our particular case, we need the concrete expression of the Levi-Civita connection associated to the Riemannian metric $\gamma_h$ on $H$. It is written in terms of the isomorphism $\psi:\mathcal{F}(H,\mathfrak{h})\rightarrow\mathfrak{X}(H)$ given by $\psi(f)(h)=TL_h(f(h))$ (resp. $\psi(f)(h)=TR_h(f(h))$). For a vector field $X\in\mathfrak{X}(H)$, the Levi-Civita covariant derivative associated to the left (resp. right)-invariant extension of $\gamma$ to $H$ is given by
\[
\nabla_{v_h}X(h)=TL_h\left(\mathbf{d}f(v_h)-\frac{1}{2}\operatorname{ad}^\dagger_vf(h)-\frac{1}{2}\operatorname{ad}^\dagger_{f(h)}v+\frac{1}{2}[v,f(h)]\right),
\]
\[
\text{resp.}\quad \nabla_{v_h}X(h)=TR_h\left(\mathbf{d}f(v_h)+\frac{1}{2}\operatorname{ad}^\dagger_vf(h)+\frac{1}{2}\operatorname{ad}^\dagger_{f(h)}v-\frac{1}{2}[v,f(h)]\right),
\]
where $v:=h^{-1}v_h$ (resp. $v:=v_hh^{-1}$), $f=\psi^{-1}(X)$, and $\operatorname{ad}^\dagger_\xi$ is the transpose of $\operatorname{ad}_\xi$ with respect to the inner product $\gamma$ on $\mathfrak{h}$, see \cite{KrMi1997}, Section 46.5.

We now specialize these formulas to the case where the vector field $X$ is given by the infinitesimal generator $\xi_H$. In the case of multiplication in the left, we have $X(h)=\xi_H(h)=\xi h$ and $f(h)=\operatorname{Ad}_{h^{-1}}\xi$. Thus we obtain
\begin{align*}
\nabla_{v_h}\xi_H(h)&=TL_h\left(-[h^{-1}v_h,f(h)]-\frac{1}{2}\operatorname{ad}^\dagger_vf(h)-\frac{1}{2}\operatorname{ad}^\dagger_{f(h)}v+\frac{1}{2}[v,f(h)]\right)\\
&=-\,\frac{1}{2}TL_h\left([v,f(h)]+\operatorname{ad}^\dagger_vf(h)+\operatorname{ad}^\dagger_{f(h)}v\right).
\end{align*}
For right-invariant metrics, we have $\xi_H(h)=h\xi$, $f(h)=\operatorname{Ad}_h\xi$ and the previous formula becomes
\[
\nabla_{v_h}\xi_H(h)=\frac{1}{2}TR_h\left([v,f(h)]+\operatorname{ad}^\dagger_vf(h)+\operatorname{ad}^\dagger_{f(h)}v\right).
\]
When condition $\rm (A)'$ is assumed, the motion equations are (see \eqref{unconstrained_A'_iso})
\begin{equation}
\label{A'_left}
\left\{
\begin{array}{l}
\vspace{0.2cm}\displaystyle\frac{d}{dt}\frac{\delta\ell}{\delta\xi}=\operatorname{ad}^*_\xi\frac{\delta\ell}{\delta\xi}-i^*\left(\frac{\partial\ell}{\partial h}h^{-1}\right),\quad \nu_h=\dot h+\xi h\\
\displaystyle\frac{D}{Dt}\nu_h-\sigma^2\frac{\partial\ell}{\partial h}^\sharp=\frac{1}{2}TL_h\left([\nu,f(h)]+\operatorname{ad}^\dagger_{f(h)}\nu+\operatorname{ad}^\dagger_\nu f(h)\right)\in T_hH
\end{array}
\right.
\end{equation}
resp.
\begin{equation}
\label{A'_right}
\left\{
\begin{array}{l}
\vspace{0.2cm}\displaystyle\frac{d}{dt}\frac{\delta\ell}{\delta\xi}=-\operatorname{ad}^*_\xi\frac{\delta\ell}{\delta\xi}-i^*\left(h^{-1}\frac{\partial\ell}{\partial h}\right),\quad \nu_h=\dot h+h\xi\\
\displaystyle \frac{D}{Dt}\nu_h-\sigma^2\frac{\partial\ell}{\partial h}^\sharp=-\frac{1}{2}TR_h\left([\nu,f(h)]+\operatorname{ad}^\dagger_{f(h)}\nu+\operatorname{ad}^\dagger_\nu f(h)\right)\in T_hH,
\end{array}
\right.
\end{equation}
and the stationarity condition \eqref{condition_1} is
\[
\frac{\delta \ell}{\delta\xi}=-\frac{1}{\sigma^2}i^*\left(\nu_h^{\,\flat} h^{-1}\right),\quad\nu_h=\dot h+\xi h,\quad\text{resp.}\quad \frac{\delta \ell}{\delta\xi}=-\frac{1}{\sigma^2}i^*\left(h^{-1}\nu_h^{\,\flat}\right),\quad\nu_h=\dot h+h\xi.
\]
If the constraint $\rm (A)$ is assumed, then we have (see \eqref{unconstrained_A_iso})
\begin{equation}
\label{A_left}
\left\{
\begin{array}{l}
\vspace{0.2cm}\displaystyle\frac{d}{dt}\frac{\delta\ell}{\delta\xi}=-\operatorname{ad}^*_\xi\frac{\delta\ell}{\delta\xi}+i^*\left(\frac{\partial\ell}{\partial h}h^{-1}\right),\quad \nu_h=\dot h-\xi h\\
\displaystyle\frac{D}{Dt}\nu_h-\sigma^2\frac{\partial\ell}{\partial h}^\sharp=-\frac{1}{2}TL_h\left([\nu,f(h)]+\operatorname{ad}^\dagger_{f(h)}\nu+\operatorname{ad}^\dagger_\nu f(h)\right)\in T_hH
\end{array}
\right.
\end{equation}
resp.
\begin{equation}
\label{A_right}
\left\{
\begin{array}{l}
\vspace{0.2cm}\displaystyle\frac{d}{dt}\frac{\delta\ell}{\delta\xi}=\operatorname{ad}^*_\xi\frac{\delta\ell}{\delta\xi}+i^*\left(h^{-1}\frac{\partial\ell}{\partial h}\right),\quad \nu_h=\dot h-h\xi\\
\displaystyle \frac{D}{Dt}\nu_h-\sigma^2\frac{\partial\ell}{\partial h}^\sharp=\frac{1}{2}TR_h\left([\nu,f(h)]+\operatorname{ad}^\dagger_{f(h)}\nu+\operatorname{ad}^\dagger_\nu f(h)\right)\in T_hH,
\end{array}
\right.
\end{equation}
and the stationarity condition \eqref{condition_1} is
\[
\frac{\delta \ell}{\delta\xi}=\frac{1}{\sigma^2}i^*\left(\nu_h^{\,\flat}h^{-1}\right),\quad\nu_h=\dot h-\xi h,\quad\text{resp.}\quad \frac{\delta \ell}{\delta\xi}=\frac{1}{\sigma^2}i^*\left(h^{-1}\nu_h^{\,\flat}\right),\quad\nu_h=\dot h-h\xi.
\]
From the general theory developed in Section \ref{sec_Lagr_approach}, these equations can be obtained by metamorphosis reduction, starting from the $G$-invariant Lagrangian $L=L\left(g,\dot g,f,\dot f\right):T(G\times H)\rightarrow\mathbb{R}$ given by
\[
L\left(g,\dot g,f,\dot f\right)=\mathcal{L}\left(g,\dot g,f\right)+\frac{1}{2\sigma^2}\|\dot f\|^2,
\]
where $\mathcal{L}:TG\times H\rightarrow\mathbb{R}$ is the $G$-invariant function associated to $\ell$. One can pass from the Lagrangian variables $(g,f)$ to the reduced variables $(\xi,\nu_h)$ via the map
\[
\left(g,\dot g,f,\dot f\right)\rightarrow (\xi,\nu_h):=\left(g^{-1}\dot g,g^{-1}\dot f\right),
\]
for example. Note the relation $h=g^{-1}f$.
\medskip

\begin{remark}\normalfont Note that if the inner product 
$\gamma$ on $\mathfrak{h}$ is bi-invariant, then  
$\operatorname{ad}^\dagger=-\operatorname{ad}$ and the 
equations above simplify.
\quad $\blacklozenge$
\end{remark} 

\subsubsection{The $N$-dimensional rigid body}
\label{subsec-RB}

We now apply the above results to the Lie groups $G=SO(N)$ and $H=GL(N)$ in order to obtain the distributed optimization approach for the $N$-dimensional rigid body. Of course, the more interesting case happens for $N=3$. We let $SO(N)$ act on $GL(N)$ by multiplication on the \textit{right}. Given $Q\in GL(N)$ and $U\in\mathfrak{so}(N)$, the associated infinitesimal generator is given by $U_{GL(N)}(Q) = QU$. The Lagrangian of the rigid body is of the form $\ell(U)=\frac{1}{4}\left\langle \mathcal{J}(U),U\right\rangle$, where $\mathcal{J}:\mathfrak{so}(N)\rightarrow\mathfrak{so}(N)$ is a symmetric positive definite operator of the form
\[
\mathcal{J}(U)=UJ+JU,
\]
where $J$ is a diagonal matrix verifying $J_i+J_j>0$ for all $i\neq j$.

\subsection*{The Clebsch optimal control approach} Using the constraint $\dot Q=QU$, the Clebsch optimal control problem yields the motion equations
\[
\dot M=[M,U],\quad M=\frac{\delta \ell}{\delta U}=\frac{1}{2}\mathcal{J}(U)
\]
of the $N$-rigid body, where we identified the dual Lie algebra $\mathfrak{so}(N)^*$ with $\mathfrak{so}(N)$ via the pairing $\langle P,V\rangle:=\operatorname{Tr}(P^TV)$. Of course, when $N=3$ and identifying $\mathfrak{so}(3)$ with $(\mathbb{R}^3,\times)$, we recover the classical Euler equations $\dot{\mathbf{M}}=\mathbf{M}\times \mathbf{U}$. Using the same pairing as above to identify the tangent and cotangent spaces, we obtain the expression $\mathbf{J}(Q,P)=\frac{1}{2}\left(Q^TP-P^TQ\right)$ for the momentum map. This yields the stationarity conditions
\begin{equation}\label{symetric_rep_rb}
U=\mathcal{J}^{-1}\left(Q^TP-P^TQ\right),\quad \dot Q=QU,\quad \dot P=PU.
\end{equation}
The two last equations are referred to as the \textit{symmetric representation of the rigid body}; see \cite{BlCr1996}, \cite{BlBrCr1997}, \cite{BlCrMaRa1998}, and \cite{GBRa2011}.

Note that one can also let $SO(N)$ act on the vector space 
$\mathfrak{gl}(N)$ instead of the group $GL(N)$, with the same results.

\subsection*{The distributed optimization approach} The penalty term reads $\|\dot Q-QU\|^2$, where the norm is associated to a Riemannian metric on $GL(N)$, and one needs to minimize the functional
\[
\int_0^T\left(\frac{1}{4}\langle \mathcal{J}(U),U\rangle+\frac{1}{2\sigma^2}\|\dot Q-QU\|^2\right)dt.
\]
If the Riemannian metric on $GL(N)$ is \textit{right}-invariant, the associated stationary conditions and equations of motion can be computed with the help of the general formula derived above. In particular, one needs to use equation \eqref{A_right}. Since the resulting equations are complicated, we do not pursue this approach here and leave it for the interested reader. This route will be taken for the ideal fluid equations below.

An alternative approach is to consider the action of $SO(N)$ on the vector space $\mathfrak{gl}(N)$ instead of the Lie group $GL(N)$. In this case we can apply the results of \S\ref{representation} and we suppose that the norm involved in the penalty term is associated to a inner product on $\mathfrak{gl}(N)$, but is not necessarily right-invariant. The minimization problem is the same and one obtains the stationary conditions
\[
\frac{1}{2}\mathcal{J}(U)=-\frac{1}{\sigma^2}\nu^\flat\diamond Q,\quad \dot Q=QU+\nu,\quad \dot \nu^\flat=\nu^\flat U,
\]
where the diamond operator is given by $P\diamond Q=-\mathbf{J}(Q,P)=-\frac{1}{2}\left(Q^TP-P^TQ\right)$. Here, $\flat:\mathfrak{gl}(N)\rightarrow\mathfrak{gl}(N)^*\simeq \mathfrak{gl}(N)$ is the flat isomorphism associated to the inner product on $\mathfrak{gl}(N)$. Recall that the dual space $\mathfrak{gl}(N)^*$ is identified with $\mathfrak{gl}(N)$ via the pairing $\operatorname{Tr}(P^TV)$, but the inner product can be different from this pairing. As usual, we define the variable
\[
P:=\frac{1}{\sigma^2}\nu^\flat
\]
and we rewrite the above conditions as
\[
U=\mathcal{J}^{-1}\left(Q^TP-P^TQ\right),\quad \dot Q=QU+\sigma^2P^\sharp,\quad \dot P=P U.
\]
These equations should be compared to the symmetric representation of the rigid body \eqref{symetric_rep_rb}. The equations of motion for $M$ take the form
\[
\dot M=[M,U]-\sigma^2P\diamond P^\sharp.
\]
Thus we get the system of equations
\begin{equation}
\label{ExSO(N)mod-eqn}
\left\{
\begin{array}{l}
\dot M=[M,U]-\sigma^2P\diamond P^\sharp
\,,\\[2mm]
\dot P=P U
\,,\\[2mm]
\dot Q=QU+\sigma^2P^\sharp,
\end{array}
\right.
\end{equation}
which is analogous to the system \eqref{matrix_representation} for a \textit{right} action and \textit{left} reduction. Therefore this system is Lie-Poisson on the dual of the Lie algebra $\mathfrak{so}(N)\,\circledS\,( \mathfrak{gl}(N)\times \mathfrak{gl}(N))$, with a symplectic 2-cocycle on the latter product.
See also  \eqref{EP-gXV*XVeqns} and \eqref{EP-so3XR3XR3eqns}.

\subsubsection{Euler fluid equations}\label{subsec-EulerEqns}

Hamilton's principle for ideal fluid flow might be summarized by saying that \emph{water moves as well as possible to get out of its own way} \cite{Mo1990}. The question pursued in \cite{Ho2009} was whether Euler's fluid equations represent optimal control, or only optimization. As it turned out, the geodesic flow represented by the Euler's fluid equations was found to arise from either formulation. An optimization method used in image-processing (metamorphosis) is found to imply Euler's equations for  incompressible flow of an inviscid fluid, without requiring that the Lagrangian particle labels exactly follow the flow lines of the Eulerian velocity vector field. That is, an optimal control formulation and an optimization formulation for incompressible ideal fluid flow both yield the \emph {same} Euler fluid equations, although their Lagrangian parcel dynamics are \emph{different}. This is a result of the \emph{gauge freedom} in the definition of the fluid pressure for an incompressible flow, in combination with the symmetry of fluid dynamics under relabeling of their Lagrangian coordinates.  
\medskip

We apply here the result of this section to the Lie group $H=\operatorname{Diff}(\mathcal{D})$ of all diffeomorphisms of the manifold $\mathcal{D}$ and its subgroup $G=
\operatorname{Diff}_{vol}(\mathcal{D})$ of volume preserving diffeomorphisms. We shall recover and extend the approach given in \cite{Ho2009}. Recall that a curve $\eta_t\in \operatorname{Diff}_{vol}(\mathcal{D})$ represents the Lagrangian motion of an ideal fluid in the domain $\mathcal{D}$, that is, the curve 
$\eta_t(x)$ in $\mathcal{D}$ is the trajectory of the fluid particle located at $x$ at time $t=0$, assuming that $\eta_0$ is the identity; $\eta_t$ is referred to as the \textit{forward map}. The Lie algebra of $G$ consists of divergence free vector fields on $\mathcal{D}$ parallel to the boundary and is denoted by $\mathfrak{g}=\mathfrak{X}_{vol}(\mathcal{D})$. The curve $\eta_t$ is the flow of the Eulerian velocity $u_t\in \mathfrak{X}_{vol}(\mathcal{D})$, that is, we have $\dot\eta_t=u_t\circ\eta_t$. 
 The curve $l_t:=\eta_t^{-1}$ is called the \textit{back-to-labels map}. (See, e.g., \cite{Constantin2001} where the name ``back-to-labels'' was introduced and the map was used as a sufficient variable to describe and analyze the incompressible Euler equations.) The back-to-labels map is related to the Eulerian velocity $u_t$ via the relation $\dot l_t+Tl_t\!\cdot\!u_t=0$.

As is well known, a curve $\eta_t\in \operatorname{Diff}_{vol}(\mathcal{D})$ is a geodesic with respect to the $L^2$ right invariant Riemannian metric if and only if $u_t$ is a solution of the Euler fluid equations
\[
\partial_t u+u\!\cdot\!\nabla u=-\operatorname{grad}p.
\]
In other words, the Euler fluid equation is given by the
Euler-Poincar\'e equation on $\mathfrak{X}_{vol}(\mathcal{D})$
associated to the Lagrangian
$\ell(u)=\frac{1}{2}\int_\mathcal{D}\|u\|^2dx$.\medskip

\noindent\textit{First approach - composition on the left:} We let the group 
$G=\operatorname{Diff}_{vol}(\mathcal{D})$ act on $H=\operatorname{Diff}(\mathcal{D})$ by composition on the \textit{left}. The infinitesimal generator is thus given by $u_{\operatorname{Diff}(\mathcal{D})}(\varphi)=u\circ\varphi$, for $\varphi\in\operatorname{Diff}(\mathcal{D})$.

\subsection*{The Clebsch optimal control approach} Using the Lagrangian $\ell(u,\varphi)=\ell(u)=\frac{1}{2}\int_\mathcal{D}\|u\|^2dx=\frac{1}{2}\|u\|^2_{L^2}$ and the constraint $\dot\varphi=u\circ\varphi$, the Clebsch optimal control problem yields the Euler equations
\[
\partial_tu+\nabla_uu=-\operatorname{grad}p.
\]
Note that here there is no dependence of $\ell$ on the variable $\varphi$, therefore the Clebsch approach yields the standard Euler-Poincar\'e equations. The stationarity conditions are
\begin{equation}\label{stat_cond_Euler_left}
u^\flat=\mathbf{J}(\varphi,\pi)=\mathbb{P}(J\varphi^{-1}(\pi\circ\varphi^{-1})),\quad \dot\varphi=u\circ\varphi,\quad \dot\pi=-(T^\ast u\circ \varphi) \cdot \pi,
\end{equation}
where $\mathbb{P}:\Omega^1(\mathcal{D})\rightarrow\Omega^1_{div}(\mathcal{D})$ is the Hodge projector and $J \varphi$ is the Jacobian determinant of $\varphi$. Here we have chosen the $L^2$ pairing between one-forms and vector fields on $\mathcal{D}$ and $\flat$ denotes the index lowering operation defined by the Riemannian metric on $\mathcal{D}$.

\subsection*{The distributed optimization approach} The penalty term is $\|\dot\varphi-u\circ\varphi\|^2_{L^2}$, where the norm is taken with respect to the \textit{left}-invariant $L^2$ metric on $\operatorname{Diff}(\mathcal{D})$, and one needs to minimize the functional
\[
\int_0^T\left(\frac{1}{2}\|u\|^2_{L^2}+\frac{1}{2\sigma^2}\|\dot\varphi-u\circ\varphi\|_{L^2}^2\right)dt
\]
among all curves $u(t), \varphi(t)$ in $\mathfrak{X}_{div}(\mathcal{D})\times \operatorname{Diff}(\mathcal{D})$. The stationarity condition \eqref{condition_1} reads
\[
\frac{\delta \ell}{\delta u}=\frac{1}{\sigma^2}\mathbb{P}(J\varphi^{-1}(\nu^\flat_\varphi\circ\varphi^{-1}))=\frac{1}{\sigma^2}J\varphi^{-1}(\nu^\flat_\varphi\circ\varphi^{-1})-\mathbf{d}k,
\]
where $\mathbb{P}:\Omega^1(\mathcal{D})\rightarrow\Omega^1_{div}(\mathcal{D})$ is the Hodge projector onto divergence free forms. In our case, the equations of motion are given by \eqref{A_left} and we get
\begin{equation}\label{equation_Euler_left}
\left\{
\begin{array}{l}
\vspace{0.2cm}\displaystyle\partial_tu+\nabla_uu=-\operatorname{grad}p,\quad \dot\varphi=u\circ\varphi+\nu_\varphi\\
\displaystyle \frac{D}{Dt}\nu_\varphi+T\varphi\circ\nabla_{\varphi^*u}\nu=-T\varphi\circ \mathcal{F}(\varphi^*u,\nu)
\end{array}
\right.
\end{equation}
where $\nu:=T\varphi^{-1} \circ  \nu_ \varphi$ and $\mathcal{F}(v,\nu)=\frac{1}{2}\left(\operatorname{grad}g(v,\nu)+\nu\operatorname{div}v+v\operatorname{div}\nu\right)$. To see this, we compute the right hand side 
\[-\frac{1}{2}TL_h\left([\nu,f(h)]+\operatorname{ad}^\dagger_{f(h)}\nu+\operatorname{ad}^\dagger_\nu f(h)\right)\] of the second equation in \eqref{A_left}. We have
\begin{align*}
[\nu,v]+\operatorname{ad}^\dagger_{v}\nu+\operatorname{ad}^\dagger_\nu v&=\nabla_v\nu-\nabla_\nu v+\nabla v^T\nu+\nabla_v\nu+\nu\operatorname{div}v+\nabla \nu^Tv+\nabla_\nu v+v\operatorname{div}\nu\\
&=\operatorname{grad}g(v,\nu)+2\nabla_v\nu+\nu\operatorname{div}v+v\operatorname{div}\nu
\end{align*}
since $\operatorname{ad}^\dagger_um=\nabla u^T m+\nabla_um+m\operatorname{div}u$.
By choosing $\nu:=TL_{\varphi^{-1}}(\nu_\varphi)=T\varphi^{-1} \circ  \nu_ \varphi$ and $v:=\operatorname{Ad}_{\varphi^{-1}}u=T\varphi^{-1}\circ u\circ\varphi=\varphi^*u$, we obtain the result. Note that
$\varphi^*u$ is an analogue of the convective velocity, but recall that $\varphi$ is not the flow of $u$.

As usual, the stationarity conditions can also be expressed in terms of the variable $\pi:=\frac{1}{\sigma^2}\nu^\flat_\varphi$. They can alternatively be written as
\[
u^\flat=\mathbf{J}(\varphi,\pi)=\mathbb{P}(J\varphi^{-1}(\pi\circ\varphi^{-1})),\quad \dot\varphi=u\circ\varphi+\sigma^2\pi^\sharp,\quad \dot{\pi}=-(T^\ast u\circ\varphi) \cdot \pi+\sigma^2\mathcal{S}(\pi),
\]
in order to be compared to \eqref{stat_cond_Euler_left}, where $\mathcal{S}$ denotes the geodesic spray of the left invariant Riemannian metric. Here $\sharp : = \flat ^{-1}$.

The equations \eqref{equation_Euler_left} can be obtained by metamorphosis reduction for the Lagrangian defined on $(u_\eta, u_f)\in T(\operatorname{Diff}_{vol}(\mathcal{D})\times\operatorname{Diff}(\mathcal{D}))$ by
\[
\frac{1}{2}\|u_\eta\|^2_{L^2}+\frac{1}{2\sigma^2}\|u_f\|^2_{L^2},
\]
where the $L^2$ norms are associated to the right and left invariant extension of the $L^2$ inner product, respectively. This Lagrangian is invariant under the tangent lift of the right $\operatorname{Diff}_{vol}$-action given by
\[
(\eta,f)\mapsto (\eta\circ h, h^{-1}\circ f).
\]
The link between the Lagrangian variables $(\eta,\dot\eta,f,\dot f)$ and the reduced variables $(u,\nu_\varphi)$ is given by the reduction map
\[
(\eta,\dot\eta,f,\dot f)\mapsto (u,\nu_\varphi):=(\dot\eta\circ\eta^{-1},T\eta\circ \dot f).
\]
In particular, we have $\varphi=\eta\circ f$.

Note that the operator $D/Dt$ denotes the covariant derivative with respect to the left-invariant $L^2$ Riemannian metric on $\operatorname{Diff}_{vol}(\mathcal{D})$ and does not have a simple expression, in general, contrary to the covariant derivative associated to the right-invariant $L^2$ Riemannian metric, which is simply given by functorial lift. As we will see below, for the penalty approach to the Euler equations, it is more convenient to work with the back-to-labels map.

Note that instead of $H=\operatorname{Diff}(\mathcal{D})$, one can use the subgroup $H=\operatorname{Diff}_{vol}(\mathcal{D})$ of volume preserving diffeomorphisms. In this case, the second equation in \eqref{equation_Euler_left} simplifies to
\[
\frac{D}{Dt}\nu_\varphi+T\varphi\circ\nabla_{\varphi^*u}\nu=0.
\]

\medskip

\noindent\textit{Second approach - composition on the right:} We now let the group 
$G=\operatorname{Diff}_{vol}(\mathcal{D})$ act on $H=\operatorname{Diff}(\mathcal{D})$ by composition on the \textit{right}. The infinitesimal generator is thus given by $u_{\operatorname{Diff}(\mathcal{D})}(l)=Tl\circ u$. Recall that the back-to-label map $l$ is related to the Eulerian velocity $u$ by the formula $\dot l=-Tl\circ u$; therefore, we need to impose condition $\rm (A)'$.

\subsection*{Clebsch optimal control approach} Using the same Lagrangian $\ell(u)=\frac{1}{2}\int_\mathcal{D}\|u\|^2dx$ as before, and imposing the condition $\dot l=-Tl\circ u$, (condition $\rm (A)'$), the Clebsch optimal control problem yields the Euler-Poincar\'e equations on $\mathfrak{X}_{vol}(\mathcal{D})$. We thus recover the Euler fluid equations
\[
\partial_tu+u\!\cdot\!\nabla u=-\operatorname{grad}p.
\]
The associated stationarity conditions are,  \cite{Ho2009}
\begin{equation}\label{stat_cond_Euler_right}
u^\flat=-\mathbb{P}\left(\pi \circ T l\right),\quad \dot l=-Tl\circ u, \quad \dot\pi=-T\pi\circ u.
\end{equation}
In analogy with equations \eqref{symetric_rep_rb} for the rigid body, these equations are referred to as the \textit{symmetric representation of the Euler fluid equations}. 

\subsection*{The distributed optimization approach} The penalty term reads $\|\dot l+Tl\circ u\|^2_{L^2}$ where the norm is taken relative to the right-invariant $L^2$ metric on the group of diffeomorphisms. Therefore, we minimize the functional
\[
\int_0^T\left(\frac{1}{2}\|u\|^2_{L^2}+\frac{1}{2\sigma^2}\|\dot l+Tl\circ u\|^2_{L^2}\right)dt,
\]
among all curves $u(t), l(t)$ in $\mathfrak{X}_{div}(\mathcal{D})\times \operatorname{Diff}(\mathcal{D})$. The stationarity condition \eqref{condition_1} reads
\[
\frac{\delta \ell}{\delta u}=-\frac{1}{\sigma^2}\mathbb{P}\left(\nu_l^ \flat \circ Tl \right)=-\frac{1}{\sigma^2}\nu_l ^\flat \circ Tl  -\mathbf{d}k,
\]
where $\nu_l: = \dot{l} + Tl \circ u $
and the associated equations of motion are
\begin{equation}\label{equation_Euler_right}
\left\{
\begin{array}{l}
\vspace{0.2cm}\displaystyle\partial_tu+\nabla_uu=-\operatorname{grad}p,\\
\displaystyle \frac{D}{Dt}\nu_l+\nabla_u\nu_l=-\,(\operatorname{grad}q)\circ l.
\end{array}
\right.
\end{equation}
These equations are obtained by computations similar to those above, but using \eqref{A'_right} instead of \eqref{A_left}. In particular, the right hand side of the second equation of \eqref{A'_right} becomes
\[
-\,(\nabla_v\nu)\circ l-\,(\operatorname{grad}q)\circ l=-\nabla_u\nu_l-\,(\operatorname{grad}q)\circ l,\quad q=\frac{1}{2}g\left(l_*u,\nu_\varphi\circ l^{-1}\right),
\]
since we need to choose $\nu:=\nu_l\circ l^{-1}$ and $v=\operatorname{Ad}_l u=l_*u$. Note that $v$ is the \textit{convective velocity} of the fluid. As usual, the stationarity conditions can also be expressed in terms of the variable $\pi:=\frac{1}{\sigma^2}\nu_l^\flat$. They can alternatively be written as
\[
u^\flat=-\mathbf{J}(l,\pi)=-\mathbb{P}\left(\pi\circ  Tl\right),\quad \dot l
=-Tl\circ u+\sigma^2\pi^\sharp,\quad \dot\pi=-T\pi\circ u+\sigma^2\mathcal{S}(\pi),
\]
in order to be compared to \eqref{stat_cond_Euler_right}, where $\mathcal{S}$ denotes the geodesic spray of the right invariant Riemannian metric.

The equations of motion \eqref{equation_Euler_right} can be obtained by metamorphosis reduction of the Lagrangian defined on $(u_\eta, u_f)\in T(\operatorname{Diff}_{vol}(\mathcal{D})\times\operatorname{Diff}(\mathcal{D}))$ by
\[
\frac{1}{2}\|u_\eta\|_{L^2}+\frac{1}{2\sigma^2}\|u_f\|^2_{L^2},
\]
where the $L^2$ norms are associated to the right invariant extension of the $L^2$ inner product. This Lagrangian is invariant under the tangent lift of the right $\operatorname{Diff}_{vol}$-action given by
\[
(\eta,f)\mapsto (\eta\circ h,f\circ h).
\]
The link between the Lagrangian variables $(\eta,\dot\eta,f,\dot f)$ and the reduced variables $(u,\nu_l)$ is given by the reduction map
\[
(\eta,\dot\eta,f,\dot f)\mapsto (u,\nu_l):=(\dot\eta\circ\eta^{-1},\dot f\circ \eta^{-1}).
\]
In particular, we have $l=f\circ\eta^{-1}$.

Working with $H=\operatorname{Diff}_{vol}(\mathcal{D})$ instead of the whole diffeomorphism group, yields the second equation of \eqref{equation_Euler_right} in the simpler form
\[
\frac{D}{Dt}\nu_l+\nabla_u\nu_l=0.
\]
These results recover Theorem 10 in \cite{Ho2009}.

\subsubsection{Optimization dynamics of a compressible fluid}
\label{compressible_section}
For the compressible fluid, we choose to minimize the functional
\[
S_d = \int\limits_0^T\left(
\ell(u,\rho)
+\frac{1}{2\sigma^2_1}\|\dot l+Tl\circ u\|^2_{L^2}
+\frac{1}{2\sigma^2_2}\|\dot\rho+\operatorname{div}(\rho u)\|^2_{L^2}
\right)dt
\]
over all curves $u(t), l(t), \rho (t)$ in $\mathfrak{X}(\mathcal{D})\times \operatorname{Diff}(\mathcal{D})
\times \operatorname{Dens}(\mathcal{D})$. 
This minimization involves penalties and tolerances at two levels. We seek the stationarity conditions implied by optimization of the functional $S_d$, subject to homogeneous endpoint and boundary conditions. We introduce the notation,
\begin{align}\label{var-note}
m&:=\frac{\delta \ell}{\delta u} \in \Omega^1( \mathcal{D}),\quad
\varpi:=\frac{\delta \ell}{\delta \rho} \in C ^{\infty}( \mathcal{D}),\\
\pi &:= \frac{1}{\sigma_1^2} (\dot l+Tl\circ u)^\flat \in T ^\ast_l \operatorname{Diff}( \mathcal{D}),\quad
\phi := \frac{1}{\sigma_2^2} \big(\dot\rho+\operatorname{div}\rho u \big) \in C^{\infty}( \mathcal{D}),
\end{align}
then write the stationarity conditions as:
\begin{eqnarray}\label{station-fluid}
\delta u:&& m+ \pi\circ T l - \rho\mathbf{d} \phi = 0 
;\nonumber\\
\delta l:&&\dot{\pi} +\operatorname{div} (\pi u) = 0 
;\nonumber\\
\delta \rho:&& \dot{\phi} + \mathbf{d}\phi\circ u - \varpi = 0 
\,.
\end{eqnarray}
Combining these equations into Hamiltonian form yields (in index notation for clarity) explicitly, in terms of indices and differential operators, 
%
\begin{equation}\label{Ham-matrix-diff}
\frac{\partial}{\partial t}
    \begin{bmatrix}
    m_i \\ \rho \\ \phi \\[1mm]  l^A \\ \pi_A
    \end{bmatrix}
= -\,\mathcal{B}
   \begin{bmatrix} 
   {\delta h_M/\delta m_j} = u^j \\
   {\delta h_M/\delta \rho} = - \varpi \\
   {\delta h_M/\delta \phi} =  \sigma_2^2 \phi \\[1mm]
   {\delta h_M/\delta l^B} = 0 \\
   {\delta h_M/\delta \pi_B} = \sigma_1^2 {\pi^\sharp}^B
   \end{bmatrix} 
\end{equation}
where
\begin{equation}\label{Ham-matrix}
\mathcal{B}=
   \begin{bmatrix}
   m_j\partial_i + \partial_j m_i &
   \rho\partial_i &
    -\phi_{,i} &
   -l^B_{,i}  &
  \pi_B\partial_i 
   \\[1mm]
   \partial_j\rho & 0 & -1 & 0 & 0
   \\[1mm]
   \phi_{,j}   & 1 & 0 & 0 & 0
   \\[1mm]
   l^A_{,j}& 0 & 0 & 0 & -1
   \\[1mm]
   \partial_j\pi_A & 0 & 0 & 1 & 0
   \end{bmatrix} .
\end{equation}
%
Here, the summation convention is enforced on repeated indices. Upper
Latin indices refer to the spatial components of the inverse map, lower 
Latin indices refer to the spatial reference frame, and subscript-comma notation 
is used for spatial derivatives. The
partial derivative $\partial_j=\partial/\partial x_j$, say, acts to the right
on all terms in a product by the chain rule. The Hamiltonian whose variations
are taken in (\ref{Ham-matrix-diff}) is given by
\[
h_M(m,\rho,\phi, l, \pi) = h(m,\rho) 
+ \frac{\sigma_1^2}{2} \|\gamma\|^2 
+ \frac{\sigma_2^2}{2} \|\phi\|^2.
\]

\subsection{$N$-dimensional Camassa-Holm equation}
\label{sec: EPDiff}

In this section we apply the distributed optimization method to the
$N$-dimensional Camassa-Holm equations 
\[
\dot v+u\!\cdot\!\nabla v+\nabla
u^T\!\cdot\!v+v\operatorname{div}u=0\,, \quad v:=(1-\alpha^2\Delta )u\,,
\]
which are the spatial representation of the geodesic spray on the group
$\operatorname{Diff}(\mathcal{D})$ of all diffeomorphisms of
$\mathcal{D}$, relative to a Sobolev $H^1$ metric; see
\cite{HoMa2004}. They are thus obtained by Euler-Poincar\'e
reduction and represent a particular case of the well known EPDiff
equations, to which the approach described here generalizes easily. For simplicity, we assume that $\mathcal{D}$ has no boundary. 

By analogy with the Euler equations, we shall give two approaches, namely, by composition on the left and on the right. However, in the case of the Camassa-Holm equations it is convenient to slightly generalize the previous setting by letting the diffeomorphism group act on a space of embeddings. More precisely, we first consider the \textit{left} action of $\operatorname{Diff}( \mathcal{D}) $ on the space of embeddings $\operatorname{Emb}(S, \mathcal{D}) $ of a manifold $S$ into $\mathcal{D}$ and obtain the distributed optimization for the cost function
\[
\int_0^T\left(\frac{1}{2}\|u\|^2_{H^1}+\frac{1}{2\sigma^2}\|\dot{\mathbf{Q}}-u\circ\mathbf{Q}\|^2\right)dt,\quad\mathbf{Q}\in\operatorname{Emb}(S,\mathcal{D}).
\]
Then, we let $\operatorname{Diff}( \mathcal{D}) $ acts on the \textit{right} on the space of embeddings $\operatorname{Emb}(\mathcal{D}, M) $ of a manifold $\mathcal{D}$ into a manifold $M$ and obtain the cost function
\[
\int_0^T\left(\frac{1}{2}\|u\|^2_{H^1}+\frac{1}{2\sigma^2}\|\dot{\mathbf{q}}+T\mathbf{q}\circ u\|^2\right)dt,\quad \mathbf{q} \in \operatorname{Emb}( \mathcal{D}, M).
\]

\subsubsection{Left action of diffeomorphisms on embedded subspaces}

Consider the left action of the configuration diffeomorphism group $G=\operatorname{Diff}(\mathcal{D})$ on $Q=\operatorname{Emb}(S,\mathcal{D})$. The infinitesimal generator associated to a Lie algebra element $u\in\mathfrak{X}(\mathcal{D})$ reads $u_{\operatorname{Emb}(S,\mathcal{D})}(\mathbf{Q})=u\circ\mathbf{Q}$ and belongs to the tangent space $T_\mathbf{Q}\operatorname{Emb}(S,\mathcal{D})$.

\subsection*{The Clebsch optimal control approach} Using the Lagrangian $\ell(u,\mathbf{Q})=\ell(u)=\frac{1}{2}\int_\mathcal{D}\|u\|_{H^1}^2dx$ and the constraint $\dot{\mathbf{Q}}=u\circ\mathbf{Q}$, the Clebsch optimal control problem yields the $N$-Camassa-Holm equation; see Section 4 in \cite{GBRa2011}.
Note that here there is no dependence of $\ell$ on the variable $\mathbf{Q}$, therefore the Clebsch approach yields the standard Euler-Poincar\'e equations. The stationarity conditions are
\begin{equation*}
\frac{\delta \ell}{ \delta u}=\mathbf{J}(\mathbf{Q},\mathbf{P})=\int_S\mathbf{P}(s)\delta (x-\mathbf{Q}(s))ds \in \Omega^1( \mathcal{D}),\quad \dot{\mathbf{Q}}=u\circ\mathbf{Q},\quad \dot{\mathbf{P}}=-(T^\ast u\circ \mathbf{Q}) \cdot \mathbf{P}.
\end{equation*}
The last equation can also be written as
\[
\frac{D}{Dt}\mathbf{P}=-\left((\nabla u)^T\circ\mathbf{Q}\right)\cdot\mathbf{P},
\]
where $D/Dt$ denotes the covariant derivative associated to the Riemannian metric on $\mathcal{D}$.

\subsection*{The distributed optimization approach} The proposed associated cost function is 
\begin{equation}\label{emb-lag1}
S_d
=
\int_0^T\Big(\ell(u)+\frac{1}{2\sigma^2}\|\dot{{\bf Q}}-u\circ{{\bf Q}}\|_{L^2}^2 \Big)dt\,.
\end{equation}
For definiteness, we rewrite this expression more explicitly as 
\begin{equation}\label{emb_lag2}
S_d
=
\int_0^T\Big(\ell(u)+\frac{1}{2\sigma^2}
\int_S |\dot{{\bf Q}}(t,s)-u(t,{\bf Q}(t,s))|^2ds \Big)dt\,,
\end{equation}
in which, for simplicity, $|\cdot|^2$ denotes the norm of vectors in $T \mathcal{D}$ defined by the Riemannian metric on $\mathcal{D}$ and $ds $ denotes the volume form on $S $. There could also be a sum on integrations over some finite number of embedded submanifolds of various dimensions, but this possibility is unimportant in the subsequent reasoning, so it will be suppressed in the notation. 

The choice of the reduced Lagrangian $\ell(u)$ will be left unspecified, except that sufficient smoothness will be assumed for the variational calculations manipulations to make mathematical sense, at least in terms of weak solutions. With these assumptions we have the following result.
\medskip

\begin{theorem}
The extremals of $S_d$ in \eqref{emb_lag2} are given by
\begin{equation}\label{emb_minima}
 \frac{\delta \ell }{\delta u}(x) = \int_S \mathbf{P}(t,s)\delta(x-{\bf Q}(t,s))ds,\quad \dot{{\bf Q}} = u \circ  {\bf Q}+ \sigma^2 \mathbf{P}^\sharp,\quad 
\frac{D}{Dt}{\mathbf{P}} = -\left(\left( \nabla u \right)^T \circ {\bf Q} \right) \cdot  \mathbf{P},
\end{equation}
where ${\bf Q} \in \operatorname{Emb}(S, \mathcal{D}) $,
$\mathbf{P}^\sharp\in T_{\bf Q} \operatorname{Emb}(S, \mathcal{D})$, and 
$D/Dt$ is the covariant derivative of the Levi-Civita connection on $\mathcal{D}$.
\end{theorem}
\medskip

\noindent\textbf{Proof.} We can obtain these conditions directly from the general equations \eqref{stationarity_cond_pi}. However, it is also instructive to derive them directly from the variational principle.

Consider the variations $\varepsilon\mapsto u_\varepsilon$ and $\varepsilon\mapsto \mathbf{Q}_\varepsilon$ and define $\mathbf{P}^\sharp$ by 
\[
\sigma^2\mathbf{P}^\sharp:=\dot{\mathbf{Q}}-u\circ\mathbf{Q}\,.
\]
For $\delta u=\left.\frac{d}{d\varepsilon}\right|_{\varepsilon=0}u_\varepsilon$ and $\delta \mathbf{Q}=\left.\frac{d}{d\varepsilon}\right|_{\varepsilon=0}\mathbf{Q}_\varepsilon$,
we have
\begin{align*}
\delta S_d&=\int_0^T\left\langle\frac{\delta \ell}{\delta u},\delta u\right\rangle dt\\
&\qquad 
+\int_0^T\!\!\int_S\left\langle \mathbf{P}(t,s),\left.\frac{D}{D\varepsilon}\right|_{\varepsilon=0}\dot{\mathbf{Q}}_\varepsilon(t,s)-\delta u(t,\mathbf{Q}(t,s))-\left.\frac{D}{D\varepsilon}\right|_{\varepsilon=0} u(t,\mathbf{Q}_\varepsilon(t,s))\right\rangle dsdt\\
&=\int_0^T\left\langle\frac{\delta \ell}{\delta u},\delta u\right\rangle dt
+\int_0^T\!\!\int_S \left\langle \mathbf{P}(t,s),\frac{D}{Dt}\left.\frac{d}{d\varepsilon}\right|_{\varepsilon=0}\mathbf{Q}_\varepsilon(t,s)  
\right\rangle dsdt\\
&\qquad
-\int_0^T\!\!\int_\mathcal{D}\!\int_S
\left\langle \mathbf{P}(s)\de (x-\mathbf{Q}(t,s)),\delta u(t,x) \right\rangle ds dx dt 
-\int_0^T\!\!\int_S
\left\langle \mathbf{P}(t,s),\nabla_{\delta\mathbf{Q}}u(t,s) \right\rangle dsdt\\
&=
\int_0^T\left\langle\frac{\delta \ell}{\delta u}-\int_S\mathbf{P}(s)\delta(x-\mathbf{Q}(t,s))ds,\delta u\right\rangle dt-\int_0^T\left\langle\frac{D}{Dt}\mathbf{P}+\left((\nabla u)^T\circ\mathbf{Q}\right)\cdot\mathbf{P},\delta\mathbf{Q}\right\rangle dt\\
&\qquad +\Big[\left\langle\mathbf{P},\delta\mathbf{Q}\right\rangle\Big]_0^T.
\end{align*}
The stationarity conditions follow immediately, upon noting that 
$\delta \mathbf{Q}(0,s)=0 = \delta \mathbf{Q}(T,x) $, so that temporal endpoint terms arising under integrations by parts may be ignored.$\qquad\blacksquare$
\medskip

Suppose the reduced Lagrangian defines a velocity norm, $\ell(u)=\frac12\|u\|^2=\frac12\left\langle u,Q_{op}(u)\right\rangle$. For example, let the norm be a Sobolev $H^1$ norm, so that it makes sense for its variational derivative in $u$ to result in a singular distribution defined on an embedded subspace. Then, the density equation 
\begin{equation}
\label{Qop_eqn}
\frac{\delta \ell}{\delta u}(t,x)= \int_S \mathbf{P}(t,s)\delta(x-{\bf Q}(t,s))ds = :(Q_{op}u) (t,x)
\end{equation}
 has a natural dual solution for the velocity, given by
\begin{equation}\label{u_def}
u(t,x) = \int_S \mathbf{P}^\sharp(t,s)G(x-{\bf Q}(t,s))ds
\end{equation}
where $G$ is the Green's function for the positive $L^2$ self-adjoint operator $Q_{op}$, that is, 
\begin{equation}\label{Green_eqn}
Q_{op\,} G(x-{\bf Q}(t,s)) = \delta(x-{\bf Q}(t,s)) .
\end{equation}
In this situation, we have enough assumptions to obtain a coupled system of equations for the momentum densities 
$\mathbf{P}(t,s)$ and $m(t,x)$. 
\medskip

\begin{theorem}
The system of variational equations \eqref{emb_minima} for the minima of $S$ in \eqref{emb_lag2} implies the following dynamics for the momentum densities $\mathbf{P}(t,s)$ and $m(t,x)$,  
\begin{eqnarray}
\pa_t v + \nabla_uv+\nabla u^T\!\cdot\! v+v\operatorname{div}(u)&=&
 -\sigma^2 \operatorname{Div} \int_S \mathbf{P}^\sharp\otimes\mathbf{P}^\sharp (t,s)\de(x-\mathbf{Q}(t,s))ds
\label{m_eqn}
\\
\frac{D}{Dt}{\mathbf{P}} + \left(\left( \nabla u \right)^T \circ {\bf Q} \right) \cdot  \mathbf{P} &=& 0,
\label{p_eqn} 
\end{eqnarray}
where $\operatorname{Div}$ denotes the divergence of a contravariant two-tensor field on $\mathcal{D}$.
The remaining decoupled equation
\[
\dot{\mathbf{Q}} = u\circ\mathbf{Q} + \sigma^2 \mathbf{P}^\sharp
\]
allows reconstruction of the Lagrangian coordinates $\mathbf{Q}(t,s)$ on the embedded surface(s) from the dynamics of the coupled equations for the momentum densities $m(t,x)$ and $\mathbf{P}(t,s)$.
\end{theorem}
\medskip

\noindent\textbf{Proof.} Substitution of equations \eqref{emb_minima} and definitions \eqref{u_def}-\eqref{Green_eqn} into the definition of the momentum $m$ in equation \eqref{Qop_eqn} verifies its evolution by \eqref{m_eqn}, upon pairing with a smooth test function and integrating appropriately by parts.

Alternatively, one can use the abstract formulation of the dynamical equations given in \eqref{unconstrained_A}. As recalled before, the Euler-Poincar\'e part of these equations gives the $N$-Camassa-Holm equation 
\[
\dot{v}+\nabla_uv+\nabla u^T\cdot v+v\operatorname{div}(u)=0
\,.
\]
Thus, it remains to compute the expression of the tensor $\mathcal{F}^\nabla$. Let $\mathbf{P}^\sharp\in T_{\mathbf{Q}}\operatorname{Emb}(S,\mathcal{D})$ and $u\in\mathfrak{X}(\mathcal{D})$, and choose $X\in\mathfrak{X}(\mathcal{D})$ such that $\mathbf{P}^\sharp(s)=X(\mathbf{Q}(s))$. Using the fact that the covariant derivative on $\operatorname{Emb}(S,\mathcal{D})$ is the functorial lift of the covariant derivative on $\mathcal{D}$, using \eqref{definition_F_nabla} we get
\begin{align*}
\left\langle\mathcal{F}^\nabla(\mathbf{P},\mathbf{P}^\sharp),u\right\rangle&=\left\langle \mathbf{P},\nabla_{\mathbf{P}^\sharp}u_{\operatorname{Emb}(S,\mathcal{D})}(\mathbf{Q})\right\rangle=\int_Sg\left(\mathbf{P}^\sharp(s),\nabla_{\mathbf{P}^\sharp(s)}u(\mathbf{Q}(s))\right)ds\\
&=\int_S\int_\mathcal{D}g\left(X(x),\nabla_{X(x)}u(x)\right)\delta(x-\mathbf{Q}(s))dxds\\
&=-\int_\mathcal{D}g\left(\int_S\operatorname{Div}\big(X(x)\otimes X(x)\delta(x-\mathbf{Q}(s))\big),u(x)\right)dxds,
\end{align*}
where we make use of the identity
\[
\int_\mathcal{D}g(X,\nabla_Yu)dx=-\int_\mathcal{D}g(\operatorname{Div}(Y\otimes X),u)dx,\quad\text{for all}\quad X,Y,u\in\mathfrak{X}(\mathcal{D}),
\]
where $\operatorname{Div}(T)^j = \nabla_i T^{ij}$, where $T = T^{ij} \frac{\partial}{ \partial x^i} \otimes \frac{\partial}{ \partial x^j}$ is a contravariant two-tensor on $\mathcal{D}$.
We thus obtain the formula
\begin{align*}
\mathcal{F}^\nabla(\mathbf{P},\mathbf{P}^\sharp)
&=-\int_S\operatorname{Div}\big(X(x)\otimes X(x)\delta(x-\mathbf{Q}(s))\big)ds\\
&=-\operatorname{Div}\int_S X(\mathbf{Q}(s))\otimes X(\mathbf{Q}(s))\delta(x-\mathbf{Q}(s)) ds\\
&=-\operatorname{Div}\int_S\left(\mathbf{P}^\sharp(s)\otimes \mathbf{P}^\sharp(s)\delta(x-\mathbf{Q}(s))\right)ds
\end{align*}
as required.$\qquad\blacksquare$
\medskip

\begin{remark} \normalfont 
Equations \eqref{m_eqn} and \eqref{p_eqn} represent a new dynamical system, whose exploration has only just begun and we expect will be a subject of future research. 
\quad $\blacklozenge$
\end{remark}

\subsubsection{Back-to-labels map for fluids}\label{btl_map}

We next present the optimal control derivation of the Camassa-Holm equation using the back-to-labels map. This means that we shall use the right action of $\operatorname{Diff}(\mathcal{D})$ on $\operatorname{Emb}(\mathcal{D}, M) $.

Recall that particles \emph{frozen} into an ideal fluid flow are represented by time-dependent vector labels $l_t$ whose components each satisfy the \emph{advection law} obtained from the time derivative of the \emph{back-to-labels map}, $l_t(x):=\eta_t^{-1}(x)=l(t,x)$, and hence it satisfies the equation
\begin{equation}
\label{label eqn}
\dot{l} + T l \circ u = 0,
\end{equation}
where $u $ is the Eulerian velocity of the fluid.

We shall slightly generalize the back-to-labels map by considering embeddings $\mathbf{q}: \mathcal{D} \rightarrow M $, where $M $ is a given Riemannian manifold, instead of diffeomorphisms $l:\mathcal{D}\rightarrow\mathcal{D}$.

\subsection*{The Clebsch approach} We recall from \cite{GBRa2011} how one can obtain the Camassa-Holm equation by Clebsch optimal control
via a generalization of the back-to-labels map.

Let the group $G=\operatorname{Diff}(\mathcal{D})$ acts freely on
the right on the manifold $\operatorname{Emb}(\mathcal{D},M)$. The associated infinitesimal generator reads $u_{\operatorname{Emb}(\mathcal{D},M)(\mathbf{Q})}=T\mathbf{q}\circ u$. Using the Lagrangian $\ell(u)=\frac{1}{2}\|u_t\|_{H^1}^2$ and the constraint $\dot {\mathbf{q}}+T\mathbf{q}\circ u=0$ we get the stationarity conditions
\[
\frac{\delta \ell}{ \delta u} = -{\bf p} \cdot T \mathbf{q}, \qquad \dot{ \mathbf{q}} + T \mathbf{q} \circ u = 0, \qquad \dot{ {\bf p}} + T {\bf p} \circ u = 0.
\]
These equations produce the Camassa-Holm equation if one uses the Hamiltonian
\[
H(\mathbf{q},\mathbf{p})=\frac{1}{2}\iint
\mathbf{p}(x)\!\cdot\!T\mathbf{q}(x)G(x-x')\mathbf{p}(x')\!\cdot\!T\mathbf{q}(x')dxdx'.
\]

\subsection*{Distributed optimization}

As opposed the Clebsch approach, we do not impose  $\dot {\mathbf{q}}_t+T\mathbf{q}_t\circ u_t=0$. Instead we use  $\|\dot {\mathbf{q}}_t+T\mathbf{q}_t\circ u_t\|^2_{L^2}$ as a penalty, that is, we consider
 the cost functional given by
\begin{equation}
S_d=
\int\limits ^T_0
\Big(\ell(u) 
+ \frac{1}{2\sigma^2} \underbrace{\
\| \dot{\mathbf{q}} + T \mathbf{q} \circ u \|_{L^2}^2\
}_{\hbox{Penalty}}
\Big)\mbox{d} t
\,.
\label{Lag-inversemap}
\end{equation}
Thus we need to minimize $S_d$ subject to spatial boundary conditions, endpoint conditions ($\mathbf{q}(0,x)$ and $\mathbf{q}(T,x)$ are prescribed), and penalize for the error in the $L^2$ norm, 
\begin{equation}\label{L2norm}
\|\dot{\mathbf{q}} + T \mathbf{q} \circ u\|_{L^2}^2=
\int_\mathcal{D}|\dot{\mathbf{q}}(x)+T\mathbf{q}(u(x))|^2dx
\end{equation}
in which, for simplicity, $|\cdot|^2$ denotes the norm of vectors in $TM$ defined by a Riemannian metric on $M$. It is important to note that the $L^2 $ Riemannian metric used in the penalty is \textit{not} invariant under the right $\operatorname{Diff}( \mathcal{D}) $-action on itself.
\medskip

\begin{remark}\normalfont (An alternative penalty term)
If $M =  \mathcal{D}$, the quantity
\[
v: = -\dot{l} \circ l ^{-1} = T l \circ u \circ l ^{-1} = l _* u = \operatorname{Ad}_l u
\] 
is called the \emph{convective velocity} \cite{HoMaRa1986} of the fluid. This is analogous to the relation $\Omega={\rm Ad}_{O^{-1}}\omega$ for $O\in SO(3)$ satisfied by body angular velocity $\Omega$ and spatial angular velocity $\omega$ for rigid body motion in $\mathbb{R}^3$, both viewed as elements of $\mathfrak{so}(3)$. Penalizing in (\ref{Lag-inversemap}) for $\|v-{\rm Ad}_{l}u\|_{L^2}^2$  is an interesting alternative approach, which will be presented, in general, in 
\S\ref{subgrp-act}. 
\quad $\blacklozenge$
\end{remark} 
\medskip

Let $\sigma^2>0$ and choose the reduced Lagrangian to be a norm $\ell(u) = \frac12\|u\|^2$. Then, when extremals of (\ref{Lag-inversemap}) exist, they will be minima. \medskip 

Later we shall specialize the reduced Lagrangian to the norm $\ell(u) = \frac12\|u\|^2_{H^1}$. For the moment, however, we leave the choice arbitrary, only assuming that sufficient smoothness is present for all functions to exist locally and be differentiable in space and time. With these assumptions we have the following result.
\medskip

\begin{theorem}
The extremals of $S_d$ in \eqref{Lag-inversemap} are given by
\begin{equation}\label{OCminima}
\frac{\delta \ell }{\delta u}+{\pi}\circ  T \mathbf{q}=0,\quad
\frac{D\pi}{D t}+{\rm Div}(\pi u)=0,\quad 
\dot{\mathbf{q}}+T\mathbf{q}\circ  u=:\sigma^2\pi^\sharp,
\end{equation}
where the expression $\operatorname{Div}(\pi u) \in T^\ast_{ \mathbf{q}} \operatorname{Emb}( \mathcal{D}, M ) $ is defined by
\[
\operatorname{Div}(\pi u) : = (\operatorname{div} u) \pi + \nabla_u \pi,\quad\text{with}\quad \nabla_{u_x} \pi : = \left.\frac{D}{D\varepsilon}\right|_{\varepsilon=0} \pi( c( \varepsilon))
\]
for $u_x \in T_x \mathcal{D}$, $\varepsilon\mapsto c(\varepsilon) $ a curve such that $\left.\frac{d}{d\varepsilon}\right|_{\varepsilon=0}c(\varepsilon) = u_x$, and  $\frac{D}{D\varepsilon} $ is the covariant derivative of the Levi-Civita connection of the Riemannian metric on $M $. 
\end{theorem}
\medskip

Note that if $M = \mathbb{R}^n$ endowed with the Riemannian metric given by the dot product then 
$\operatorname{Div}( \pi u)_i = \operatorname{div}( \pi_i u)$.
\medskip

\noindent\textbf{Proof.} Define $\pi\in T_\mathbf{q}\operatorname{Emb}(\mathcal{D},M)$ by $\sigma^2\pi:=\dot{\mathbf{q}}+T\mathbf{q}\circ u$. For variations $\varepsilon\mapsto u_\varepsilon$ and $\varepsilon\mapsto \mathbf{q}_\varepsilon$, we compute
\begin{align*}
\delta S_d&=\int_0^T\left\langle\frac{\delta\ell}{\delta u},\delta u\right\rangle dt+\int_0^T\left\langle \pi,\left.\frac{D}{D\varepsilon}\right|_{\varepsilon=0}\left(\dot{\mathbf{q}}_\varepsilon+T\mathbf{q}_\varepsilon\circ u_\varepsilon\right)\right\rangle\\
&=\int_0^T\left\langle\frac{\delta\ell}{\delta u},\delta u\right\rangle dt+\int_0^T\left\langle \pi,\frac{D}{Dt}\delta\mathbf{q}+T\mathbf{q}\circ\delta u+\nabla_u\delta \mathbf{q}\right\rangle\\
&=\int_0^T\left\langle\frac{\delta\ell}{\delta u}+\pi\circ T\mathbf{q},\delta u\right\rangle
-
\int_0^T\left\langle \frac{D}{Dt}\pi + \operatorname{Div}(\pi u),\delta\mathbf{q}\right\rangle,
\end{align*}
where in the last equality, we used integration by parts and the definition of $\operatorname{Div}$.$\qquad\blacksquare$

\begin{theorem}
The system of variational equations \eqref{OCminima} for the minima of $S_d$ yields the following dynamical system for the momentum $\pi$ and momentum 1-form $v^\flat:=\de \ell/\delta u=-\pi\circ T\mathbf{q}$,  
\begin{eqnarray}
\pa_t v + \nabla_uv+\nabla u^T\!\cdot\!v+v\operatorname{div}(u)&=& \sigma^2 (\nabla\pi)^T\!\cdot\!\pi^\sharp\label{OCmin_Lie}\\
\pa_t {\pi}+\operatorname{Div}(\pi u)&=&0,\label{OCmin_Lie2}
\end{eqnarray}
where $\operatorname{Div}(\pi u)$ is defined above. The decoupled equation $\sigma^2\pi^\sharp=\dot{\mathbf{q}}+T\mathbf{q}\circ u$ allows reconstruction of the labels $\mathbf{q}$ from the dynamics of the coupled equations for $v$ and $\pi$.
\end{theorem}
\medskip

\noindent
\textbf{Proof.} One can directly obtain these equations from the stationarity condition given in \eqref{OCminima}. We shall however use the abstract formulation \eqref{unconstrained_A'} and compute the tensor field $\mathcal{F}^\nabla$ defined in \eqref{definition_F_nabla}. Given $\pi\in T^\ast_\mathbf{q}\operatorname{Emb}(\mathcal{D},M)$, $u\in\mathfrak{X}(\mathcal{D})$, and a curve $\varepsilon \mapsto \mathbf{q}_ \varepsilon \in \operatorname{Emb}( \mathcal{D}, M) $ such that $\left.\frac{d}{d\varepsilon}\right|_{\varepsilon=0}\mathbf{q}_\varepsilon = \pi^\sharp$, we have
\begin{align*}
\left\langle\mathcal{F}^\nabla(\pi,\pi^\sharp),u\right\rangle
&=\left\langle\pi,\nabla_{\pi^\sharp} u_{\operatorname{Emb}(\mathcal{D},M)}(\mathbf{q})\right\rangle
=\int_\mathcal{D} g\left(\pi^\sharp(x),\left.\frac{D}{D\varepsilon}\right|_{\varepsilon=0}T\mathbf{q}_\varepsilon(u(x))\right)dx\\
&=\int_\mathcal{D} g\left(\pi^\sharp(x),\nabla_u\pi(x)\right)dx
=\left\langle (\nabla\pi^T)\!\cdot\!\pi^\sharp,u\right\rangle
\end{align*}
which proves \eqref{OCmin_Lie}. Equation \eqref{OCmin_Lie2} is part of the system \eqref{OCminima}. $\qquad \blacksquare$
\medskip

\begin{remark}\normalfont (Two-component Camassa-Holm equation)
If $\mathcal{D}=\mathbb{R}$, $M=\mathbb{R}$, and we assume appropriate decay properties at infinity such that all boundary terms appearing in integration by parts vanish, specializing the reduced Lagrangian to 
\[
\ell(u) = \frac12\|u\|^2_{H^1} = \frac{1}{2} \int_ \mathbb{R} \left(u^2 + \alpha^2 u_x^2\right)d{x},
\]
with homogeneous boundary conditions on the infinite real line or on a periodic spatial interval, yields variational derivative $(\de \ell/\delta u)^\sharp=v=u-\alpha^2u_{xx}$, for a length scale $\alpha$. In this case,
equations \eqref{OCmin_Lie}, \eqref{OCmin_Lie2} recover the \textit{two-component Camassa-Holm equations}, 
\begin{eqnarray}
\pa_t v + (uv)_x + vu_x
 &=&
\sigma^2\pi\pi_x
\,,
\label{CH2-motion-eqn}\\
\pa_t \pi + (u\pi)_x  &=& 0
 \,.
 \label{CH2-continuity-eqn}
\end{eqnarray}
This system forms a completely integrable Hamiltonian system with soliton solutions associated to an isospectral linear eigenvalue problem, so it may be solved analytically by using the inverse scattering transform method \cite{ChLiZh2006,Ku2007}. These equations are also known to be the spatial representation of geodesics on the semidirect product $\operatorname{Diff}(\mathbb{R})\,\circledS\,\mathcal{F}(\mathbb{R})$; see \cite{HoTr2008}, \cite{GBTrVi2009}.
\quad $\blacklozenge$
\end{remark}

\subsection{Metamorphosis dynamics}\label{optimage-reg}

Consider a Lie group $G$ acting on the \textit{left} on a manifold $N$. The Lie group $G$ is the group of deformations and the manifold $N$ contains what are called ``deformable objects''. In imaging applications we take $G$ to be the group of diffeomorphisms of $N$.
\medskip

\begin{definition}\label{def-metamorf}
A {\bfi metamorphosis} {\rm (\cite{TrYo2005,HoTrYo2009})} is a pair of curves $(g_t,\,\eta_t)\in G \times N$ parameterized by time $t$, with $g_0 = \id$. 
Its {\bfi image} is the curve $n_t\in N$ defined by the action $n_t = g_t\eta_t$ denoted by concatenation from the left.
The quantities $g_t$ and $\eta_t$ are called, respectively, the {\bfi deformation part} of the
metamorphosis, and its {\bfi template part}. When $\eta_t$ is
constant, the metamorphosis is said to be a {\bfi pure
deformation}. In the general case, the image is a combination of a
deformation and template variation.
\end{definition}
\medskip

A metamorphosis may be determined as an optimal curve $(g_t,\eta_t)$ with $g_t \in G$ and $\eta_t \in N$ with respect to a metric that is invariant under the {\bfi right action} of $G$ on $G\times N$ defined by
\begin{equation}
(g,\eta)h=\left(gh,h^{-1}\eta\right)
\label{r-action}
\end{equation}
for any $g,h\in G$ and $\eta\in N$. 
More specifically, a metamorphosis $(g, \eta)$ may be obtained by seeking a stationary point $\de S=0$ of a right-invariant cost function $S$ on $T(G\times N)$. This general situation has been considered in detail in the first sections of the paper.

The present conventions are those of equations \eqref{unconstrained_A} with the upper sign chosen. Recall in particular, that we start from a right $G$-invariant Lagrangian of the form
\[
L(g,\dot g,\eta,\dot\eta)=\mathcal{L}(g,\dot g,\eta)+\frac{1}{2\sigma^2}\|g\dot\eta\|^2,
\]
where the norm involved in the penalty is associated to a Riemannian metric $g$ on $N$. The corresponding reduced Lagrangians on $\mathfrak{g}\times TN$ read
\[
\ell_{EP}( u,\nu_n)=\ell( u,n)+\frac{1}{2\sigma^2}\|\nu_n\|^2,\quad\ell_M( u,n,\dot n)=\ell( u,n)+\frac{1}{2\sigma^2}\|\dot n- u_N(n)\|^2,
\]
where the reduced variables are
\[
 u=\dot gg^{-1} \in \mathfrak{g} ,\quad n=g\eta \in N,\quad \nu_n=g\dot \eta \in T_n N
\]
with $g \in G$, $\eta \in N$.
\medskip

\subsubsection{Subgroup actions}\label{subgrp-act}\rm

We shall discuss in this paragraph the particular case in which $N$ is also a Lie group that contains $G$ as subgroup and on which $G$ acts by multiplication on the left. We also assume that the Riemannian metric $g$ on $N$ is left invariant (relative to left translations by elements of $N $). In this case, one can make use of left trivialization of the tangent bundle $TN$ to get the diffeomorphism
\[
\mathfrak{g}\times TN\rightarrow \mathfrak{g}\times N\times \mathfrak{n},\quad ( u,n,\dot n)\mapsto ( u,n,n^{-1}\dot n)=:( u,n,\zeta).
\]
The reduced Lagrangian in terms of the new variables is denoted $\ell_L$ and reads
\[
\ell_L( u,n,\zeta)=\ell( u,n)+\frac{1}{2\sigma^2}\|\zeta-\operatorname{Ad}_{n^{-1}} u\|^2
\]
since we have the relations
\[
n^{-1}\nu_n=n^{-1}(\dot n- u_N(n))=n^{-1}(\dot n- u n)=\zeta-\operatorname{Ad}_{n^{-1}} u.
\]
We now rewrite the stationarity conditions relative to these new variables. Consider variations $\varepsilon\mapsto u_\varepsilon$ and $\varepsilon\mapsto n_\varepsilon$ of the curves $u$ and $n$. We have as usual
\begin{equation}\label{xi-prime}
\delta\zeta=\dot\Sigma+[\zeta,\Sigma]
\end{equation}
where $\Sigma=n^{-1}\delta n$.
Likewise,
\begin{align*}
\delta(\Ad_{n^{-1}} u)&= 
\Ad_{n^{-1}}\left(\delta  u + [ u,\delta n n^{-1}]\right)=\Ad_{n^{-1}}\left(\delta  u + [ u,\operatorname{Ad}_n\Sigma]\right)\\
&=\Ad_{n^{-1}}\delta  u + \left[\operatorname{Ad}_{n^{-1}} u,\Sigma\right].
\end{align*}
For simplicity, we suppose that $\ell$ does not depend on $n$. Substituting these relations into the variation of the action integral we get
\begin{align*}
\de S_d
&=\de \int_0^T \ell_L( u,n,\zeta)dt= \delta \int_0^T\left(\ell( u)+\frac{1}{2\sigma^2}\|\zeta-\operatorname{Ad}_{n^{-1}} u\|^2\right) dt\\
&=
\int_0^T\left(\left\langle\frac{\delta\ell}{\de  u},\delta u\right\rangle + \left\langle\pi, \delta\ze-\delta(\Ad_{n^{-1}} u)\right\rangle\right) dt\\
&=
\int_0^T \left(\left\langle\frac {\delta\ell}{\delta u},\delta u\right\rangle
+ \left\langle\pi,\dot{\Sigma} + \ad_\zeta\Sigma-\Ad_{n^{-1}}\delta u - \ad_{\left(\Ad_{n^{-1}} u\right)}\Sigma\right\rangle\right) dt
\\ &=
\int_0^T\left(\left\langle\frac{\delta\ell}{\delta  u}-\Ad^*_{n^{-1}}\pi,\delta u\right\rangle - \left\langle \dot{\pi} - \ad^*_\ze\pi +  \ad^*_{\left(\Ad_{n^{-1}} u\right)}\pi, \Sigma\right\rangle\right)dt
+ \Big[\left\langle\pi,\Sigma\right\rangle\Big]_0^T,
\end{align*}
where $\pi\in\mathfrak{n}^*$ is the image momentum dual to the left-invariant image velocity $\ze\in\mathfrak{n}$, that is,
\[
\pi:=\frac{\delta\ell_L}{\delta\ze}=\frac{1}{\sigma^2}\left(\zeta-\operatorname{Ad}_{n^{-1}} u\right)^\flat=\frac{1}{\sigma^2}\left(n^{-1}\nu_n^{\,\flat}\right)=\frac{1}{\sigma^2}n^{-1}\left(\nu_n^{\,\flat}\right)= \in\mathfrak{n}^*.
\]

Stationarity $\de S=0$ and $\Sigma(0) = \Sigma(T) = 0$ then implies
\begin{equation}
\label{stat-eqns}
\frac{\delta\ell}{\delta  u}=\Ad^*_{n^{-1}}\pi
\quad\hbox{and}\quad
\dot{\pi} =  \ad^*_\ze\pi -  \ad^*_{(\Ad_{n^{-1}} u)}\pi  
= \ad^*_{\sigma^2\pi^\sharp}\pi = \sigma^2\ad^*_{\pi^\sharp}\pi.
\end{equation}
From the general theory, since the $G $-action on $N $ is by isometries it follows that $\mathcal{F}^\nabla=0$, and thus these equations imply the Euler-Poincar\'e equations. It is also instructive to obtain them directly. Taking the time derivative and using general results relating the $\Ad^*$ and $\ad^*$ operations yields
\begin{eqnarray}
\frac{d}{dt}\frac{\delta\ell}{\delta  u}
&=&\frac{d}{dt}\Big(\Ad^*_{n^{-1}}\pi\Big)
\nonumber\\&=&
\Ad^*_{n^{-1}}\Big(\dot{\pi} - \ad^*_\ze\pi\Big)
\quad\Big(\hbox{with $\ze=n^{-1}\dot{n}$}\Big)
\nonumber \\
\hbox{by (\ref{stat-eqns}b)}
&=&-\,\Ad^*_{n^{-1}}\ad^*_{\left(\Ad_{n^{-1}} u\right)}\pi
\nonumber\\&=&-\,\ad_ u^*\Big(\Ad^*_{n^{-1}}\pi\Big)
\nonumber\\
\hbox{by (\ref{stat-eqns}a)}
&=&-\,\ad_ u^*\frac{\delta\ell}{\delta  u}
\,. \nonumber
\end{eqnarray}
In turn, using $u = \dot{g} g ^{-1}$ and $n= g \eta$, from
the Euler-Poincar\'e equation we get the conservation law,
\begin{align*}
0&=\Ad^*_{g}\left(
\frac{d}{dt}\frac{\delta\ell}{\delta  u}
+\ad_ u^*\frac{\delta\ell}{\delta  u}\right)
=\frac{d}{dt}\left(\Ad^*_{g}\frac{\delta\ell}{\delta  u}\right) 
=
\frac{d}{dt}\left(\Ad^*_{g}\Ad^*_{n^{-1}}\pi\right)\\
&=
\frac{d}{dt}\left(\Ad^*_{\eta^{-1}}\pi\right)
=
\Ad^*_{\eta^{-1}}\left(\dot{\pi} - \ad^*_\upsilon\pi\right),
\end{align*}
where $\upsilon: = \eta^{-1}\dot{\eta}$ is the left-invariant template velocity. 
\medskip

\begin{remark}\normalfont (Interpretation of the equations) 
$\quad$ 
\begin{enumerate}
\item
The conservation laws for $\Ad^*_{g}(\delta\ell/\delta  u)$ 
and $\Ad^*_{\eta^{-1}}\pi$ provide the interpretations of the 
momentum dynamics. Namely, the momentum 
$\delta\ell/\delta  u$ (resp. $\pi$) undergoes coadjoint 
motion with respect to $g$ (resp. $\eta^{-1}$).
\item
The peculiar form of the momentum equation (\ref{stat-eqns}b) 
is then understood, because the template velocity $\upsilon$ 
is proportional to image momentum $\pi$ by a factor of the 
penalty constant, which also maps it from the dual of the Lie 
algebra, back to Lie algebra, namely, 
\[
\upsilon := \eta^{-1}\dot{\eta} = n^{-1}\nu_n = \sigma^2\pi^\sharp.
\]
Perhaps not unexpectedly, when $\sigma^2\to0$ the template velocity vanishes and the remaining image motion reduces to a pure deformation governed by the Euler-Poincar\'e equation.

\item
The metamorphosis $(g_t,\eta_t)$ is determined as an initial value problem, as follows. Given the Lagrangian $\ell(u)$, the  Euler-Poincar\'e equation 
\[
\frac{d}{dt}\frac{\delta\ell}{\delta  u}+\ad_ u^*\frac{\delta\ell}{\delta  u}=0\,,
\]
determines the velocity $ u=\dot{g}g^{-1}$ which then yields $g_t$ by reconstruction from solving $\dot{g}_t=u_tg_t$. Next, the relations 
\[ 
\dot{\pi} =  {\ad}^*_{\left(\ze-\Ad_{n^{-1}} u\right)}\pi  
\quad\hbox{and}\quad
\sigma^2\pi^\sharp = \ze - \Ad_{n^{-1}} u
\,,
\]
with $\ze=n^{-1}\dot{n}$ and 
$\dot{n} =  u n+\nu_n$
need to be negotiated to obtain the image curve $n_t$. Finally, the template curve is obtained from $\eta_t=g_t^{-1}n_t$. This process is worth discussing in an example. \quad $\blacklozenge$
\end{enumerate}
\end{remark}

\subsubsection{Example: Metamorphosis equations on $SE(2)$} 

In $SE(2)$ the manifold of ``deformable objects'' $N=\mathbb{R}^2$ is acted upon by the Lie group of ``deformations'' $G= SO(2)$ on the left.
The situation simplifies in this case because $N$ is a vector space and we recover the setting described in \S\ref{representation}.
Hence,
\[
\ell_M( u,n,\nu) 
= \ell( u) + \frac{1}{2\sigma^2}  \|\nu\|^2
= \ell( u) + \frac{1}{2\sigma^2}  \| \dot{n}- u n\|^2=\ell_{LP}( u,n,\dot n)
\]
and the cost function becomes 
\[
S_d =  \int_0^T \left(\ell(u) + \frac{1}{2\sigma^2}  \|\nu\|^2\right) dt=\int_0^T \left( \ell(u) + \frac{1}{2\sigma^2}  \| \dot{n}-un\|^2\right) dt
\,,
\]
where the $\mathfrak{se}(2)$ Lie algebra action $un$ may be written on $\mathbb{R}^2$ as a cross product of vectors \cite{Ho2008}
\[
un= u\zhat\times n.
\]
Consequently, the SDP metamorphosis equations (see \eqref{stationarity_conditions_p_representation})
\begin{eqnarray*}
\frac{\de \ell}{\de u} + \pi \diamond n &=& 0,\\
\dot{\pi} - u \pi
&=& 0,\\
\dot{n} - un 
&=& \sigma^2 \pi^\sharp = \nu,
\end{eqnarray*}
may be written in vector form as
\begin{eqnarray*}
\frac{\de \ell}{\de u} \zhat + \pi \times n &=& 0,\\
\dot{\pi} - u \zhat \times \pi 
&=& 0,\\
\dot{n} - u\zhat \times n 
&=& \sigma^2 \pi = \nu.
\end{eqnarray*}

A few statements may be made about the qualitative properties of the solutions of this system.
\begin{enumerate}
\item We first observe that $|\pi|$ is constant because by the second equation above, we have
$\frac{d}{dt}| \pi|^2 = 2 \pi\cdot \dot{ \pi} = 2\pi \cdot ( u \zhat \times \pi )=0 $. So $\pi$ executes circular motion in the plane at constant rotation frequency $\pi\times\dot{\pi}/|\pi|^2=u\zhat$. 

\item
Substituting the second and third equations into the time derivative of the first one yields the conservation law, 
\[
\frac{d}{dt} \frac{\delta\ell}{\delta u} = 0,
\]
for the planar motion. In particular, we obtain the constant of motion $\pi\times n=$ const.

\item The other two equations are closed provided one may solve ${\delta\ell}/\delta u$ for $u$, which of course we shall assume is possible. More precisely, we now assume that the Legendre transformation $u \mapsto \delta \ell/ \delta u$ is a diffeomorphism. In this case, since ${\delta\ell}/\delta u$ is constant, $u$ is also constant. \item
It remains to determine the effects of $\sigma^2\ne0$ on the dynamics of $n$. A short computation shows that:
\[
\frac{d}{dt}(\pi\cdot n) = \sigma^2 |\pi|^2
\quad\hbox{and}\quad
\frac{d}{dt}|n|^2 = 2\sigma^2 (\pi\cdot n),
\]
so, since $|\pi|^2=$ const, $|n|^2(t)$ increases quadratically with scaled time $\sigma^2t$ and the motion may be visualized as taking place in $\mathbb{R}^3$ with coordinates $(x_1,x_2,x_3)=(|\pi|^2,|n|^2,\pi\cdot n)$ along the parabolas formed by intersections of level sets of the two integrals of motion $|\pi|^2=$ constant and $|\pi\times n|^2=$ const.
The rotation frequency of $n$ is found as
\[
\frac{n\times\dot{n}}{|n|^2} = \zhat \left(u + \frac{\sigma^2}{|n|^2}\frac{\de l}{\de u}\right).
\]
As $\sigma^2t\to\infty$, the directions of the vectors $\pi$ and $n$ tend toward a state of alignment, rotating together at frequency $u\zhat$. In contrast, for $\sigma^2=0$, the vectors $\pi$ and $n$ keep their magnitudes and rotate together at frequency $u\zhat$ with constant relative orientation.
\end{enumerate}

\subsubsection{Lie-Poisson Hamiltonian formulation of metamorphosis for right  action}

In this example we particularize the system of motion 
equations \eqref{M_left_right} to the case of a representation but without imposing the endpoint condition 
at $t=1$. The resulting equations are obtained by metamorphosis reduction from an arbitrary Lagrangian 
$L:T(G\times V) \rightarrow \mathbb{R}$, where $V$ is a vector space. Thus, the equations below are more general 
that those obtained in the penalty approach. 

As explained in Section \ref{sec_Lagr_approach}, the variational problem optimizes over metamorphoses 
$(g_t, \eta_t)$ by minimizing $S=\int_0^1 L\,dt$, for 
a  Lagrangian $L$ of the form
\[
L(g_t, \dot g_t, \eta_t, \dot\eta_t)
=
\mathcal{L}(g_t,\dot g_t,\eta_t)
+
\frac{1}{2\sigma^2}\|g_t\dot\eta_t\|^2,
\]
with fixed boundary conditions for the initial and final 
images $n_0$
and $n_1$, with image $n_t = g_t\eta_t$ for template 
$\eta_t$ and $g_0 = \id_G$; thus only the images are constrained at the endpoints.

For the concrete metamorphosis example, the group $G$ of 
diffeomorphisms $\operatorname{Diff}(\mathcal{D})\ni g$ of 
the domain $\mathcal{D}$ is taken to act on the space of 
smooth maps (images) $V=\mathcal{F}(\mathcal{D})\ni\eta$ by 
the left action $g\eta: = \eta \circ g ^{-1}$ of $G$ on $V$. 
Therefore, the right action  \eqref{r-action} of $G$ on 
$G\times V $ is given in this case by $(g, \eta)h : = 
(g \circ h, \eta\circ h ) $ for $g, h \in 
\operatorname{Diff}(\mathcal{D}) $ and 
$\eta\in \mathcal{F}( \mathcal{D}) $. 
The reduced Lagrangians $\ell_{LP}(u_t,n_t,\dot n_t)$ and 
$\ell_M(u_t,n_t,\nu_t)$ are defined on the space 
$\mathfrak{g}\times V \times V$. In imaging applications, $u_t=\dot{g}_tg_t^{-1}$ is the velocity along the optimal path $g_t$ sought between two images; $n_t :=g_t\eta_t$ 
is the path in the image space; and $\nu_t:=
g_t \dot{\eta}_t$ is the image velocity.

From a visual point of view, image metamorphoses
are similar to what is usually called ``morphing'' in computer
graphics. The evolution of the image over time, 
$t\mapsto n_t$, is a
combination of deformations and image intensity variation. Algorithms
and experimental results 
for the solution of the boundary value problem (minimize the time-integrated Lagrangian
between two images) can be found in 
\cite{MiYo2001, GaYo2005}.

\medskip

From the general metamorphosis equations \eqref{M_left_right} (with the minus sign corresponding to the right action of 
$G$ on $G \times  V$) we obtain the dynamical system 
\begin{equation}
\label{eq:meta.2}
\left\{
\begin{array}{l}
\displaystyle
\frac{\partial}{\partial t} \frac{\delta \ell_M}{\delta u} + \ad^*_{u_t}
\frac{\delta \ell_M}{\delta u} 
+ \frac{\delta \ell_M}{\delta n} \diamond n_t
+ \frac{\delta \ell_M}{\delta \nu} \diamond \nu_t = 0
\,,\\
\\
\displaystyle
\frac{\partial}{\partial t} \frac{\delta \ell_M}{\delta \nu} - u_t \frac{\delta \ell_M}{\delta \nu} - \frac{\delta \ell_M}{\delta n}= 0
\,,\\
\\
\displaystyle
\dot n_t = \nu_t + u_tn_t \,,\\ \\
\displaystyle
\frac{\delta \ell_M}{\delta u}(1) + \frac{\delta \ell_M}{\delta \nu}(1) \diamond
n_1 = 0
\,,\\
\end{array}
\right.
\end{equation}    
where for $n\in V$, $a\in V^*$, and $u\in\mathfrak{X}(\mathcal{D})=\mathfrak{g}$, the infinitesimal actions and the diamond operators are given by
\begin{align*}
u n &= -\mathbf{d} n \cdot u\in V=\mathcal{F}(\mathcal{D})
\,,\\
u a&=\operatorname{div}(a u)\in 
V^*=\mathcal{F}(\mathcal{D})^\ast \cong \mathcal{F}(\mathcal{D})
\,,\\
n\diamond a &= -\, a\, \mathbf{d} n \in \mathfrak{g}^\ast= \Omega^1(\mathcal{D}).
\end{align*}
Even though we fixed the standard volume form on $\mathcal{D}
\subset \mathbb{R}^n$ so densities on $\mathcal{D}$ are identified with functions and one-form densities with one-forms, we recall that one should
think of $ua$ as a density and $n \diamond a $ as a one-form density.

In contrast to earlier sections, fixed endpoints at $t=1$ are not assumed in metamorphosis. This difference leads to the last equation in the system \eqref{eq:meta.2}. For details of the derivation of the system \eqref{eq:meta.2} and discussions of the regularity of its solutions, see \cite{HoTrYo2009}.

System \eqref{eq:meta.2} describes coadjoint motion
\begin{equation}
\frac{\partial}{\partial t} \bigg(\frac{\delta \ell_M}{\delta u} +
\frac{\delta \ell_M}{\delta \nu}\diamond n\bigg) 
+ \ad_{u_t}^*\bigg(\frac{\delta \ell_M}{\delta u} +
\frac{\delta \ell_M}{\delta \nu}\diamond n\bigg) = 0
\,,
\label{eqn:ad-star}
\end{equation}
or, equivalently, 
\begin{equation}
\frac{\partial}{\partial t} 
\Bigg( {\rm Ad}_{g_t}^*\bigg(\frac{\delta \ell_M}{\delta u} 
+
\frac{\delta \ell_M}{\delta \nu}\diamond n\bigg) \Bigg)
= 0
\,,
\label{eqn:Ad-star-cst}
\end{equation}
so that
\begin{equation}
\bigg(\frac{\delta \ell_M}{\delta u} 
+
\frac{\delta \ell_M}{\delta \nu}\diamond n \bigg) \bigg |_{t}
=
{\rm Ad}_{g_t^{-1}}^*\bigg(\frac{\delta \ell_M}{\delta u} 
+
\frac{\delta \ell_M}{\delta \nu}\diamond n\bigg) \bigg |_{t=0}
\,,
\label{coad-mot}
\end{equation}
for the coadjoint action of the Lie group $G$ on the dual of its Lie algebra $\mathfrak{g}$.

\subsubsection*{Hamiltonian formulation}
One passes from the Euler-Poincar\'e metamorphosis equations on the Lagrangian side
to their Lie--Poisson Hamiltonian formulation via the {\bfi Legendre
transformation}; see the presentation and general formulas at the end of Section \ref{section_HP_metamorphosis}.
The Legendre transformation of the reduced Lagrangian $\ell_M(u, n, \nu): \mathfrak{g} \times  V  \times V \rightarrow \mathbb{R}$ in its variables $ u $ and $\nu$ defines the Hamiltonian,
\begin{equation}
h(\mu, n, \beta)
 = \left\langle \mu, u \right\rangle  + \left\langle \beta , \nu \right\rangle
- \ell_M(u, n, \nu),
\label{legendre-xform}
\end{equation}
on $ \mathfrak{g} ^\ast \times  V  \times V ^\ast$, where 
\begin{equation}\label{mom-var-derivs}
\mu = \frac{\delta \ell_M}{\delta u}
\quad \text{and} \quad
\beta  = \frac{\delta \ell_M}{\delta \nu}
\end{equation}
are given by the Legendre transformation.
The variational derivatives of the Hamiltonian $h$ are
\begin{equation}\label{dual-var-derivs}
\frac{\delta h}{\delta \mu} 
=
u,
\quad
\frac{\delta h}{\delta \beta} 
=
 \nu ,\quad
\frac{\delta h}{\delta n} 
=
- \frac{\delta \ell_M}{\delta n}.
\end{equation}
Consequently, the Euler-Poincar\'e equations  
\eqref{eq:meta.2} for metamorphosis in the Eulerian
description imply the following equations, for the
Legendre-transformed variables, $(\mu,n, \beta)$, written as a matrix operation, symbolically as
{\small
\begin{equation} \label{LP-Ham-struct-symbol}
\frac{\partial}{\partial t}
\left[ \begin{array}{c} 
\mu  \\ n \\ \beta 
\end{array}\right]
= -
\left[ \begin{array}{ccc} 
\quad {\rm ad}^\ast_\Box\,\mu \quad  & 
-\Box \diamond n  \quad     &\beta \diamond \Box      
\\ 
-\Box\, n  &  0 & -1
\\
-\Box\, \beta&  1 & 0
\end{array} \right]
\left[ \begin{array}{c} 
\delta h/\delta\mu \\ 
\delta h/\delta n\\
\delta h/\delta \beta
\end{array}\right]
=:
\mathcal{B}
\left[ \begin{array}{c} 
\delta h/\delta\mu \\ 
\delta h/\delta n \\
\delta h/\delta \beta 
\end{array}\right],
\end{equation}
}
$\!\!$with boxes $\Box$ indicating where the substitutions occur. These equations can also be obtained from the system \eqref{M_Hamiltonian_A} (with minus sign chosen in $\mp$) by explicitly computing every term for this situation.
The Poisson bracket defined by the $L^2$ skew-symmetric Hamiltonian matrix $\mathcal{B}$ is given by
\begin{equation}
\big\{f,h \big\}( \mu, n, \beta)
=
\mathlarger{\int}
\left[ \begin{array}{c} 
\delta f/\delta\mu \\ 
\delta f/\delta n \\
\delta f/\delta \beta  
\end{array}\right]^T
\mathcal{B}
\left[ \begin{array}{c} 
\delta h/\delta\mu \\ 
\delta h/\delta n\\
\delta h/\delta \beta  
\end{array}\right]
{\rm d}x 
\,.
\end{equation}
The pair $(n, \beta)$ satisfies canonical Poisson-bracket relations. The other parts of the Poisson bracket are linear in the variables $(\mu, n, \beta)$. This linearity is the signature of the Lie-Poisson bracket on  the dual of the semidirect product Lie algebra of vector fields $\mathfrak{X}( \mathcal{D})$ acting on functions $\mathcal{F}(\mathcal{D}, W)$ and its dual $\mathcal{F}(\mathcal{D}, W^\ast)$  with a canonical cocycle between them. The semidirect product Lie algebra bracket on $\mathfrak{g} \times V \times V$ is 
\[
\left[(u, n, \nu),(\bar{u},\bar{n},\bar{\nu}) \right]
=
\left([u,\bar{u}], u\bar{n}-\bar{u}n , u\bar{\nu}-\bar{u}\nu\right).
\]

A similar Lie-Poisson bracket was found for complex fluids in \cite{Holm2002}. Ongoing work in this direction includes a Lagrange-Poincar\'e formulation of these equations (\cite{GBTr2010}). 

\section{Conclusions and outlook}\label{sec-ConclusionsOutlook}

This paper has begun the development of the family of dynamical systems associated with optimal control and optimization problems. The theory was developed in the context of many examples inspired by control theory and optimization, particularly in the new area of applications in imaging analysis of the theory of \emph{metamorphosis}, a means of optimally tracking the changes of shape necessary for registration of images of various types, or data structures, without requiring that the transformations of shape be diffeomorphisms. The main idea was to soften the exact dynamical constraint by replacing it with a quadratic penalty term. The resulting \emph{optimization dynamics} was studied by using  methods that originated in geometric mechanics. In particular, Lagrange-Poincar\'e reduction and its associated variational formulations were adapted to this sort of \emph{optimal inexact reduction}. This approach allowed us to obtain the equations of metamorphosis dynamics that are 
 naturally generated by the stationarity conditions, then study their properties from both the Lagrangian and Hamiltonian points of view. 

This geometric setup for optimization dynamics was illustrated in diverse examples in Section \ref{examples_section}.
Besides metamorphosis (\S\ref{optimage-reg}), these examples included optimally reduced versions of the heavy top (\S\ref{subsec-HT}), the double bracket equations (\S\ref{adj_repr}), the $N$-dimensional free rigid body (\S\ref{subsec-RB}),  the Euler equations for an inviscid ideal fluid both incompressible and compressible (\S\ref{subsec-EulerEqns}), and the $N$-dimensional Camassa-Holm equation (\S\ref{sec: EPDiff}).  For the one-dimensional Camassa-Holm equation the optimal reduction process produced its integrable Hamiltonian extension, the two-component Camassa-Holm equations in (\ref{CH2-motion-eqn}) and (\ref{CH2-continuity-eqn}). 

We plan to continue the investigation  of the relationships among problems
in imaging, optimal control, and geometric mechanics. In particular, we plan to continue developing the dynamical systems framework for designing and interpreting  methods of large deformation matching for image registration in computational anatomy.

\subsection*{Acknowledgements}
We are grateful to A. M. Bloch, M. Bruveris, P. Constantin, C. J. Cotter, D. C. P. Ellis, B. A. Khesin, J. E. Marsden, D. Meier,  A. Trouv\'e, F.-X. Vialard and L. Younes for many useful and pleasant conversations about these and related matters. The work by DDH was partially supported by a Wolfson Award from the Royal Society of London and an Advanced Grant from the European Research Council. FGB acknowledges the partial support of Swiss National Science Foundation grants 200020-117511 and of a Swiss National Science Foundation Postdoctoral Fellowship. TSR acknowledges the partial support of Swiss National Science Foundation grants 200020-117511 and 200020-126630.

{\footnotesize

\bibliographystyle{new}

}
\end{document}